\documentclass[a4paper,11pt]{article}

\usepackage{latexsym}
\usepackage{amsmath}
\usepackage[dvips]{graphics}

\newtheorem{theorem}{Theorem}
\newtheorem{lemma}[theorem]{Lemma}

\newtheorem{prop}[theorem]{Proposition}

\newtheorem{definition}[theorem]{Definition}

\begin{document}

\def\Cx{{\bf C}}
\def\R{{\bf R}}
\def\Z{{\bf Z}}
\def\x{{\bf x}}
\def\o{{\bf o}}
\def\del{\partial}
\def\Lap{\bigtriangleup}
\def\Tr{{\rm Tr}}
\def\^{\wedge}
\def\goinf{\rightarrow\infty}
\def\goes{\rightarrow}
\def\bm{\boldmath}
\def\-{{-1}}
\def\inv{^{-1}}
\def\sqr{^{1/2}}
\def\isqr{^{-1/2}}

\def\reff#1{(\ref{#1})}
\def\vb#1{{\partial \over \partial #1}} 
\def\vbrow#1{{\partial/\partial #1}} 
\def\Del#1#2{{\partial #1 \over \partial #2}}
\def\Dell#1#2{{\partial^2 #1 \over \partial {#2}^2}}
\def\Dif#1#2{{d #1 \over d #2}}
\def\Lie#1{ {\cal L}_{#1} }
\def\diag#1{{\rm diag}(#1)}
\def\abs#1{\left | #1 \right |}
\def\rcp#1{{1\over #1}}
\def\paren#1{\left( #1 \right)}
\def\brace#1{\left\{ #1 \right\}}
\def\bra#1{\left[ #1 \right]}
\def\angl#1{\left\langle #1 \right\rangle}
\def\vector#1#2#3{\paren{\begin{array}{c} #1 \\ #2 \\ #3 \end{array}}}
\def\svector#1#2{\paren{\begin{array}{c} #1 \\ #2 \end{array}}}

\def\GL#1{{\rm GL}(#1)}
\def\SL#1{{\rm SL}(#1)}
\def\PSL#1{{\rm PSL}(#1)}
\def\O#1{{\rm O}(#1)}
\def\SO#1{{\rm SO}(#1)}
\def\IO#1{{\rm IO}(#1)}
\def\ISO#1{{\rm ISO}(#1)}
\def\U#1{{\rm U}(#1)}
\def\SU#1{{\rm SU}(#1)}

\def\diffeos{diffeomorphisms}
\def\diffeo{diffeomorphism}
\def\Teich{{Teichm\"{u}ller}}
\def\Poin{{Poincar\'{e}}}

\def\Gam{\mbox{$\Gamma$}}
\def\d{{d}}
\def\VII#1{\mbox{VII${}_{#1}$}}
\def\VI#1{\mbox{VI${}_{#1}$}}
\def\Nil{{\rm Nil}}
\def\Sol{{\rm Sol}}
\def\Isom{{\mathrm{Isom}}}

\def\hh{{h}}
\def\ggg{{\rm g}}
\def\uh#1#2{\hh^{#1#2}}
\def\dh#1#2{\hh_{#1#2}}
\def\mh#1#2{\hh^{#1}{}_{#2}}
\def\ug#1#2{\ggg^{#1#2}}
\def\dg#1#2{\ggg_{#1#2}}
\def\uug#1#2{\tilde{\ggg}^{#1#2}}
\def\udg#1#2{\tilde{\ggg}_{#1#2}}
\def\udh#1#2{\tilde{\hh}_{#1#2}}

\def\c#1{\chi_{#1}}
\def\cc#1#2{\chi_{#1}{}^{#2}}
\def\uc#1{\chi^{#1}}
\def\s#1{\sigma^{#1}}
\def\ss#1#2{\sigma^{#1}{}_{#2}}
\def\om#1#2#3{\omega_{#1#2}{}^{#3}}
\def\omd#1#2#3{\omega_{#1#2#3}}
\def\CC{C}
\def\C#1#2#3{\CC^{#1}{}_{#2#3}}
\def\Sig#1{\Sigma^{#1}}
\def\Sigg#1#2{\Sigma^{#1}{}_{#2}}
\def\Chi#1{X_{#1}}
\def\Chii#1#2{X_{#1}{}^{#2}}
\def\dug#1#2{g_{#1}{}^{#2}}
\def\dgm#1#2{\gamma_{#1#2}}
\def\ugm#1#2{\gamma^{#1#2}}
\def\covsset{\mathcal{S}_2}
\def\consset{\mathcal{S}^2}
\def\X{X}
\def\stdh{h_{\mathrm{s}}{}}
\def\dd#1#2{\frac{\del^2{#1}}{\del {#2}^2}}

\def\wa{\!\!\!\!&=&\!\!\!\!}
\def\wb{\!\!\!\!&\equiv &\!\!\!\!}
\def\ws{&=}
\def\nd{\noindent}
\def\D{{\mathcal{D}}}
\def\tr{\mathrm{tr}}
\def\nnpar{\nonumber \\}
\def\xnnpar{\\}

\def\proofmark{\textit{Proof}: }
\def\defmark{{\bf Definition}\hspace{1em}}
\def\defsmark{{\bf Definitions}\hspace{1em}}
\def\notemark{{\bf Note}\hspace{1em}}
\def\remarkmark{{\bf Remark}\hspace{1em}}
\def\conventionmark{{\bf Convention}\hspace{1em}}
\def\endofproofmark{\hfill\rule{.5em}{.5em}}

\def\hD{{\hat D}}
\def\E#1#2#3{(E_{#1})_{#2#3}}
\def\uE#1#2#3{(E_{#1})^{#2#3}}
\def\dpi#1{\pi_{#1}}
\def\dq#1{q_{#1}}
\def\lmx{\sqrt{\frac{\lambda^2+2}{2}}}

\def\title#1{{\Large {#1}}}
\def\author#1{\textsc{#1}}
\def\address#1{\textit{#1}}
\def\abstract#1{{\bf Abstract}\par\medskip \parbox{.9\linewidth}{#1}}

\def\pubnum{Preprint}

\hfill
\parbox{6cm}{{\pubnum} \par Oct. 2002}
\parbox{6cm}{}
\par
\vspace{1.5cm}
\begin{center}
\Large
Perturbations of Spatially Closed Bianchi III Spacetimes
\end{center}

\author{Masayuki Tanimoto ${}^*$ \footnote{JSPS Research Fellow.},
  Vincent Moncrief ${}^\dagger$,
  Katsuhito Yasuno ${}^\ddagger$}

\medskip

\address{${}^*$ 
  Department of Physics,
  Yale University,
  New Haven, CT 06511--8499, USA}

\address{${}^\dagger$
  Departments of Physics and Mathematics,
  Yale University, 
  New Haven, CT 06511--8499, USA}

\address{${}^\ddagger$ 
  Department of Physics,
  Tokyo Institute of Technology, 
  2--12--1 Oh-okayama, Meguro-ku,
  Tokyo 152--8551, Japan}
\vskip 1 cm

\abstract{Motivated by the recent interest in dynamical properties of
  topologically nontrivial spacetimes, we study linear perturbations
  of spatially closed Bianchi III vacuum spacetimes, whose spatial
  topology is the direct product of a higher genus surface and the
  circle. We first develop necessary mode functions, vectors, and
  tensors, and then perform separations of (perturbation)
  variables. The perturbation equations decouple in a way that is
  similar to but a generalization of those of the Regge--Wheeler
  spherically symmetric case. We further achieve a decoupling of each
  set of perturbation equations into gauge-dependent and independent
  parts, by which we obtain wave equations for the gauge-invariant
  variables. We then discuss choices of gauge and stability
  properties. Details of the compactification of Bianchi III manifolds
  and spacetimes are presented in an appendix. In the other appendices
  we study scalar field and electromagnetic equations on the same
  background to compare asymptotic properties.}

\def\tm{t_-}
\def\tp{t_+}
\def\TT{\textsc{tt}}

\section{Introduction}

Astronomical observations strongly support the picture of an isotropic
and homogeneous universe which is expanding at such a rate, relative
to its energy density, that it will continue to expand
forever. Isotropy and homogeneity must be interpreted as valid only
when the universe is viewed in a coarse-grained sense --- on a
sufficiently large scale that fine details such as clusters of
galaxies are blended into a uniform background. For this reason most
studies of cosmological perturbations have begun with a homogeneous
and isotropic background and thus with a
Friedman-Robertson-Walker (FRW) metric to perturb.

There are however several logical limitations to this point of
view. First of all since one cannot really observe the entire universe
it is perfectly conceivable that the actual universe is only locally
but not globally homogeneous and isotropic. For example the negatively
curved $k=-1$ FRW models can be spatially compactified in infinitely
many ways to yield spacetimes which, though indistinguishable from
the conventional FRW models on sufficiently small scales (which might
however exceed the observable universe in spatial extent)
are nevertheless inhomogeneous in the global sense. In particular, at
a given instant of cosmic time, the length of the shortest closed
spacelike geodesic through a given point (roughly the shortest
distance around the universe from there) would depend upon the
location of that point in a compactified model. This is a clear
(though perhaps unobservable) violation of global
homogeneity. Furthermore since astronomical observations are limited
to (past) lightlike directions for any given observer, any
hypothetical breakdown of global homogeneity might be unobservable,
even in principle,
at a given epoch of cosmic time but became observable at a later
epoch.

Both classes of ``open'' FRW models (the $k=-1$ hyperbolic and the
$k=0$ flat models) are amenable to spatial compactification but the
$k=0$ models are somewhat unstable towards perturbation into locally
homogeneous but non-isotropic, Kasner-like behavior in which the
effective Hubble constant would be directionally dependent and thus
(in principle at least) observable from any vantage point in the
universe. By contrast, the compactified $k=-1$ models are known
to be stable in the direction of cosmological expansion \cite{FM02b}
and thus do not seem to require an excessive ``fine-tuning'' of the
initial conditions (or perhaps an inflationary scenario) to protect
them from the development of non-isotropic expansion.

An even more radical departure from the conventional FRW point of view
is conceivable however and even suggested by recent results and
conjectures on possible 3-manifold topologies and recent insights into
the dynamics of expanding cosmological models implied by Einstein's
equations. Aside from the compact hyperbolic manifolds
mentioned above which can support $k=-1$ locally (but not globally)
homogeneous and isotropic metrics there is a more general class of
prime compact 3-manifolds that one can form by ``glueing'' finite
volume hyperbolic manifolds across certain special surfaces (called
``incompressible 2-tori'') to so-called graph manifolds of infinite
fundamental group. This set of graph manifolds includes for example
the trivial and non-trivial circle bundles over compact higher genus
surfaces. This glueing construction (which is distinct from the
connected sum operation in which the glueing occurs across embedded
spheres instead of tori)
leads to a class of manifolds which no longer admit even an everywhere
locally homogeneous and isotropic metric since the graph manifolds
pieces are incompatible with the extension of such a structure from
the hyperbolic pieces.

At first sight it would seem that such manifolds are ruled out, as
models for the actual universe, by the most basic of astronomical
observations. However some recent studies of the so-called reduced
Hamiltonian for Einstein's equations and their implications for
cosmological dynamics suggest a different point of view. They support
the scenario that,
under cosmological expansion, the spatial metric (after rescaling by
the cube of the mean curvature to take out the overall effect of
expansion) converges to the hyperbolic metric on the hyperbolic
components of the manifold but collapses the graph manifold components
to zero rescaled volume asymptotically. This picture is suggested by
the known behavior of the reduced Hamiltonian (and its quasi-local
variant) which monotonically decreases for every non-selfsimilar
solution of the field equations and whose value at any instant (in a
constant mean curvature or CMC time slicing)
is the rescaled volume at that instant. The infimum of the reduced
Hamiltonian is known to be zero for a pure graph manifold of the type
we are considering and the asymptotic vanishing of this Hamiltonian
corresponds to the Cheeger--Gromov style collapse of the rescaled
volume mentioned above. This total ``collapse of the graphs'' is seen
to occur for the explicitly known, spatially compactifiable Bianchi
vacuum spacetimes of types I, II, III, VI${}_0$, and VIII \cite{FM02a}
which are all graph manifolds of the type we want. By contrast the
infimum of the reduced Hamiltonian for compact hyperbolic manifolds is
conjectured
(through its direct connection to the topological invariant known as
the $\sigma$-constant) to be achieved by convergence to the hyperbolic
metric and it is known that this canonical metric provides at least an
isolated local minimum to the reduced Hamiltonian
\cite{FM99,FM00a,FM00b,FM01}.

Thus one is led to consider a model of the universe in which the
spatial topology is a composite of hyperbolic and graph manifold
pieces but for which the Einsteinien dynamics predicts that the
spatial volume is asymptotically dominated by the (locally homogeneous
and isotropic) hyperbolic components. Whether a physically plausible
cosmological model could be developed
along these lines (e.g., whether structure formation could be largely
confined to the hyperbolic components) is not known but, from the
mathematical point of view, it is worth noting that such manifolds
which are composite with respect to the torus decomposition described
above can still be prime (i.e., indecomposible into more fundamental
prime manifolds through the sphere decomposition). As such they are
still among the most basic building blocks of compact 3-manifolds and
indeed, taken together with the special case of pure compact
hyperbolic manifolds, form perhaps the largest distinguished
subset of such prime manifolds. Together with the (pure) graph
manifolds of infinite fundamental group they are conjectured to
exhaust the family of so-called prime, $K(\pi,1)$ manifolds. If
current conjectures (including the still unproven Poincar\'{e}
conjecture) on 3-manifold topology are correct the most general
compact, connected, orientable 3-manifold is constructable as the
connect sum of a finite number of $K(\pi,1)$ factors, a finite number
of $S^2\times S^1$ handles and a finite number of spherical space
forms (i.e., quotients of $S^3$ by finite subgroups of $\SO4$ which
act freely and properly discontinuously on $S^3$).

Of course there is no known exact solution of Einstein's equations on
any of these more exotic torus decomposible or sphere decomposible
manifolds. However there are known solutions available for many of the
constituents of such composite manifolds. In particular, there are
known Bianchi solutions, both vacuum and fluid filled, for most of the
graph manifold examples mentioned above. Since these could conceivably
play an important role in the structure and evolution of a plausible
universe model we are motivated to study their stability properties
for much the same reason that people have studied the stability
properties of the FRW models.

Going beyond the FRW models and the Bianchi I, V and IX families which
include such models the simplest manifolds to consider would seem to
be those of the Bianchi III (or Thurston's $H^2\times\R$) type which
can be compactified to trivial $S^1$-bundles over higher genus
surfaces. Starting with a family of explicitly known vacuum solutions
we here develop the tensor harmonics needed for the separation of the
spatial coordinates on such backgrounds and carry out this separation
explicitly for the linearized Einstein equations as well as for the
Maxwell and massless scalar wave equations. These harmonics
have much in common with those developed by Regge and Wheeler
\cite{RW} for the study of spherically symmetric backgrounds (such as
Schwarzschild or Nariai-Kantowski-Sachs solutions) but with
eigenfunctions of the scalar Laplacian on the higher genus surface
playing the role of the ordinary spherical harmonics on $S^2$ in the
Regge--Wheeler analysis. For completeness though we develop also the
harmonics based on the transverse-traceless tensors admitted by the
higher genus surfaces as well as those based on the (Hodge-) harmonic
one-forms allowed by such surfaces. Since the first Betti number
of $S^2$ is vanishing and since $S^2$ has a trivial \Teich\ space
these latter classes of tensor harmonics had no analogue in the
Regge--Wheeler analysis. Furthermore even the scalar harmonics vary
with the \Teich\ parameters determining the higher genus surface and
thus depend upon the particular compactification of $H^2$ chosen in
the construction. As we shall see the asymptotic decay properties of
the gravitational, electromagnetic or scalar field perturbations can
vary significantly with the \Teich\ parameters of the chosen compact
surface. 

To obtain a further simplification of the linearized field equations,
we also carry out a decomposition of the metric perturbations into
gauge invariant, gauge dependent and constrained variables, the last
two subsets being essentially canonically conjugate to each other. We
then derive the decoupled wave equations satisfied by the various
gauge invariant perturbation modes and discuss the behavior of their
solutions on the chosen backgrounds. By choosing a particular gauge
for the remaining, unconstrained variables we show how the full
spacetime metrics of the perturbed models can be expressed in terms of
the solutions of the gauge invariant wave equations.

Our analysis lays the groundwork for the study of a number of
physically interesting questions about these models. For example do
the perturbed models decay in some natural, invariant sense towards
local homogeneity in the asymptotic regime or do the perturbations get
driven away from asymptotic homogeneity by the Cheeger--Gromov
``collapse of graph'' effect of the background? Does the reduced
Hamiltonian for the perturbed solutions asymptotically achieve the
infimum attained by the background solutions (corresponding to the
vanishing of the $\sigma$-constant for these manifolds) or does it
stay bounded away
from its infimum for a generic perturbations? Can the methods
developed here be generalized to treat the perturbations of Bianchi II
(Thurston Nil), Bianchi VI${}_0$ (Thurston Sol), and Bianchi VIII
(Thurston $\widetilde\SL{2,\R}$) backgrounds and, if so, can one
answer for those models the same questions raised above for the
Bianchi III (Thurston $H^2\times\R$) models studied here? Finally can
one recover the essential features of the asymptotic behavior of these
models from energy arguments for the linearized field equations, which
avoid the need for an explicit separation of variables and, if so, can
one
develop these into energy arguments for the fully non-linear equations
in the case of sufficiently small perturbations? These are some of the
issues we hope to study in the future.

The plan of this paper is as follows. In the next section we briefly
describe the background spacetime we consider and set up the
background (canonical) variables. The background spacetime is a
generalization of the usual Bianchi III model that is spatially
compactified. Details of the compactification are discussed in
Appendix \ref{apx:compact}. In Sec.\ref{sec:he} we construct mode
functions, vectors, and tensors on the spatial manifold. Using these
in the subsequent section we perform separations of (perturbation)
variables and write down the (separated) constraint equations. It is
claimed that we have four kinds of perturbations. The subsequent four
sections are devoted to detailed analysis of them, which includes
decoupling of each set of perturbation equations into gauge-dependent
and independent parts, and deriving the wave equation for the
gauge-invariant variable. In Sects.\ \ref{sec:prefgauge} and
\ref{sec:evendiffeo} we discuss choices of gauge.  The last section is
devoted to a summary with some discussions of the stability
properties. In Appendices \ref{apx:scalar} and \ref{apx:elemag} we
study scalar field and electromagnetic equations on the same
background. All the appendices deal with subjects that have not
appeared elsewhere and are substantial parts of this paper.

We employ the abstract index notation \cite{Wa} and use leading Latin
letters $a,b,\cdots$ to denote abstract indices for vectors and
tensors. When useful however we also write a vector or tensor without
abstract indices, as long as no confusion occurs. Greek indices
$\mu,\nu,\cdots$ are used to denote spacetime coordinates, running $0$
to $3$. Middle Latin letters $i,j,\cdots$ are used for general
purposes, e.g. for equations, variables, or generators for infinite
groups, as well as for spatial coordinates, running $1$ to
$3$. Capital Latin letters $I,J,\cdots$ are used to label an invariant
basis for a Bianchi group, and run from $1$ to $3$.

\section{The background}

Our spatial manifold $M$ is assumed to be the direct product $M\simeq
\Sigma_g\times S^1$, where $\Sigma_g$ is the closed surface with genus
$g\geq2$, and $S^1$ the circle. The geometry is of Bianchi type III
(or $H^2\times\R$ in Thurston's terminology). Let $\tilde M\simeq
\R^3$ be the universal cover of $M$. We can define on $\tilde M$ the
invariant 1-forms $\ss Ia$ ($I=1\sim3)$ of Bianchi III.  Also, we can
define their duals (invariant vectors) $\cc Ia$: $\ss Ia\cc
Ja=\delta^I_J$. In local coordinates they are expressed as
\begin{equation}
  \s 1=dx/y,\quad   \s 2=dy/y,\quad   \s 3=dz,
\end{equation}
and
\begin{equation}
  \c 1=y\del_x,\quad   \c 2=y\del_y,\quad   \c 3=\del_z.
\end{equation}
Since $M$ is compact, the spatial metric $q_{ab}$ on $M$ should be
locally rotationally symmetric (LRS) to have $H^2$ symmetry (otherwise
we cannot make $\tilde M$ compact by discrete actions of
isometry). The background spatial metric is therefore:
\begin{equation}
  \label{eq:defq}
  q_{ab}= q_1\dh ab+q_2l_{ab}, \quad \mbox{(the background metric)}
\end{equation}
where
\begin{eqnarray}
  \label{eq:defh}
  \dh ab\wa \ss 1a\ss 1b+\ss 2a\ss 2b \quad
  \mbox{(the standard metric on $H^2$)}, \\
  l_{ab}\wa \ss 3a\ss3b \quad
  \mbox{(the standard metric on $\R$)},
\end{eqnarray}
and the scale factors $q_i$ $(i=1,2)$ are constants on $M$, and will
be functions of time $t$. We may view the metric $q_{ab}$ as one on
the closed manifold $M$, as well as one on the universal cover $\tilde
M$. We understand that an appropriate set of identifications of points
is specified on $\tilde M$ when $q_{ab}$ is viewed as a metric on the
closed manifold $M$. Similarly, we may view $\dh ab$ as a metric on
$\Sigma_g$ as well as one on $H^2$, and view $l_{ab}$ as a metric on
$S^1$ as well as one on $\R$.  (For details of the compactification,
see Appendix \ref{apx:compact}, where however we explicitly
distinguish metrics on universal covers from ones on compactified
manifolds by writing tilde on the metrics of the universal covers.)

We should note that $\s1$ and $\s2$ (or $\c1$ and $\c2$)
\textit{cannot} be defined on the compact quotient $M$ by themselves,
since they are not invariant under the isometries of $H^2$. However,
they will always appear in a combined form as in the metric
\reff{eq:defh} or in the area 2-form $d\mu_h=\s1\^\s2$, which are both
well defined on $M$. On the other hand, $\s3$ and $\c3$ are well
defined on both $M$ and $\tilde M$ by themselves. Note that $\c3$ is
the natural $S^1$-fiber generator when viewing $M$ as a $S^1$-bundle
over $\Sigma_g$. We regard $\c3=\del_z$ not only as a vector but as an
explicit differential operator acting on the functions on $M$ (or
$S^1$), i.e.,
\begin{equation}
 \c3:\mathcal{C}^\infty\goes \mathcal{C}^\infty.
\end{equation}
We also note that in
our convention, the scalar curvature for the hyperbolic metric
$h_{ab}$ gives $R_h=-2$ (in particular, not normalized to give $-1$).

We think of $q_1$ and $q_2$ as the configuration variables in the
background phase space. To find the conjugate momentum $\pi^{ab}$, let
$\mu_{q_0}$ be the determinant of the standard (i.e.,
time-independent) metric $q_{0ab}=\dh ab+l_{ab}$ with respect to the
coordinates $(x,y,z)$. For later convenience we also define
three-forms $\d\mu_{q_0}$ and $\d^3x$ by
\begin{equation}
  \d\mu_{q_0}\equiv \mu_{q_0}\d^3x= \s1\^\s2\^\s3.
\end{equation}
The determinant $\mu_q$ of the metric $q_{ab}$ is also defined
similarly. The two determinants are related with
$\mu_q=q_1\sqrt{q_2}\mu_{q_0}$.  Now, it is easy to see that the
conjugate momentum can be written in the form
\begin{equation}
  \pi^{ab}= \mu_{q_0}(\frac{\pi_1}2\uh ab+\pi_2l^{ab}),
  \quad \mbox{(the conjugate momentum)}
\end{equation}
where $\uh ab= \cc 1a\cc 1b+\cc 2a\cc 2b$ and $l^{ab}= \cc 3a\cc 3b$
are the the inverses of $\dh ab$ and $l_{ab}$, respectively.  In fact,
the functions of time $\pi_1$ and $\pi_2$ are chosen so that they are
canonically conjugate to $q_1$ and $q_2$, as seen from
\begin{equation}
  \Theta_0\equiv \rcp{C_M} \int_M  \d^3x
  \pi^{ab}\dot q_{ab}= \pi_1\dot q_1+\pi_2 \dot q_2,
\end{equation}
where $C_M\equiv \int\d\mu_{q_0}$ is a constant. 

We can calculate the Hamiltonian constraint for the background
canonical variables $q_i$ and $\pi_i$ ($i=1,2$), which is
\begin{eqnarray}
  \label{eq:backH}
  \mathcal{H}(q,\pi)\wb \mu_q^{-1}(\pi^{ab}\pi_{ab}-\rcp2
  (\tr\pi)^2)-\mu_q R_q \nonumber \\
  \wa \mu_q(-\pi_1\pi_2 q_1\inv +\rcp2 (\pi_2)^2q_1^{-2}q_2+2q_1\inv)
  \\
  &\approx& \!\!\!\! 0. \nonumber
\end{eqnarray}
On the other hand the momentum constraint is found to be trivial: $
\mathcal{H}^a(q,\pi)\equiv -2D_b\pi^{ab}=0$.  The background
Hamiltonian $H_0$ is therefore given by\footnote{It is well known that
  one cannot obtain a correct reduced Hamiltonian by the
  straightforward procedure from Bianchi class B models including
  Bianchi III. This is however not the case for the present LRS case,
  which is spatially compactifiable. (In general, any spatially
  compactifiable Bianchi models have natural Hamiltonians.)}
\begin{eqnarray}
  H_0\wa \rcp{C_M} \int_M \d^3x N\mathcal{H} \nonumber \\
  \wa N\sqrt{q_2}
  (-\pi_1\pi_2 +\rcp2 (\pi_2)^2q_1^{-1}q_2+2),
\end{eqnarray}
where $N$ is the lapse function.
With this Hamiltonian we obtain the following Einstein equations:
\begin{equation}
\label{eq:backeom}
\begin{split}
  \dot q_1 =& \frac{\del H_0}{\del \pi_1} \\
  =& -N\sqrt{q_2}\pi_2, \\
  \dot \pi_1 =& -\frac{\del H_0}{\del q_1} \\
  =& \rcp2N\pi_2^2{q_2}^{3/2}q_1^{-2}, \\
  \dot q_2 =& -N\sqrt{q_2}(\pi_1-\pi_2q_1\inv q_2), \\
  \dot \pi_2 =& \rcp4N{q_2}^{-1/2}
  (2(\pi_1\pi_2-2)-3\pi_2^2q_1\inv q_2),
\end{split}
\end{equation}
where $\dot{}\equiv d/dt$.
We will refer to $q_i$ and $\pi_i$ $(i=1,2)$ (and $N$) as the
\textit{background variables}, and think of them as given functions of
time which satisfy the above evolution equations and the constraint
$\mathcal{H}=0$.

The general vacuum solution of Bianchi III is given, up to isometry,
by a one parameter family of spacetime metrics (\S 26 of \cite{Pe}),
for which we write
\begin{equation}
  g^{(0)}_{ab}=-({t_+}/{t_-}) (dt)_a(dt)_b+t_+^2 \dh ab
  +({t_-}/{t_+}) l_{ab},
\end{equation}
where
\begin{equation}
  \begin{split}
  \tp &\equiv t+k, \\
  \tm &\equiv t-k,
  \end{split}
\end{equation}
and $k$ is a real parameter. This metric corresponds to the following
solution for the background variables:
\begin{equation}
  \label{eq:BIIIsol}
  q_1= t_+^2,\;
  q_2= {t_-}t_+\inv,\;
  \pi_1={-2t}\,{t_+^{-2}},\;
  \pi_2=-2\,t_+,\;
  N=q_2^{-1/2}.
\end{equation}

\section{Harmonic expansion of the perturbation variables}
\label{sec:he}

Since our manifold $(M,q_{ab})$ is naturally a trivial $S^1$-fiber
bundle $\Sigma_g\times S^1$ any function on $M$ may be expanded by the
products of mode functions on $(\Sigma_g,\dh ab)$ and
$(S^1,l_{ab})$. This is in fact the case when the natural fiber and
base are \textit{orthogonal} to each other. (See Appendix
\ref{apx:compact} for details of our background solution.) We only
consider such backgrounds, and call them \textit{spatially closed
  orthogonal Bianchi III backgrounds}.  \footnote{The orthogonal form
  universal cover metric \reff{eq:defq} does \textit{not}
  automatically imply this orthogonality of $(M,q_{ab})$. See Appendix
  \ref{apx:compact}.}  In the following therefore we first construct
each of the sets of mode quantities, i.e., functions, vectors, and
symmetric tensors,  for $S^1$ and for $\Sigma_g$, and then construct
those for $M\simeq S^1\times\Sigma_g$ by taking products of them.

\textbf{Mode Quantities on $S^1$}

The Laplacian on $(S^1,l_{ab})$ is $(\c3)^2=\del^2/\del z^2$. The mode
functions are therefore, taking the period of $z$-coordinate as
$2\pi$, simply given by $e^{imz}$, where the eigenvalue $m$ takes
values
\begin{equation}
 m=0,\pm1,\cdots.
\end{equation}
We refer to $m$ as the \textit{fiber eigenvalue}. It is
convenient to redefine the eigenfunctions to be explicitly real so
that we can directly calculate perturbation Hamiltonians which are
second order functions.  This can be done by taking the real or
imaginary part of the complex eigenfunction. Let us define pairs
of real eigenfunctions $c_m$ and $\bar c_m$ by the relations
\begin{equation}
  \chi_3c_m=-m\bar c_m,\quad \chi_3\bar c_m=m c_m.
\end{equation}
We assume that these functions are normalized so that the square
integrals on $(S^1,l_{ab})$ give unity:
\begin{equation}
  \int_{S^1} (c_m)^2 dz=\int_{S^1} (\bar c_m)^2 dz=1.
\end{equation}
(For example, we can take $c_m=(1/\sqrt\pi)\cos mz$, $\bar
c_m=(1/\sqrt\pi)\sin mz$.)  Mode vectors and symmetric tensors
can be constructed as $c_m \ss3a$, $\bar c_m \ss3a$, $c_ml_{ab}$, and
$\bar c_ml_{ab}$. 

\textbf{Mode Quantities on $\Sigma_g$}

The mode functions on the hyperbolic surface $(\Sigma_g,\dh ab)$ must
satisfy
\begin{equation}
  \Lap_h\hat S_\lambda=-\lambda^2\hat S_\lambda,
\end{equation}
where $\Lap_h$ is the Laplacian with respect to $\dh ab$, and
$-\lambda^2$ is its eigenvalue. We will refer to $\lambda$ as the
\textit{base eigenvalue}. (For simplicity we omit the subscript
$\lambda$ in $\hat S_\lambda$ from now on unless it is necessary.) We
assume that $\hat S$ is real and normalized so that the square
integral on $(\Sigma_g,\dh ab)$ gives unity, i.e.,
\begin{equation}
  \int_{\Sigma_g}\d\mu_h \hat S^2=1,
\end{equation}
where $\d\mu_h=\s1\^\s2$ is the area two-form associated with the
metric $\dh ab$.

By analogy with the Regge--Wheeler spherically symmetric case
\cite{RW}, we make ``even'' and ``odd'' vectors on the $\Sigma_g$
from the scalar $\hat S$ when $\lambda>0$:
\begin{eqnarray}
  \hat S_a\wa
  \rcp\lambda\hD_a \hat S,
  \quad \mbox{(the ``even'' vector on $\Sigma_g$)} \\
  \hat V_a\wa 
  \rcp\lambda\epsilon_a{}^b\hD_b \hat S,
  \quad \mbox{(the ``odd'' vector on $\Sigma_g$)}
\end{eqnarray}
where $\hD_a$ and $\epsilon_{ab}=\ss1{[a}\ss2{b]}$ are, respectively,
the covariant derivative operator and volume form associated with the
metric $\dh ab$.  We always write hats ( $\hat{}$ ) on quantities on
$\Sigma_g$ (except $\dh ab$ and $\epsilon_{ab}$), and raise and lower
indices of hatted quantities (and $\epsilon_{ab}$) by $\uh ab$ and
$\dh ab$, so, e.g., $\epsilon_a{}^b\equiv\epsilon_{ac}\uh cb$. The
coefficients $1/\lambda$ are determined so that the square integrals
give unity, i.e.,
\begin{equation}
  \int_{\Sigma_g} \d\mu_h \uh ab\hat S_a\hat S_b=1,
\end{equation}
and the same for $\hat V_a$. When $\lambda=0$, we have $\hat S_a=\hat
V_a=0$, since $\hat S=\mathrm{constant}$.

\def\hv{U}
Other than these even and odd vectors we need to consider the
harmonic vectors $\hat \hv_a$:
\begin{equation}
  \label{eq:LB}
 \Lap_\mathrm{LB} \hat \hv_a=0,
\end{equation}
where $\Lap_\mathrm{LB}=d\delta+\delta d$ is the Laplace-Beltrami
operator. This condition is equivalent, in the notation of standard
differential geometry, to $d\hat\hv=\delta\hat\hv=0$ when the one-form
$\hat\hv=\hat\hv_a$ is on a closed surface, and they can be, in our
language using the covariant derivatives, expressed as
\begin{equation}
  \label{eq:LBcov}
  \hat D_{[a}\hat \hv_{b]}=0,\quad \hat D^a\hat \hv_a=0.
\end{equation}
(This is not the same as $\Lap_h\hat \hv_a=0$.)  It is a special case
of the Hodge decompositions that the space $A^1(\hat M)$ of one-forms on
a closed Riemannian surface $\hat M$ can be decomposed into the direct
sum as $A^1(\hat M)=H^1(\hat M)\oplus dA^0(\hat M)\oplus \delta
A^2(\hat M)$ with respect to the standard $L^2$-norm, where $H^1(\hat
M)$ is the space of harmonic forms, $dA^0(\hat M)$ the space of exact
forms, and $\delta A^2(\hat M)$ the space of dual exact forms. It can
be easily checked that the spaces of even and odd vectors on
$\Sigma_g$ can be identified with, respectively, $dA^0(\Sigma_g)$ and
$\delta A^2(\Sigma_g)$. Hence the harmonic vectors definitely complete
the possible (smooth) vectors on $\Sigma_g$. Recall that harmonic
forms are always closed ($d\hat\hv=0$). From Hodge and de Rham's
theorems we know that $H^1(\hat M)$ is isomorphic to the real-valued
cohomology group $H^1(\hat M,\R)$. Regge and Wheeler did not have to
consider harmonic vectors in the spherically symmetric case, because
the cohomology group $H^1(S^2,\R)$ of the sphere is trivial. In the
present hyperbolic case, however, $H^1(\Sigma_g,\R)$ is nontrivial and
given by $\R^{2g}$, so we have $2g$ independent harmonic vectors on
$\Sigma_g$. Let us choose these vectors so that they are
$L^2$-orthonormal to each other (using the method
of Schmidt, if necessary):
\begin{equation}
  \int_{\Sigma_g}\d\mu_h \uh ab\hat\hv_a^{(r)}\hat\hv_b^{(r')}
  =\delta_{rr'},
\end{equation}
where $r,r'=1,\cdots,2g$. We call parameter $r$ \textit{the cohomology
  label}. All the properties we need for the harmonic vectors can be
deduced from the definition \reff{eq:LBcov}, which does not depend on
$r$. In this sense $r$ can be viewed as a label that distinguishes
``degenerated states''.  We omit writing these labels when we do not
need to specify them explicitly.

Some of the key relations of the three vectors, which we can easily
confirm from the definition, are:
\begin{equation}
  \label{eq:keyrelforvecs}
  \Lap_h \hat Y_a=
  - (\mathrm{eig}_{\Lap_h})\,\hat Y_a,\quad
  \mathrm{eig}_{\Lap_h}\equiv
  \begin{cases}
    \lambda^2+1, & (\hat Y_a=\hat S_a) \\
    \lambda^2+1, & (\hat Y_a=\hat V_a) \\
    1, & (\hat Y_a=\hat \hv_a)
  \end{cases}
\end{equation}
and
\begin{equation}
  \hat D^b\hat D_a \hat Y_b= 
  - (\mathrm{eig}_{\hat D \hat D})\,\hat Y_a,\quad
  \mathrm{eig}_{\hat D \hat D}\equiv
  \begin{cases}
    \lambda^2+1, & (\hat Y_a=\hat S_a) \\
    1, & (\hat Y_a=\hat V_a) \\
    1. & (\hat Y_a=\hat \hv_a)
  \end{cases}
\end{equation}

Let us consider symmetric tensors on $\Sigma_g$. They can be split
into scalar, vector, and tensor parts. The scalar part is the part
that is constructed from the scalar mode functions $\hat S$, which
part is subdivided into three parts. One of them is the trace part
given by $\hat S\dh ab$, while the others are the (traceless) even and
odd parts which are given, when $\lambda>0$, by
\begin{eqnarray}
  \hat S_{ab}\wa \rcp\lambda\sqrt{\frac{2}{\lambda^2+2}}
  \paren{\hD_a\hD_b\hat S+\frac{\lambda^2}2\hat S\dh ab}, \\
  \hat V_{ab}\wa \sqrt{\frac{2}{\lambda^2+2}}
  \hD_{(a}\hat V_{b)}.
\end{eqnarray}
Again, the coefficients have been chosen so that the square integrals
give unity, i.e,
\begin{equation}
  \label{eq:tensorsq}
  \int_{\Sigma_g}\d\mu_h \uh ac\uh bd\hat S_{ab}\hat S_{cd}=1,
\end{equation}
and the same for $\hat V_{ab}$. When $\lambda=0$, we define that
$\hat S_{ab}=\hat V_{ab}=0$.

The vector part is the part that cannot be constructed from the mode
functions $\hat S$ but from the harmonic vectors $\hat U_a$. It is
given by \footnote{This tensor is not a zero tensor, as opposed to the
  flat torus case.}
\begin{equation}
  \hat\hv_{ab}=\hat D_a\hat\hv_b.
\end{equation}
The coefficient ($=1$) is as always determined so that the
square integrals give unity like $\hat S_{ab}$ and $\hat V_{ab}$.
An explicit symmetrization in the right hand side of the above
equation is not necessary because of the closedness of $\hat \hv_a$
(See the first equation of Eqs.\reff{eq:LBcov}).

The remaining tensor part is the part that consists of all other
possible symmetric tensors on $(\Sigma_g,\dh ab)$. From York's
decomposition \cite{Yo73}, which is analogous to the Hodge
decomposition, we can identify that they are transverse (i.e.,
divergenceless) and traceless (\TT) tensors $\hat W_{ab}$:
\begin{eqnarray}
  \hD^b\hat W_{ab}\wa 0, \\
  \uh ab\hat W_{ab}\wa 0.
\end{eqnarray}
($\hat S\dh ab$ corresponds to York's trace part, while $\hat S_{ab}$,
$\hat V_{ab}$, and $\hat\hv_{ab}$ all correspond to York's vector
part. Beware that York's usage of the term ``vector'' part (or type)
is different from ours.)  TT tensors appear as variations of the
metric $\dh ab$ towards the directions that generate \Teich\
deformations \cite{Tro}. The freedom of $\hat W_{ab}$ therefore
corresponds to the linearized \Teich\ deformations of $\Sigma_g$,
which means we have $6g-6$ independent symmetric \TT-tensors on
$(\Sigma_g,\dh ab)$. We define $\hat W^{(s)}_{ab}$, $(s=1,\cdots,
6g-6)$ so that they are orthonormal to each other, i.e.,
\begin{equation}
  \int_{\Sigma_g}\d\mu_h \hat W^{(s)}_{ab}\hat W^{(s')ab}
  =\delta_{ss'}.
\end{equation}
We call parameter $s$ \textit{the \Teich\ label}. Like the cohomology
label we can think that this is a label that distinguishes
``degenerated states''. We therefore omit writing these labels
whenever they are not important.

We can now summarize the symmetric tensors on $(\Sigma_g,\dh
ab)$ as in the table below. It is straightforward to check that they
are orthogonal to each other with respect to the $L^2$-norm.
\begin{table}[hbtp]
\begin{center}
\begin{tabular}{|cl|l|l|l|} \hline
  $\hat S\dh ab$  & : Even scalar (trace) prt. & Trace part &  & 
  Generated from \\
  \cline{1-3}
  $\hat S_{ab}$   & : Even vector prt. &   & Scalar type & 
  the eigenfunction \\
  \cline{1-2}
  $\hat V_{ab}$   & : Odd vector prt.  & Traceless part &  &
     $\hat S=\hat S_\lambda$ \\
  \cline{1-2}
  \cline{4-5}
  $\hat \hv_{ab}$ & : Harmonic vec. prt. & & Vector type & 
  Topological \\
  \cline{1-2}
  \cline{4-4}
  $\hat W_{ab}$   & : \TT\ part      &  & Tensor type & 
  origin \\
  \hline
\end{tabular}
\end{center}
\caption{The symmetric tensors on
  $(\Sigma_g,\dh ab)$. \label{tab:tensors}}
\end{table}

Some of the key relations for these tensors are:
\begin{equation}
  \Lap_h\hat Y_{ab}=-(\mathrm{eig}_{\Lap_h})\hat Y_{ab},\quad
  \mathrm{eig}_{\Lap_h}\equiv
  \begin{cases}
    \lambda^2, & (\hat Y_{ab}=\hat S\dh ab) \\
    \lambda^2+4, & (\hat Y_{ab}=\hat S_{ab}) \\
    \lambda^2+4, & (\hat Y_{ab}=\hat V_{ab}) \\
    4, & (\hat Y_{ab}=\hat \hv_{ab}) \\
    2, & (\hat Y_{ab}=\hat W_{ab})
  \end{cases}
\end{equation}
and
\begin{equation}
  \hat D^c\hat D_{(a} \hat Y_{b)c}=
  \begin{cases}
    -\frac{\lambda^2}{2}\hat Y_{ab}
    +\lambda\lmx\hat S_{ab},
    & (\hat Y_{ab}=\hat S\dh ab) \\
    -\paren{\frac{\lambda^2}{2}+3}\hat Y_{ab}
    +\frac{\lambda}{2}\lmx\hat S\dh ab
    & (\hat Y_{ab}=\hat S_{ab}) \\
    -\paren{\frac{\lambda^2}{2}+3}\hat Y_{ab}, 
    & (\hat Y_{ab}=\hat V_{ab}) \\
    -3\,\hat Y_{ab}, 
    & (\hat Y_{ab}=\hat \hv_{ab}) \\
    -2\, \hat Y_{ab}. & (\hat Y_{ab}=\hat W_{ab})
  \end{cases}
\end{equation}

\textbf{Mode Quantities on $M\simeq\Sigma_g\times S^1$}

We are now in a position to define the symmetric mode tensors on $M$.
Let $S, \bar S\in\mathcal{C}^\infty(M)$ be the products
\begin{eqnarray}
  S\wb c_m \hat S, \\
  \bar S\wb \bar c_m \hat S,
\end{eqnarray}
and similarly we make vectors and tensors on $M$ by letting
\begin{eqnarray}
  \bar S_a\wb \bar c_m \hat S_a, \\
  \bar V_a\wb \bar c_m \hat V_a, \\
  \bar \hv_a\wb \bar c_m \hat \hv_a, \\
  S_{ab}\wb c_m\hat S_{ab}, \\
  V_{ab}\wb c_m\hat V_{ab}, \\
  \hv_{ab}\wb c_m\hat \hv_{ab}, \\
  W_{ab}\wb c_m \hat W_{ab}.
\end{eqnarray}
Then, we define our symmetric mode tensors $E_1\sim E_9$ as
\begin{alignat}{2}
    \E1ab =& S \dh ab, 
     & \quad \mbox{(the even-trace part)} \\
    \E2ab =& S_{ab} ,
     & \quad \mbox{(the even-base part)} \\
    \E3ab =& S l_{ab} ,
     & \quad \mbox{(the even-fiber part)} \\
    \E4ab =& 2 \bar S_{(a}\ss3{b)} ,
     & \quad \mbox{(the even-cross part)} \\
    \E5ab =& V_{ab}  ,
     & \quad \mbox{(the odd-base part)} \\
    \E6ab =& 2 \bar V_{(a}\ss3{b)},
     & \quad \mbox{(the odd-cross part)} \\
    \E7ab =& \hv_{ab}  ,
     & \quad \mbox{(the harmonic-base part)} \\
    \E8ab =& 2 \bar \hv_{(a}\ss3{b)},
     & \quad \mbox{(the harmonic-cross part)} \\
    (E_{9})_{ab} =& W_{ab}.
     & \quad \mbox{(the \TT\ part)}
\end{alignat}
Here, each of $E_1$ to $E_6$ are labeled by the eigenvalues $\lambda$
and $m$, $E_7$ or $E_8$ is only by $m$ with the cohomology label $r$,
and $E_{9}$ is only by $m$ with the \Teich\ label $s$. When the
Laplacian $\Lap_q$ on $M$ acts on these tensors their response depends
in general on the eigenvalues  $\lambda$ and $m$, but not on the
cohomology labels $r$ and \Teich\ label $s$, which merely
distinguish ``degenerated states'' for each $m$, as already remarked.

For convenience we categorize these basis tensors in two ways,
``parts'' and ``types'', which we summarize in Table
\ref{tab:basis}. Note that the types are determined with respect to
the associated quantities on the surface $\Sigma_g$, not to those on
the whole $M$. This usage, which seems natural in the present case,
does therefore not directly correspond to the one adopted in the usual
perturbation theory of the isotropic cosmologies.
\begin{table}[hbtp]
\begin{center}
\begin{tabular}{|c|l|l|l|} \hline
  $E_1$  &  &  &     \\
  \cline{1-1}
  $E_2$   &  Even part & & Associated with: \\
  \cline{1-1}
  $E_3$   &    &  & the eigenfunctions \\
  \cline{1-1}
  $E_4$ &   & Scalar type & $\hat S_\lambda$ on $\Sigma_g$, \\
  \cline{1-2}
  $E_5$   &  Odd part      &  & 
    and $c_m$ and $\bar c_m$ on $S^1$,  \\
  \cline{1-1}
  $E_6$   &        &  &   \\
  \cline{1-4}
  $E_7$   &  Harmonic part &  Vector type & 
  the harmonic vectors on $\Sigma_g$, \\
  \cline{1-1}
  $E_8$   &  &  & with  $c_m$ and $\bar c_m$ on $S^1$, \\
  \hline
  $E_9$   &  \TT\ part  & Tensor type & 
  and the \TT-tensors on $\Sigma_g$, with $c_m$  \\
  \hline
\end{tabular}
\end{center}
\caption{The symmetric tensors on
  $(M,q_{ab})$. \label{tab:basis}}
\end{table}

For convenience we define a standard $L^2$-inner product in
$\mathcal{S}_2(M)$ (the space of covariant symmetric tensors on $M$) by
\begin{equation}
  F,G\in \mathcal{S}_2(M),\quad
  (F,G)_0 \equiv \int_M\d\mu_{q_0}q^{ac}q^{bd}F_{ab}G_{cd}.
\end{equation}
The subscript $0$ in the bracket is a reminder that the integral is
taken with respect to the standard volume element
$\mu_{q0}$. \footnote{Similarly, we could define another inner product
  $(E_i,E_j)$, of which integral is taken with respect to the
  time-dependent element $\mu_{q}$. The two are simply related by
  $(E_i,E_j)=q_1\sqrt{q_2}(E_i,E_j)_0$. In this note, however, we only
  use the standard inner product $(E_i,E_j)_0$.}  Contractions are on
the other hand taken with the time-dependent metric $q^{ab}$, since we
stick to the rule that we raise and lower tensors and vectors on the
spatial manifold $M$ with the metrics $q^{ab}$ and
$q_{ab}$.

The basis tensors $E_i$ defined above are found to be orthogonal with
respect to this inner product for different kinds:
\begin{equation}
  \label{eq:EE0}
  (E_i,E_j')_0=0, \quad (i\neq j;i,j=1\sim 9),
\end{equation}
where the dash signifies that different values in the labels can be
taken. Among the same kind we have the following relations: \footnote{
  For notational simplicity, we assume that $\lambda$ and $\lambda'$
  are different even if their numerical values are the same if they
  correspond to distinct eigenfunctions.}
\begin{equation}
  \label{eq:EE}
  (E_i,E_i')_0=
  \begin{cases}
     (f_i)\inv \delta_{mm'}\delta_{\lambda\lambda'}, &
     (i=1\sim 6) \\
     (f_i)\inv \delta_{mm'}\delta_{rr'}, &
     (i=7,8) \\
     (f_i)\inv \delta_{mm'}\delta_{ss'}, &
     (i=9)
  \end{cases}
\end{equation}
where $f_i$ are functions of time given explicitly by
\begin{equation}
\label{eq:deffi}
\mbox{
\begin{tabular}{|c|c|c|c|c|c|c|c|c|c|}
\hline
$i$  & $1$ & $2$ & $3$ & $4$ & $5$ & $6$ & $7$ & $8$ & $9$  \\ \hline
$f_i$ & 
$\rcp2 \dq1^2$ &
$\dq1^2$ &
$\dq2^2$ &
$\rcp2 \dq1\dq2$ &
$\dq1^2$  &
$\rcp2 \dq1\dq2$ &
$\dq1^2$  &
$\rcp2 \dq1\dq2$ &
$\dq1^2$ \\ \hline
\end{tabular}
}.
\end{equation}

\section{Perturbation variables and constraints}
\label{sec:constraints}
\def\oddGamma{\mathcal{O}} \def\evenGamma{\mathcal{E}}
\def\harGamma{\mathcal{U}} \def\TTGamma{\mathcal{W}}

Let us apply the basic tools of mode expansion presented in the
previous section to the gravitational perturbation
problem.\footnote{Before proceeding to the perturbation problem
  however it is also beneficial to investigate simpler field equations
  on the same background, especially the massless scalar and
  source-free electromagnetic fields, to obtain insights for the
  properties of the perturbation problem. These materials are given in
  Appendices \ref{apx:scalar} and \ref{apx:elemag}.}  Now that we have
the complete set of basis tensors $E_i$, we can expand perturbation
variables in terms of them. Let $\gamma_{ab}\equiv \delta q_{ab}$ be a
perturbation of the spatial metric $q_{ab}$, and let $p^{ab}\equiv
\delta \pi^{ab}$ be its canonical conjugate. From the relations
\reff{eq:EE} we find that we should put
\begin{equation}
  \label{eq:expantion}
  \begin{split}
    \gamma_{ab}=& \sum \gamma^i \E iab, \\
    p^{ab}=& \mu_{q_0}\sum f_i\, p_i\, (E_i)^{ab}
  \end{split}
\end{equation}
to keep that the components $\gamma^i$ and $p_i$ are canonical. In
fact, the above relations show that
\begin{equation}
  \Theta\equiv \int_M \d^3x p^{ab}\dot\gamma_{ab}
  = \sum p_i\dot\gamma^i.
\end{equation}
Here, $\gamma^i=\gamma^i(t)$ and $p_i=p_i(t)$ are functions of time
$t$, and specified by the same labels as those of the corresponding
basis. The sums are all taken from $i=1$ to $9$, and all possible
eigenvalues of and all possible values of other labels (i.e., $r$ and
$s$).

\subsubsection*{Constraints}

In general, perturbation variables are constrained by the first
variations of the Hamiltonian and momentum constraints:
\begin{align}
  \delta \mathcal{H} & \approx 0, \\
  \delta \mathcal{H}^a & \approx 0.
\end{align}
(See Ref.\cite{Mo75-1} for a general formula for $\delta
\mathcal{H}\equiv D\mathcal{H}(q,\pi)\cdot(\gamma,p)$ and $\delta
\mathcal{H}^a\equiv D\mathcal{H}^a(q,\pi)\cdot(\gamma,p)$ .) For our
variables they are all obtained by straightforward computations using
Eqs.\reff{eq:expantion}. It is, however, worth noting that since the
variation of the Hamiltonian constraint itself is a scalar (density)
only the ``scalar type'' variables (associated with $E_1$ to $E_6$)
can contribute to it, and it should become a linear combination of
$\gamma^i$ and $p_i$ $(i=1\sim6)$. However, the odd variables (i.e.,
$\gamma^i$ and $p_i$ for $i=5$ and $6$) do not actually contribute to
this constraint, due to the zeros $\uh ab(E_i)_{ab}=l^{ab}(E_i)_{ab}=
D^aD^b(E_i)_{ab}=0$, for $i=5$ and $6$. Here, $D_a$ is the covariant
derivative operator associated with the spatial (background) metric
$q_{ab}$. (See Eq.\reff{eq:D=hatD+chi} for a useful formula.)
Therefore this constraint takes contributions only from the even
variables $\gamma^i$ and $p_i$ $(i=1\sim4)$. We find after an explicit
computation that the constraint for each mode is, when $\lambda>0$,
given by $\delta\mathcal{H}=\mu_0(q_1\sqrt{q_2})\inv(\mathcal{D}H)S$,
where
\begin{equation}
  \begin{split}
    \mathcal{D}H \equiv&
    -\left( 2\,m^2\,{\dq1} 
      + \frac{{\dq2}\,\left( 2\,{{\lambda}}^2\,{\dq1} 
          + {{\dpi2}}^2\,{\dq2} \right) }{2\,{\dq1}} \right){\gamma^1}
    - \lambda\sqrt{ \frac{2 + \lambda^2}{2} }\, \dq2\gamma^2 \\
    &
    - \left( (\lambda^2-1) \dq1
      + \frac{{\dpi1}\,{\dpi2}\,{\dq1}}{2} 
      - \frac{3\,{{\dpi2}}^2\,{\dq2}}{4} \right){\gamma^3}
    +2{\lambda}\,m\,{\dq1} \gamma^4 \\
    &
    - {\dpi2}{\dq1}{\dq2} {p_1}
    - {\dq2}\left( {\dpi1}{\dq1}-{\dpi2}{\dq2} \right) {p_3}
    \approx 0.
  \end{split}
\end{equation}
In the case $\lambda=0$, since $(E_2)_{ab}$ and $(E_4)_{ab}$ vanish
identically, the above expression is not valid, but still the correct
expression can be obtained from the same formula by setting
$\gamma^2=\gamma^4=p_2=p_4=0$, as well as setting $\lambda=0$. Thus,
it is given by
\begin{equation}
  \label{eq:DH:0}
  \begin{split}
    \mathcal{D}H^{(\lambda=0)}= & 
    -\left( 2\,m^2\,{\dq1} 
      + \frac{ {\dpi2}^2\,{\dq2}^2 }{2\,{\dq1}} \right){\gamma^1} \\
    &
    - \left( -\dq1
      + \frac{{\dpi1}\,{\dpi2}\,{\dq1}}{2} 
      - \frac{3\,{{\dpi2}}^2\,{\dq2}}{4} \right){\gamma^3} \\
    &
    - {\dpi2}{\dq1}{\dq2} {p_1}
    - {\dq2}\left( {\dpi1}{\dq1}-{\dpi2}{\dq2} \right) {p_3}
    \approx 0.
  \end{split}
\end{equation}
In particular, we denote this function when $m=0$ as $\D
H^{(\lambda=m=0)}$.

Since the first variation of the momentum constraint is a vector
constraint it has four kinds of components with respect to the basis
covariant vectors $S_a$, $V_a$, $\hv_a$, and $\bar S\ss3a$. Let us
write the corresponding components as, respectively, $\mathcal{D}H_S$,
$\mathcal{D}H_V$, $\mathcal{D}H_\hv$, and $\mathcal{D}H_l$, i.e.,
raising the indices, the variation of the momentum constraint
function is of the form
\begin{equation}
\delta\mathcal{H}^a=-\mu_0 \bra{ \sum(\mathcal{D}H_S)
S^a+\sum(\mathcal{D}H_V) V^a+\sum(\mathcal{D}H_\hv)
\hv^a+\sum(\mathcal{D}H_l)q_2\inv \bar S\cc3a },  
\end{equation}
where the sum is taken over
all possible eigenvalues and other labels. By straightforward
computations the scalar type components, $\mathcal{D}H_S$,
$\mathcal{D}H_V$, and $\mathcal{D}H_l$ are, when $\lambda>0$, found to
be
\begin{align}
  \label{eq:DHS}
\mathcal{D}H_S &=
   {\lambda}(\dq1p_1 
   - \dpi2 \gamma^3 )
   - \sqrt{\frac{\lambda^2+2}{2}}(\dpi1 \gamma^2 
   +2\dq1 p_2 )
   + {m}(2 \dpi2 \gamma^4
   + \dq1p_4 )
   \approx 0,
\\
  \label{eq:DHl}
\mathcal{D}H_l &=
    - {\lambda}(\dpi1 \gamma^4
    + \dq2 p_4)
    +{m}(\dpi1 \gamma^1
    - \dpi2 \gamma^3-2\dq2 p_3)
    \approx 0,
\\
  \label{eq:DHV}
  \mathcal{D}H_V &=
  - \sqrt{\frac{\lambda^2+2}{2}}(2q_1p_5+\pi_1\gamma^5)
  +{m}(q_1p_6+2\pi_2\gamma^6)\approx 0.
\end{align}
Again, when $\lambda=0$, since $E_i$ ($i=2,4,5$ and $6$) vanish
identically the above expressions are not valid, but still the correct
forms are obtained from the same expressions by setting the associated
variables zeros, at the same time setting $\lambda=0$. As a result,
$\mathcal{D}H_S$ and $\mathcal{D}H_V$ are found to vanish identically,
while $\mathcal{D}H_l$ gives a nontrivial contribution:
\begin{equation}
  \label{eq:DHl:0}
  \mathcal{D}H_{l}^{(\lambda=0)}=
  {m}(\dpi1 \gamma^1
  - \dpi2 \gamma^3-2\dq2 p_3)
  \approx 0.
\end{equation}
When $m=0$, however, this also vanishes.

Finally, the vector type component $\mathcal{D}H_\hv$ is given by
\begin{equation}
  \label{eq:DHU}
  \mathcal{D}H_\hv=
  - (2q_1p_7+\pi_1\gamma^7)
  +{m}(q_1p_8+2\pi_2\gamma^8)\approx 0.
\end{equation}
Note that this is the same as the one obtained by formally setting
$\lambda=0$ in $\mathcal{D}H_V$ (and changing the indices as $5\goes
7$ and $6\goes 8$).

The \TT-basis $E_9$ does not contribute to any of the constraints, so
the associated variables $\gamma^9$ and $p_9$ are not constrained.

\subsubsection*{Decouplings}

Before going into further analysis of the evolutions of the
perturbations we present in advance how the perturbations decouple. It
is direct computations to confirm the following.
\begin{theorem}
  \label{th:1}
  Perturbations of spatially closed orthogonal Bianchi III vacuum
  spacetimes (with fixed lapse and shift) are excited and evolved
  independently by either
  \begin{description}
  \item[(i)] the set of the even tensors $E_1\sim E_4$ belonging to
    the mode for a given pair $(\lambda,m)$,
  \item[(ii)] the set of the odd tensors $E_5$ and $E_6$ belonging to
    the mode for a given pair $(\lambda,m)$,
  \item[(iii)] the set of harmonic tensors $E_7$ and $E_8$ belonging
    to the mode for given $m$ and $r$, or
  \item[(iv)] the \TT\ tensor $E_9$ belonging to the mode for
    given $m$ and $s$.
  \end{description}
  Here, $\lambda$ is the base eigenvalue, and $m$
  is the fiber eigenvalue of the spatial manifold.
  $r$ is the cohomology label, and $s$ is the \Teich\ label.
  (When two different eigenstates have the same numerical value(s) of
  $\lambda$ and/or $m$ we understand that their eigenvalues
  are distinguished.)
\end{theorem}
(When we allow the lapse and shift to perturb, the appropriate kind,
i.e., even, odd, harmonic, or \TT\ type of mode vectors and scalars
are also needed to express the perturbation and those kinds of
perturbations still decouple from each other.) We call these kinds of
decoupled perturbations, respectively, the \textit{even, odd,
  harmonic, and \TT\ mode perturbations}. We denote the even mode
phase space for given $\lambda$ and $m$ as
$\evenGamma_{\lambda,m}$. Similarly, $\oddGamma_{\lambda,m}$ is the
odd phase space, $\harGamma_{m}$ the harmonic one, and $\TTGamma_{m}$
the \TT\ one.

We summarize the mode phase spaces and their constraints in the table
below. The Hamiltonian for each phase space will be shown in
subsequent sections.
\begin{table}[hbtp]
\begin{center}
\begin{tabular}{|l|l|c|} \hline
   Part & Mode phase space & Constraint function(s) \\
  \hline
   & $ \hspace{.6em} \evenGamma_{\lambda,m}$
   \hfill $ (\lambda>0,m\in\Z)$ & $\D H,\D H_S,\D H_l $ \\
  \cline{2-3}
   Even & $ \hspace{.6em} \evenGamma_{0,m}$ \hfill
   $ (\lambda=0,m\neq0)$ & $ \D H^{(\lambda=0)},\D H_l^{(\lambda=0)} $ \\
  \cline{2-3}
   & $ \hspace{.6em} \evenGamma_{0,0}$ \hfill 
   $ (\lambda=0,m=0)$ & $ \D H^{(\lambda=m=0)}$ \\
  \hline
   Odd & \hspace{.6em} $\oddGamma_{\lambda,m}$
   \hfill $ (\lambda>0,m\in\Z)$ & $ \D H_V $  \\
  \hline
   Harmonic & \hspace{.6em} $\harGamma_{m}$ \hfill 
   $ (m\in\Z) $ & $ \D H_U$ \\
  \hline
   \TT & $\hspace{.6em} \TTGamma_{m}$ \hfill
   $ (m\in\Z)$ & $\emptyset$ \\
  \hline
\end{tabular}
\end{center}
\caption{Mode phase spaces and their constraints. Each phase space
  is tacitly assumed to be endowed with an appropriate Hamiltonian
  function, too. These six kinds of mode phase spaces should be
  treated separately. \label{tab:phasespaces}}
\end{table}

For future convenience let us define some terminology here. We call
perturbations corresponding to all zero eigenvalues (i.e.,
$\lambda=m=0$ for even or odd, or $m=0$ for the harmonic or \TT)
\textit{zero mode perturbations}. We call even and odd perturbations
for which either $\lambda=0$ or $m=0$, or all harmonic and \TT\
perturbations \textit{global perturbations}. We call the generic
perturbations that are not zero mode nor global ones \textit{local
  perturbations}. Perturbations with $m=0$ is also called
\textit{$U(1)$-symmetric perturbations}.

\section{Odd Perturbations}
\label{sec:odd}
\def\gf{\theta}
\def\Hodd{\tilde H_{\mathrm{odd}}}
\def\Hoddtot{H_\mathrm{odd}}
\def\HoddGI{H_\mathrm{odd}^\mathrm{GI}}
\def\cxodd{c_{\mathrm{odd}}}
\def\Nqq#1{\frac{N#1}{q_1\sqrt{q_2}}}
\def\oQ{Q_\mathrm{B}}
\def\os{s} 
\def\Jname{multiplier function}

In this section we deal with the basic problems of the odd
perturbations, mainly the separation of gauge freedom.
The odd perturbations are by definition associated with the odd
tensors $\E5ab$ and $\E6ab$. We assume $\lambda>0$ as these tensors
identically vanish for $\lambda=0$. In the following we work with each
$\oddGamma_{\lambda,m}$ for a fixed pair of $\lambda>0$ and $m\in\Z$,
and assume that the perturbation canonical tensors are given by (See
Eq.\reff{eq:expantion})
\begin{equation}
  \label{eq:oddgamma}
  \begin{split}
    \gamma_{ab} &= \gamma^5\E5ab+\gamma^6\E6ab, \\
    p^{ab} &= \mu_{q_0}(f_5p_5\uE5ab+f_6p_6\uE6ab),
  \end{split}
\end{equation}
where $f_5$ and $f_6$ are functions of time defined in
Eq.\reff{eq:deffi}. The variation of the (covariant) shift vector
$\delta N_a=\delta (q_{ab}N^b)$ should be an odd vector of the form
\begin{equation}
  \label{eq:oddshiftdef}
  \delta N_a=\os V_a,
\end{equation}
where $\os=\os(t)$. We call $\os$ the \textit{odd shift function}. The
odd perturbations do not contribute to variations of the lapse;
$\delta N=0$.

The canonical variables which span the odd phase space
$\oddGamma_{\lambda,m}$ are given by
$\gamma^5$, $\gamma^6$, $p_5$, and $p_6$.
As we have seen in the previous section the odd variables are
constrained by a single constraint $\mathcal{H}_V\approx 0$, which
defines a three-dimensional subspace in
$\oddGamma_{\lambda,m}$. Moreover, this constraint function generates
gauge transformations in this subspace, so the true (i.e.,
gauge-invariant) dynamics is lying in the two dimensional space which
is obtained by contracting those gauge orbits. We denote this two
dimensional phase space as $\bar\oddGamma_{\lambda,m}$, and call it
the \textit{gauge-invariant odd phase space}.  We call the canonical
variables in $\bar\oddGamma_{\lambda,m}$ the \textit{gauge-invariant
  (canonical) variables}, and denote them as $Q$ and $P$.
\footnote{Although a constraint function is also gauge-invariant, we
  will, unless otherwise stated, only refer to $Q$ or $P$ as a
  gauge-invariant variable.}

One of the main tasks in this section is to find a canonical
transformation such that it defines the gauge-invariant canonical
variables, and then apply it to the original Hamiltonian in
$\oddGamma_{\lambda,m}$, which is presented right below. As a result
we obtain a gauge-decoupled form of Hamiltonian $\Hoddtot$, for which the
gauge-invariant part $\HoddGI(Q,P)$ determine the dynamics of the
gauge-invariant variables. We present the wave equation for the
gauge-invariant variable $Q$. The evolution of the other gauge
variable will be solved in terms of the gauge-invariant variables in
Sec.\ref{sec:prefgauge}.

\subsubsection*{The evolution equations and Hamiltonian 
  for the odd variables} 

The evolution equations for the odd variables can be reduced from the
general equations of motion by substituting Eqs.\reff{eq:oddgamma} and
\reff{eq:oddshiftdef}. See Ref.\cite{Mo75-1} for the general
perturbation Hamilton equations in the case of $N=1$, $N_a=0$, and
$\delta N=0=\delta N_a$.  Note that, since the lapse function $N$ for
our background is not unity we need to generalize the formula by
putting $d/dt\goes (1/N)d/dt$. We also add the contributions of the
perturbed shift. After a straightforward computation we
obtain:
\begin{equation}
  \label{eq:oddeqns}
  \begin{split}
  \dot \gamma^5&= \frac{N}{q_1\sqrt{q_2}}\bigg[
  (\pi_1q_1-\pi_2q_2){\gamma^5}
  +2q_1^2{p_5} \bigg] +\sqrt{2(\lambda^2+2)}\os,
  \\
  \dot \gamma^6&= \frac{N}{q_1\sqrt{q_2}}\bigg[
  \pi_2q_2{\gamma^6}
  +q_1q_2{p_6} \bigg] -m\os,
  \\
  \dot p_5&= \frac{q_1\sqrt{q_2}}{N}\bigg[ -\rcp2
  \paren{m^2+\pi_1^2+\rcp2\pi_2^2q_1^{-2}q_2^2-\pi_1\pi_2q_1^{-1}q_2}
  {\gamma^5}
  \\
  & \hspace{4em} -m\lmx{\gamma^6}
  \\
  & \hspace{4em} -(\pi_1q_1-\pi_2q_2){p_5} \bigg] 
  -\sqrt{\frac{\lambda^2+2}{2}}\pi_1q_1\inv \os,
  \\
  \dot p_6&= \frac{N}{q_1\sqrt{q_2}}\bigg[
  -m\lmx{\gamma^5}
   \\
  & \hspace{4em} -\paren{\lambda^2+\pi_1\pi_2+\rcp2\pi_2^2q_1^{-1}q_2}
  {\gamma^6}
   \\
  & \hspace{4em} -\pi_2q_2{p_6} \bigg]+2m\pi_2 q_1\inv\os.
  \end{split}
\end{equation}
We can easily reconstruct the Hamiltonian
$\Hodd(\gamma^5,\gamma^6,p_5,p_6)$ for the odd variables from the above
equations of motion; It is \footnote{One can in principle get the same
  Hamiltonian by computing the second variation of the Hamiltonian
  constraint function. (See, e.g., Ref.\cite{Mo75-2} for a general
  formula of the second variation.)  It would, however, require
  tremendous calculations, due to the second order terms.}:
\begin{equation}
  \label{eq:Hoddtilde}
  \begin{split}
     \Hodd = \Nqq{} & \bigg[
    \rcp4 \paren{m^2+\pi_1^2
      +\rcp2\pi_2^2q_1^{-2}q_2^2-\pi_1\pi_2q_1^{-1}q_2}(\gamma^5)^2 \\
    & \quad
    +\rcp2\paren{\lambda^2+\pi_1\pi_2+\rcp2\pi_2^2q_1^{-1}q_2}(\gamma^6)^2
    \\
    & \quad
    +m\lmx\gamma^5\gamma^6 \\
    & \quad
    +q_1^2 \, p_5^2
    +\rcp2 q_1q_2 \, p_6^2 \\
    & \quad
    +(\pi_1q_1-\pi_2q_2)\gamma^5p_5
    +\pi_2q_2 \gamma^6p_6
    \bigg] -\frac{\os}{q_1}\mathcal{D}H_V,
  \end{split}
\end{equation}
where $\mathcal{D}H_V=\mathcal{D}H_V(\gamma^5,\gamma^6,p_5,p_6)$ is
the constraint function for the odd perturbation \reff{eq:DHV}. 

We comment that the terms in the shift function $\os$ in the
Hamiltonian and evolution equations can easily be added \textit{after
  computing them with} $\os=0$. The easiest way to do this is to note
the fact that the perturbation Hamiltonian is obtained by computing
(half of) the second variation of the background Hamiltonian:
\begin{equation}
  \begin{split}
  H &= \rcp2 \delta\delta H_0 \\
  &= \rcp2 \delta\delta \int_Md^3x(N\mathcal{H}+N_a\mathcal{H}^a) \\
  &= \int_Md^3x(\rcp2N\delta\delta\mathcal{H}+\delta N\delta\mathcal{H}
  +\delta N_a \delta\mathcal{H}^a).
  \end{split}
\end{equation}
(We dropped the term $N_a\delta\delta\mathcal{H}^a$, which is zero
for our background with $N_a=0$.) The first term in the integration
contributes to the Hamiltonian for which $\os=0$. The second term
is zero for the odd perturbations (but not for the even ones). The
last term is the one we want to calculate, which is
\begin{equation}
  \begin{split}
  \int_Md^3x\delta N_a \delta\mathcal{H}^a
  &= -\int_Md\mu_0\os V_a (\mathcal{D}H_V)V^a \\
  &= -{\os}q_1\inv\mathcal{D}H_V\int_Md\mu_0 (c_m)^2\uh ab\hat V_a\hat V_b  \\
  &= -{\os}q_1\inv\mathcal{D}H_V.
  \end{split}
\end{equation}
Thus, we obtain the final form of the Hamiltonian with an arbitrary
$\os$. (Note that we already have the explicit form of
$\mathcal{D}H_V$.) The evolution equations with $\os$ may be obtained
from that Hamiltonian. This technique will be more helpful for the
even perturbation case where much longer computations are required.

\def\ssnameCT{Canonical Transformation}
\subsubsection*{\ssnameCT}

To decouple the Hamiltonian into gauge-dependent and independent parts
and obtain the gauge-invariant phase subspace
$\bar\oddGamma_{\lambda,m}$ we need to find a canonical transformation
such that one of the new momentum variables coincides with the
constraint; let ($\gamma_*,p_*$) be the canonical pair thereof, for
which the momentum is constrained $p_*=0$. Once such a canonical
transformation has been found, the other canonical pair, denoted as
$(Q,P)$, automatically becomes gauge-invariant
\cite{Mo74-1,Mo75-2}. We find such a canonical transformation by the
method of generating function \cite{Go} (also \cite{La}). Let
$\mathcal{S}(\gamma^i,P_i)$ be the generating function we are looking
for, which is a function of the original configuration variables
$\gamma^i=(\gamma^5,\gamma^6)$ and new momenta $P_i=(P,p_*)$. Since
both the original and transformed perturbation equations should be
linear, it is sufficient for us to consider linear transformations
\cite{La}. We can therefore assume that $\mathcal S$ is second order:
\begin{equation}
  \label{eq:defgenfun}
  \mathcal{S}=\rcp2\sum_{i=5}^6
  \sum_{j=5}^6\alpha_{ij}\gamma^i\gamma^j
  +\sum_{i=5}^6\beta_i\gamma^i,
\end{equation}
where $\alpha_{ij}=\alpha_{(ij)}$. Each $\beta_i$ should be a linear
function of $P_i$, while $\alpha_{ij}$ depend only on the background
variables. 

We remark that we can redefine the constraint function as
\begin{equation}
  \label{eq:defCV}
  C_V\equiv\frac{\gf}{q_1}\mathcal{D}H_V,
\end{equation}
using a (nonzero) arbitrary function of time $\gf=\gf(t)$. We call
this parameter function the \textit{mock shift function}. The factor
$q_1$ in the above definition is just for convenience. This
redefinition is in fact possible, since the original constraint
surface specified by $\mathcal{D}H_V=0$ is equivalent to that for
$C_V=0$. Since we define a new momentum variable $p_*$ to coincide
with $C_V$, this function is one of those parameters which control the
canonical transformation we consider.

We set the following Hamilton-Jacobi-like equation
\begin{equation}
  p_*=C_V(\gamma^i,p_i=\frac{\del\mathcal{S}}{\del\gamma^i}).
\end{equation}
Substituting Eq.\reff{eq:defgenfun}, we obtain the following equations
for the generating function coefficients $\alpha_{ij}$ and $\beta_i$:
\begin{eqnarray}
  \label{eq:oHJ3}
  -\gf(\sqrt{2(\lambda^2+2)}\beta_5-m\beta_6)\wa p_*,
  \\
  \label{eq:oHJ2}
  -\sqrt{2(\lambda^2+2)}\alpha_{56}+m(\alpha_{66}+2\pi_2q_1\inv)\wa0,
  \\
  \label{eq:oHJ1}
  -\sqrt{\frac{\lambda^2+2}{2}}(2\alpha_{55}+\pi_1q_1\inv)
  +m\alpha_{56}\wa0.
\end{eqnarray}

These equations do not completely determine $\alpha_{ij}$ and
$\beta_i$; for example, we can regard $\beta_6$ and $\alpha_{66}$ as
free functions. This freedom corresponds to the freedom of performing
canonical transformations \textit{among} the gauge-invariant variables
like $(Q,P)\goes (Q',P')$. As we will see below,  $\beta_6$ solely
determines the configuration variable $Q$, and once $\beta_6$ is
fixed, then $\alpha_{66}$ becomes a parameter that controls the
canonical transformations of the form $(Q,P)\goes (Q,P')$. We exploit
this freedom to make the final form of the gauge-invariant Hamiltonian
neat, especially ``diagonal''. For later convenience we ``solve'' the
above equations for $\alpha_{ij}$ in terms of a single $\alpha_{66}$:
\begin{equation}
  \label{eq:aij2a66}
  \begin{split}
    \alpha_{55} &= 
    \frac{\nu^2}{2}(\alpha_{66}+2\pi_2q_1\inv)-\rcp2\pi_1q_1\inv, \\
    \alpha_{56} &= 
    \frac{\nu}{\sqrt{2}}(\alpha_{66}+2\pi_2q_1\inv),
  \end{split}
\end{equation}
where we have defined a normalized fiber eigenvalue,
\begin{equation}
\nu\equiv \frac{m}{\sqrt{\lambda^2+2}}.
\end{equation}

We simply set
\begin{equation}
  \label{eq:decideb6}
  \beta_6=P+p_*.
\end{equation}
Then from Eq.\reff{eq:oHJ3}, we have
\begin{equation}
  \beta_5=\frac{\nu}{\sqrt2} P
  +\paren{\frac{\nu}{\sqrt2}-\rcp{\sqrt{2(\lambda^2+2)}\, \gf}}p_*.
\end{equation}
The gauge-invariant variable $Q$, conjugate to the momentum $P$ is
now obtained from the generating function as
\begin{equation}
  \begin{split}
  \label{eq:oddQ}
  Q & =\frac{\del\mathcal{S}}{\del P}
  =\sum_{i=5}^6 \frac{\del{\beta_i}}{\del P}\gamma^i \\
  & =\frac{\nu}{\sqrt2}\gamma^5+\gamma^6.
  \end{split}
\end{equation}
On the other hand, the gauge variable $\gamma_*$, conjugate to
$p_*$ is given by
\begin{equation}
  \label{eq:oddgamma*}
  \begin{split}
  \gamma_* &= \frac{\del\mathcal{S}}{\del p_*}
  =\sum_{i=5}^6 \frac{\del{\beta_i}}{\del p_*}\gamma^i \\
  &= \paren{\frac{\nu}{\sqrt2}-\rcp{\sqrt{2(\lambda^2+2)}\, \gf}}
  \gamma^5
  +\gamma^6.
  \end{split}
\end{equation}
The inverse is therefore
\begin{equation}
  \label{eq:gamma56Q*}
  \begin{split}
    \gamma^5 &= \sqrt{2(\lambda^2+2)}\, \gf\, (Q-\gamma_*), \\
    \gamma^6 &= (1-m\gf)Q+m\gf\gamma_*.
  \end{split}
\end{equation}
We also obtain the following transformation
between the original and new momentum variables:
\begin{equation}
  \label{eq:newm}
  \begin{split}
  p_5 &= \frac{\del\mathcal{S}}{\del\gamma^5}
  = \alpha_{55}\gamma^5+\alpha_{56}\gamma^6 +\beta_5 \\
  &= \paren{\alpha_{56}-\sqrt{\frac{\lambda^2+2}{2}}\pi_1q_1\inv \gf}Q
  +\frac{\nu}{\sqrt2}P \\
  & \hspace{2em} +\sqrt{\frac{\lambda^2+2}{2}}\pi_1q_1\inv \gf\gamma_*
  +\paren{\frac{\nu}{\sqrt2}-\rcp{\sqrt{2(\lambda^2+2)}\, \gf}}p_*, \\
  p_6 &= \frac{\del\mathcal{S}}{\del\gamma^6}
  =\alpha_{56}\gamma^5+\alpha_{66}\gamma^6+\beta_6 \\
  &= (\alpha_{66}+2m\pi_2q_1\inv \gf)Q+P-2m\pi_2q_1\inv \gf \gamma_*+p_*.
  \end{split}
\end{equation}
Equations \reff{eq:newm} and \reff{eq:gamma56Q*} complete the
canonical transformation that takes the Hamiltonian into a
desired gauge-decoupled form. The functions $\gf$ and $\alpha_{66}$
are free parameter functions that can be specified at any time one
wants. 

It is easy to see that the variable $Q=(\nu/\sqrt2)\gamma^5+\gamma^6$
is certainly gauge-invariant. See the induced map
\reff{eq:oddindN56}.

\def\ssnameGDH{The Gauge-Decoupled Form of Hamiltonian}
\subsubsection*{\ssnameGDH}

The total Hamiltonian $\Hoddtot$ for the new set of canonical
variables should be of the following (semi-)decoupled form
\begin{equation}
  \label{eq:oddtotHam}
  \Hoddtot=\HoddGI(Q,P)+J p_*,
\end{equation}
where $J=J(Q,P,\gamma_*,p_*,\os;\gf)$ is a function that possibly
depends on all the new variables $Q$, $P$, $\gamma_*$, and $p_*$, and
the shift function $\os$ (and also parametrically on the mock shift
$\gf$). We call $J$ the \textit{\Jname} or
\textit{renormalized shift function} for the odd variables. This form
of Hamiltonian is a consequence of our system being a
first-constrained system (e.g., \cite{Go}).  In fact, the Hamilton
equations for the gauge variables
\begin{equation}
  \label{eq:oddgaugeevol}
  \begin{split}
  \dot\gamma_* &= 
  J+\frac{\del J}{\del p_*}p_*, \\
  \dot p_* &= 
  -\frac{\del J}{\del \gamma_*}p_*
  \end{split}
\end{equation}
are consistent with the constraint
\begin{equation}
  p_*=0.
\end{equation}
Moreover, we can easily see that the evolution of the gauge-invariant
variables $Q$ and $P$ is determined, up to constraint, only from the
gauge-invariant part Hamiltonian $\HoddGI(Q,P)$. In fact, while we have
\begin{equation}
  \begin{split}
    \dot Q &= 
    \frac{\del \Hoddtot}{\del P}
    =\frac{\del \HoddGI}{\del P}+\frac{\del J}{\del P}p_*, \\
    \dot P &= 
    -\frac{\del \Hoddtot}{\del Q}
    =-\frac{\del \HoddGI}{\del Q}-\frac{\del J}{\del Q}p_*
  \end{split}
\end{equation}
from the total Hamiltonian, the last term for each equation
vanishes if the constraint $p_*=0$ is imposed.

We can find the decoupled Hamiltonian $\Hoddtot(Q,P,\gamma_*,p_*)$
from the original Hamiltonian $\Hodd(\gamma^5,\gamma^6,p_5,p_6)$ using
the definition of the new variables \reff{eq:oddQ},
\reff{eq:oddgamma*} and \reff{eq:newm}. Since our canonical
transformation is time-dependent, however, we must add the
time-derivative of the generating function $\mathcal{S}$ according to
the standard prescription. The correct relation is therefore
\begin{equation}
  \Hoddtot=\Hodd+\frac{\del\mathcal{S}}{\del t},
\end{equation}
where the time derivative operator $(\del/\del t)$ acts only on the
background variables (i.e., they do not act on the perturbation
variables like $Q$ or $P$). From a direct computation we find that the
the gauge-invariant part $\HoddGI(Q,P)$ is given by
\begin{equation}
  \label{eq:HoddGI}
\begin{split}
  \HoddGI(Q,P) &=
  \bigg( \lambda^2+\pi_2(\pi_1+4\nu^2\pi_2+\rcp2\pi_2q_1\inv q_2) \\
  & \quad \quad +2\pi_2(2\nu^2q_1+q_2)\alpha_{66}+q_1u(\alpha_{66})^2
  \bigg) \frac{NQ^2}{2q_1\sqrt{\dq2}} \\
  & \quad  +\rcp2 \dot\alpha_{66}Q^2 +\frac{NuP^2}{2\sqrt{q_2}}
  +{\cxodd}\frac{NQP}{\dq1\sqrt{\dq2}},
\end{split}
\end{equation}
where we have defined a combined scale function
\begin{equation}
  u\equiv \nu^2 q_1+q_2,
\end{equation}
and a cross term function
\begin{eqnarray}
  \label{eq:oddcPQ}
  \cxodd\wb
  \pi_2\paren{2\nu^2q_1+q_2}+q_1u\,\alpha_{66},
\end{eqnarray}
for convenience.  We have used Eqs.\reff{eq:aij2a66} to express
$\alpha_{ij}$ in terms of $\alpha_{66}$ only, so $\alpha_{66}$ is now
considered as a free function of time. Also, observe that the mock
shift $\gf$ and shift function $\os$ do not appear in $\HoddGI$ as
they should not.

As for the \Jname\ $J$, we find that
\begin{equation}
  \label{eq:oddF}
  \begin{split}
    J(Q,P,\gamma_*,p_*,\os;\gf) &=
    \paren{ (2 \pi_2 + q_1 \alpha_{66})
      \paren{u-\frac{m}{\lambda^2+2}\frac{q_1}{\gf}}
      \frac{N}{q_1\sqrt{q_2}}
      + \frac{\dot \gf}{\gf}
    }Q \\
  & \quad
  +\paren{u-\frac{m}{\lambda^2+2}\frac{q_1}{\gf}}\frac{NP}{\sqrt{q_2}} \\
  & \quad
  -\paren{\frac{\pi_2\sqrt{q_2}N}{q_1}+\frac{\dot \gf}{\gf}}\gamma_*
  \\
  & \quad
  +\paren{u+\frac{q_1(1-2m\gf)}{(\lambda^2+2)\gf^2}}
  \frac{Np_*}{2\sqrt{q_2}}-\frac{\os}{\gf}.
  \end{split}
\end{equation}
This function determines the evolution of the gauge variable
$\gamma_*$ through the equation of motion
\begin{equation}
  \label{eq:dgam*=J}
  \dot\gamma_*=J|_0\equiv J(Q,P,\gamma_*,p_*=0,\os;\gf).
\end{equation}
Note that we can think of the gauge-invariant variables $Q$ and $P$ in
the above equation as \textit{given functions}, since the evolution of
the gauge-invariant variables is determined independently of the gauge
variable $\gamma_*$.  We will need the above equation to determine the
actual evolution of the metric functions $\gamma^5$ and $\gamma^6$,
especially the evolution expressed in terms of $Q$ and $P$ (or
$\dot Q$).

It is interesting to note that even if we leave the shift function
unperturbed (i.e., $\os=0$) we can arbitrarily specify the ``gauge
velocity'' $J$ by suitably specifying $\gf(t)$, and thereby we can
virtually have any desired profile of the gauge variable
$\gamma_*(t)$. In this sense we can say that the mock shift function
$\gf$ mimics the role of the shift function $\os$. We will see however
that this feature is superficial in the sense that the profiles of the
metric functions $\gamma^5$ and $\gamma^6$ are not affected by the
choice of $\gf$ as long as $\os=0$. In other words, this arbitrariness
is an artifact contained in the canonical transformation we
consider. If we allow the shift function $\os$ to vary, however, the
gauge function $\gf$ automatically gets related, after solving the
gauge equation of motion, to $\os$. (cf. Sec.\ \ref{sec:prefgauge}.)

\subsubsection*{The Wave Equation for the Odd Perturbation}

To eliminate unnecessary ambiguities, let us choose
\begin{equation}
  \label{eq:a66}
  \alpha_{66}=\frac{\pi_2}{q_1}\paren{\frac{q_2}{u}-2}
\end{equation}
to make $\cxodd=0$, with which the gauge-invariant Hamiltonian
$\HoddGI$ becomes a diagonal form. In this case, the equations for the
gauge-invariant odd variables (up to constraint) become
\begin{equation}
  \label{eq:oddHamEqns}
  \begin{split}
  \dot Q =&  \frac{NuP}{\sqrt{q_2}}, \\
  \dot P =& -
  \bigg((\lambda^2+2)-\frac{\dq2}{u}
    +\frac{\dpi2\dq2}{4u\dq1}(5\dpi2\dq2-2\dpi1\dq1)
    +\frac{\dpi2\dq2^2}{u^2\dq1}(\dpi1\dq1-2\dpi2\dq2)
    \bigg)
  \frac{NQ}{q_1\sqrt{\dq2}}.
  \end{split}
\end{equation}
Then, eliminating $P$ we obtain the following wave equation for $Q$:
\begin{equation}
  \label{eq:ddQ}
  \begin{split}
  \ddot Q &- 
    \paren{\frac{\dot N}{N}
      -N\sqrt{\dq2}
      \paren{\frac{-\dpi1\dq1+3\dpi2\dq2}{2q_1q_2}
        +\frac{\dpi1\dq1-2\dpi2\dq2}{uq_1}}
      }\dot Q \\
  &+ \frac{N^2}{q_1{q_2}}
  \bigg((\lambda^2+2)u-{\dq2}
    +\frac{\dpi2\dq2}{4\dq1}(5\dpi2\dq2-2\dpi1\dq1)
    +\frac{\dpi2\dq2^2}{u\dq1}(\dpi1\dq1-2\dpi2\dq2)
    \bigg)Q=0.
  \end{split}
\end{equation}
Here, we have used the background Einstein equations \reff{eq:backeom}
to express the time-derivatives of the background variables. This
formal equation is valid for any spatially closed orthogonal Bianchi
III vacuum background (with vanishing shift). (We also remark that
since the definition of $Q$ is independent of the choice of
$\alpha_{ij}$, the above wave equation for $Q$ does not depend upon
the choice \reff{eq:a66}.)

If we substitute the exact solution \reff{eq:BIIIsol} into
the above wave equation we obtain an explicit form of wave equation:
\begin{equation}
  \label{eq:owaveex}
  \ddot Q-\frac{2\nu^2(t-2k)\tp}{\tm u}\,\dot Q
  +\paren{
    (\lambda^2 +2)\frac{u}{\tm^2}
    +\frac{2(t-3k)}{\tp^2 \tm}
    -\frac{4(t-2k)}{\tp^3u}}Q=0,
\end{equation}
where $u= \nu^2t_+^2+{t_-}{t_+\inv}$.

\section{Harmonic Perturbations}
\label{sec:ham}

We see in this section that the harmonic perturbations can be
(formally) viewed as a limit of the odd perturbations and
because of that, we are required to perform almost no new calculations
on top of those for the odd ones. 

Remember that the harmonic perturbations are those associated with the
harmonic vectors on $\Sigma_g$. The corresponding basis tensors are
$E_7$ and $E_8$, which are specified by the fiber eigenvalue
$m$. (More precisely, for each $m$ there are $2g$ independent sets of
$E_7$ and $E_8$, in accordance with the $2g$ independent harmonic
vectors $\hat \hv_{a}^{(r)}$ ($r=1,\cdots,2g)$.)  In the following we
work with each mode phase space $\harGamma_{m}=\harGamma_{m}^{(r)}$
for a given $m$ (and $r$). (We hereafter omit writing the cohomology
label $r$ for simplicity). We assume that the canonical perturbation
tensors are given by (cf. Eq.\reff{eq:expantion})
\begin{eqnarray}
  \label{eq:hamgamma}
  \gamma_{ab} \wa \gamma^7\E7ab + \gamma^8 \E8ab, \\
  p^{ab}\wa \mu_{q_0}(f_7p_7\uE7ab+f_8p_8\uE8ab),
\end{eqnarray}
where the functions $f_7$ and $f_8$ are defined in
Eq.\reff{eq:deffi}. Observe that $\harGamma_{m}$ is spanned by
$\gamma^7$, $\gamma^8$, $p_7$, and $p_8$, and therefore
$\dim\harGamma_{m}=4$.
The variation of the shift vector
$\delta N_a$ should be a harmonic type vector of the form
\begin{equation}
  \label{eq:hamshiftdef}
  \delta N_a=\os \hv_a,
\end{equation}
where $\os=\os(t)$. We call $\os$ the \textit{harmonic shift
  function}. (Using the same notation of the odd shift function $\os$
may not cause confusion.) The harmonic perturbations do not contribute
to variations of the lapse; $\delta N=0$.

The variables are constrained by a single constraint
$\mathcal{D}H_\hv$ (See Eq.\reff{eq:DHU}). A similar count to the odd
case gives $4-1-1=2$ as the dimension of the gauge invariant phase
space $\bar\harGamma_{m}$.

What we have to do to obtain a gauge-decoupled Hamiltonian and thereby
obtain a set of gauge invariant equations is completely parallel to
those for the odd perturbations. Remember that both the odd symmetric
tensor and harmonic symmetric tensor are traceless (on $\Sigma_g$);
$\uh ab \hat V_{ab}=\uh ab\hat\hv_{ab}=0$.  Notice also that almost
all relations for the harmonic vector $\hat\hv^a$ related to the
covariant derivative are the same for the odd vector $\hat V^a$ if we
assume $\lambda=0$. For example (see Eq.\reff{eq:keyrelforvecs}),
$\Lap_h \hat \hv^a=-\hat\hv^a$ is the same as $\Lap_h \hat
V^a=-(\lambda^2+1)\hat V^a$ if we put $\lambda=0$. (We are talking
about ``formal'' resemblances here, forgetting the fact that the odd
vector is not defined for $\lambda=0$.) The only difference is that
the odd vector does not satisfy the closed condition that the harmonic
vector does. That is, we have $\hat D_{[a}\hat\hv_{b]}=0$ (see
Eq.\reff{eq:LBcov}), but $\hat D_{[a}\hat V_{b]}\neq 0$. This
difference however does not affect our calculations in any way, since
these vectors contribute to the perturbations only through the
symmetrized tensors $\hat V_{ab}\propto \hat D_{(a}\hat V_{b)}$ or
$\hat \hv_{ab}= \hat D_{(a}\hat \hv_{b)} (=\hat D_{a}\hat
\hv_{b})$. Due to these properties, \textit{all the necessary formulas
  (e.g., Hamiltonians) for the harmonic perturbations can be obtained
  from the ones for the odd perturbations by (formally) putting
  $\lambda=0$, as well as changing the indices $5\goes 7$ and $6\goes
  8$, if necessary.}  So, we are not trying to re-present most of the
results.

One of our interests may however be the wave equation for the
background solution \reff{eq:BIIIsol}, which is given by
\begin{equation}
  \ddot Q-\frac{m^2(t-2k)\tp}{\tm u}\,\dot Q
  +\paren{
    m^2\frac{\tp^2}{\tm^2}+\frac{4}{\tp^2}
    -\frac{4(t-2k)}{\tp^3u}}Q=0,
\end{equation}
where $u= (m^2/2)\tp^2+\tm/\tp$. (This can be obtained from
Eq.\reff{eq:owaveex} by putting $\lambda=0$ and therefore
$\nu=m/\sqrt2$.) Here, the gauge-invariant variable $Q$ for the
harmonic perturbation is defined by (cf. Eq.\reff{eq:oddQ})
\begin{equation}
  Q=\frac m2\gamma^7+\gamma^8.
\end{equation}
When $m=0$ (the zero mode), this equation becomes
\begin{equation}
  \ddot Q+\frac{4k}{\tp^2\tm}Q=0,
\end{equation}
for which the general solution is given by
\begin{equation}
  Q=\frac{\tm}{\tp}\paren{C_1+C_2\paren{t-\frac{4k^2}{\tm}+4k\log\tm}},
\end{equation}
where $C_1$ and $C_2$ are integration constants.

\section{Transverse-Traceless Perturbations}
\label{sec:TT}
\def\ttQ{Q_\mathrm{B}}

Remember that the variables for the transverse-traceless (\TT)
perturbations are not constrained, so we do not need any extra steps
to get gauge decoupling.

The \TT-perturbation variables can be written
\begin{equation}
  \begin{split}
  \gamma_{ab} &= \gamma^9\E9ab, \\
  p^{ab} &= \mu_{q_0}f_9p_9\uE9ab.
  \end{split}
\end{equation}
We omit writing the subscripts $s$ that distinguish the $6g-6$
independent \TT-tensors. We remark again that each mode (for a fixed
$s$) is parameterized by the fiber eigenvalue $m$, but not by the base
one $\lambda$. The \TT\ perturbations do not contribute to
perturbations of the lapse function and shift vector; $\delta
N=0$, $\delta N_a=0$.

It is not difficult to find by a straightforward calculation the
evolution equations for the \TT-variables $\gamma^7$ and $p_7$, which
are given by
\begin{equation}
  \label{eq:TTeom}
  \begin{split}
    \dot\gamma^9 &= N\paren{2q_1^2\frac{p_9}{q_1\sqrt{q_2}}
      +(\pi_1q_1-\pi_2q_2)\frac{\gamma^9}{q_1\sqrt{q_2}}},
    \\
    \dot p_9 &= -N\paren{(m^2+\pi_1^2+\rcp2
      \pi_2^2q_1^{-2}q_2^2-\pi_1\pi_2q_1\inv
      q_2)\frac{\gamma^9}{2q_1\sqrt{q_2}}
      +(\pi_1q_1-\pi_2q_2)\frac{p_9}{q_1\sqrt{q_2}}}.
  \end{split}
\end{equation}
Let us define for coherent notation
\begin{equation}
  Q\equiv \gamma^9,\quad P\equiv p_9.
\end{equation}
Eliminating $P$, we find that the wave equation for $Q$ is given by
\begin{equation}
  \begin{split}
  \ddot Q- 
  & 
  \paren{
    \frac{\dot N}{N}+\frac{N}{2q_1\sqrt{q_2}}(\pi_1q_1-3\pi_2q_2)
    } \dot Q
  \\ &
  +N^2\paren{
    {m^2}{q_2\inv}
    -q_1\inv(1+\rcp2\pi_1\pi_2-\frac{5}{4}\pi_2^2q_1\inv q_2)
    } Q=0.
  \end{split}
\end{equation}
Here, we used the background Einstein equation \reff{eq:backeom} to
eliminate time-derivatives of the background variables. 

Note however that since $\gamma^9$ contributes to the perturbation of
metric in the form $\delta \dg ab=\gamma^9 \E9ab$ and the base $\E9ab$
is tangent to the base ($\E9ab\cc3b=0$), we can say that $\gamma^9$
belongs to the `base' (not `fiber') part, and the base part of the
background metric has unique scale function $q_1$.  A more natural
choice of variable is therefore to take the rescaled
variable\footnote{Similar arguments for the other kinds of
  perturbations are not trivial, since a gauge-invariant variable is
  in general a linear combination of variables that belong to
  different parts `base', `fiber', and `cross' ones.}
\begin{equation}
  \ttQ\equiv \gamma^9/q_1= Q/q_1.
\end{equation}
(The character B stands for `base.')
Let us rewrite the wave equation for $\ttQ$, which is given
by
\begin{equation}
  \begin{split}
  \ddot \ttQ- 
  & 
  \paren{\frac{\dot N}{N}
    +\frac{N}{2q_1\sqrt{q_2}}(\pi_1q_1+\pi_2q_2)}\dot \ttQ
  +N^2(m^2 q_2\inv)\ttQ =0.
  \end{split}
\end{equation}
It is interesting to observe that \textit{this wave equation for the
  \TT\ perturbation is exactly the same as the wave equation for the
  scalar field in case of $\lambda=0$}. (See Eq.\reff{eq:swaveeq} in
Appendix \ref{apx:scalar})

On the exact background solution \reff{eq:BIIIsol}, we have
\begin{equation}
  \label{eq:weqnttQk}
  \ddot\ttQ+\frac{2t}{\tp\tm}\dot\ttQ
  +{m^2\,\frac{t_+^2}{t_-^2}}\ttQ=0.
\end{equation}

\textbf{Solutions in some special cases}

(i) The flat background ($k=0$) solution with $m\neq 0$ is
\begin{equation}
  \ttQ=\rcp t \paren{C_1\sin mt+C_2\cos mt},
\end{equation}
where $C_1$ and $C_2$ are integration constants.

(ii) The zero mode ($m=0$) solution with arbitrary $k$ is
\begin{equation}
  \ttQ=
  \begin{cases}
    C_1 + C_2\log\frac{\tm}{\tp}, & (k\neq 0) \\
    C_1 + C_2\rcp t. & (k=0)
  \end{cases}
\end{equation}

\section{Even Perturbations}
\label{sec:ev}

\def\Heven{\tilde H_{\mathrm{even}}}
\def\Heventot{H_\mathrm{even}}
\def\HevenGI{H_\mathrm{even}^\mathrm{GI}}
\def\cxeven{c_{\mathrm{even}}}
\def\sqlmx{\frac{\lambda^2}{\lambda^2+2}}
\def\lmpp{{\lambda^2+2}}
\def\lmm{{\lambda^2-1}}

The remaining perturbations are the even perturbations, which are
generated by the even mode tensors $\E iab$ ($i=1\sim4$). It is worth
bearing in mind that, although any kind of computation becomes much
longer and sometimes hideous due to the increased number of basic
variables and the complicated couplings among them, the basic
procedure to get a decoupled Hamiltonian and wave equation is
completely systematic and the same as that for the odd case. (We
however do not introduce {mock shift and lapse functions} similar to
the odd mock shift function (cf. Eq.\reff{eq:defCV}) to simplify the
computations and arguments.)

Remember that two of the basis tensors, $\E2ab$ and $\E4ab$,
identically vanish when $\lambda=0$, so this case should be treated
separately. We say that the perturbation is \textit{generic} when
$\lambda>0$, and \textit{nongeneric} when $\lambda=0$. We first deal
with the generic case.

\subsection{Generic ($\lambda>0$) perturbations}

In the following we work with the generic even mode phase space
$\evenGamma_{\lambda,m}$ for given $\lambda>0$ and $m\in\Z$, and
suppose that the perturbation canonical tensors are given by (see
Eq.\reff{eq:expantion})
\begin{equation}
  \label{eq:evengamma}
  \begin{split}
    \gamma_{ab} &= \sum_{i=1}^{4}\gamma^i\E{i}ab, \\
    p_{ab} &= \mu_{q_0}\sum_{i=1}^{4} f_ip_i\uE iab,
  \end{split}
\end{equation}
where the functions $f_i$ ($i=1\sim 4$) are defined in
Eq.\reff{eq:deffi}. The eight (canonical) variables
$\gamma^i$, $p_i$ ($i=1\sim4$) span the mode phase space
$\evenGamma_{\lambda,m}$, so $\dim \evenGamma_{\lambda,m}=8$.
The variation of the lapse function and shift vector should be of the
form
\begin{equation}
  \label{eq:evenlapseandshiftform}
  \begin{split}
  \delta N &= \os_0 S, \\
  \delta N_a &= \os_1 S_a+\os_3 \bar S\ss3a,
  \end{split}
\end{equation}
where $\os_i$ ($i=0,1,3$) are functions of time.  We call $\os_0$ the
\textit{even lapse function}, $\os_1$ the \textit{even base shift
  function}, $\os_3$ the \textit{even fiber shift function}.  The
gauge-invariant dynamics is of course not affected by these functions.

As we have seen in Sec.\ref{sec:constraints}, the even variables are
constrained by three constraints $\mathcal{D}H\approx 0$,
$\mathcal{D}H_S\approx 0$, and $\mathcal{D}H_l\approx 0$, which define
a five $(=8-3)$ dimensional subspace in
$\evenGamma_{\lambda,m}$. Moreover, each function generates
independent gauge transformations in this subspace. The
gauge-invariant dynamics is therefore in a two $(=5-3)$ dimensional
subspace, which we denote $\bar\evenGamma_{\lambda,m}$, and call the
\textit{gauge-invariant (even) phase space}.  We call the canonical
variables in $\bar\evenGamma_{\lambda,m}$ the \textit{gauge-invariant
  (canonical) variables}, and denote them as $Q$ and $P$.

Before performing a canonical transformation to get a decoupled
Hamiltonian, we need to find the evolution equations and Hamiltonian
for the original variables.

\subsubsection*{The evolution equations and Hamiltonian
  for the even variables} 

It is a lengthy but straightforward computation to obtain evolution
equations, using Eqs.\reff{eq:evengamma}. For simplicity we only
present the evolution equations \textit{for the vanishing even lapse
  and shift functions (i.e., synchronous gauge)
  $\os_0=\os_1=\os_3=0$}. (The general equations with arbitrary even
lapse and shift functions may be easily recovered from the general
Hamiltonian given after the following evolution equations.)
\begin{equation}
  \begin{split}
  \frac{q_1\sqrt{q_2}}{N}\dot \gamma^1&= 
  -\rcp2 \pi_2q_1\gamma^3 -q_1q_2{p_3},
  \\
  \frac{q_1\sqrt{q_2}}{N}\dot \gamma^2&= 
  (\pi_1 q_1-\pi_2q_2){\gamma^2}+2q_1^2p_2,
  \\
  \frac{q_1\sqrt{q_2}}{N}\dot \gamma^3&= 
  -\pi_2q_1\inv q_2^2\gamma^1-\rcp2(\pi_1 q_1-3\pi_2q_2)\gamma^3
   \\
  & \quad  -q_2(q_1p_1-q_2p_3),
  \\
  \frac{q_1\sqrt{q_2}}{N}\dot \gamma^4&= 
  \pi_2q_2\gamma^4+q_1q_2 p_4,
  \\
  \frac{q_1\sqrt{q_2}}{N}\dot p_1&= 
   -( \pi_2^2q_1^{-2}q_2^2 - m^2) {\gamma^1}
  + \rcp4\left( 2{{\lambda}^2} 
    + {3{{\pi_2}}^2{{q_1\inv}}{q_2}} \right){\gamma^3} \\
  & \quad - m\lambda{\gamma^4}
  + {{\pi_2}{q_1\inv}{{q_2}}^2} {p_3},
  \\
  \frac{q_1\sqrt{q_2}}{N}\dot p_2&= 
  - \rcp2{\left( m^2 + {{\pi_1}}^2 - {\pi_1}{\pi_2}{q_1\inv}{q_2} + 
      \rcp2 \pi_2^2q_1^{-2}q_2^2 \right) } {\gamma^2} \\
  & \quad + \lmx\left( \frac{\lambda}{2}{\gamma^3}
  - m{\gamma^4}\right) \\
  & \quad  - \left(  {\pi_1}{q_1} -{\pi_2}{q_2} \right){p_2},
  \\
  \frac{q_1\sqrt{q_2}}{N}\dot p_3&= 
  \rcp4\left( 2{{\lambda}^2} + 
      {3\pi_2^2{q_1\inv}{q_2}} \right){\gamma^1}
  + \frac{\lambda}{2}\lmx{\gamma^2}
  \\
  & \quad  
  - \rcp2{\left( \frac{3}{4}\,\pi_2^2  + \rcp2\,{\pi_1}{\pi_2}{q_1}q_2\inv
      -{q_1}q_2\inv \right) }{\gamma^3}
  \\
  & \quad
  + \rcp2{{\pi_2}{q_1}} {p_1}
  + \rcp2{\left( {\pi_1}\,{q_1}-3{\pi_2}{q_2} \right) } {p_3},
  \\
  \frac{q_1\sqrt{q_2}}{N}\dot p_4&= 
  - m \lambda {\gamma^1}
  - m\lmx{\gamma^2}
  \\
  &  \quad
  - \rcp2{{\pi_2}{{q_1\inv}} \left( 2{\pi_1}{q_1} + 
        {\pi_2}{q_2} \right) }{\gamma^4}
  - {\pi_2}\,{q_2} {p_4}.
  \end{split}
\end{equation}

It is at once possible to read off the Hamiltonian which gives rise to
these equations, and also easy to generalize it to include the even
lapse and shift functions (see the comment below
Eq.\reff{eq:Hoddtilde}). The Hamiltonian with arbitrary even lapse and
shift functions is given by
\begin{equation}
  \begin{split}
    \label{eq:Heventilde}
     \Heven = \Nqq{} & \bigg[
    \rcp2 \paren{ \pi_2^2q_1^{-2}q_2^2 - m^2}(\gamma^1)^2 \\
    & \quad
    +\rcp4 \paren{m^2+\pi_1^2
      -\pi_1\pi_2q_1^{-1}q_2+\rcp2\pi_2^2q_1^{-2}q_2^2}(\gamma^2)^2 \\
    & \quad
    +\rcp4\paren{\frac{3}{4}\,\pi_2^2  + \rcp2\,{\pi_1}{\pi_2}{q_1}q_2\inv
      -{q_1}q_2\inv}(\gamma^3)^2
    \\
    & \quad
    +\rcp4{\pi_2}{q_1\inv} \left( 2{\pi_1}{q_1} + 
        {\pi_2}{q_2} \right)(\gamma^4)^2
    \\
    & \quad
    - \rcp4\left( 2{{\lambda}^2} 
      + {3{{\pi_2}}^2{{q_1\inv}}{q_2}} \right){\gamma^1}{\gamma^3}
    \\
    & \quad
    + m\lambda{\gamma^1}{\gamma^4}
    \\
    & \quad
    - \lmx\left( \frac{\lambda}{2}{\gamma^3}
      - m{\gamma^4}\right){\gamma^2}
    \\
    & \quad
    +q_1^2p_2^2
    +q_2^2p_3^2
    \\
    & \quad
    +\rcp2 q_1q_2 \paren{p_4^2-2 p_1{p_3}}
    \\
    & \quad
    -\pi_2q_1\inv q_2^2\gamma^1p_3
    +(\pi_1 q_1-\pi_2q_2){\gamma^2}p_2
    \\
    & \quad
    -\rcp2 \pi_2q_1\gamma^3p_1
    -\rcp2(\pi_1 q_1-3\pi_2q_2)\gamma^3p_3
    +\pi_2q_2\gamma^4p_4
    \bigg] \\
    & \hspace{-1em}
    +\frac{\os_0}{q_1\sqrt{q_2}} \mathcal{D}H
    -\frac{\os_1}{q_1} \mathcal{D}H_S
    -\frac{\os_3}{q_2} \mathcal{D}H_l,
  \end{split}
\end{equation}
where $\mathcal{D}H$, $\mathcal{D}H_S$, and $\mathcal{D}H_l$ are the
constraint functions for the even variables, given in
Sec.\ref{sec:constraints}.

\subsubsection*{\ssnameCT}

We need to find a canonical transformation so that three of the new
momenta coincide with the three constraint functions. \footnote{ We
  remark that the constraint functions in linear perturbation theory
  automatically strongly commute, which is the condition that is
  necessary (and sufficient) to assure that they can be new momentum
  variables after a canonical transformation. In fact, since the
  constraint functions are linear, their brackets do not depend on the
  perturbation variables, and \cite{Go} since the brackets at least
  weakly (i.e., up to the constraints) commute with each other, this
  in turn implies that they must strongly commute.}  The remaining new
momentum is identified with the gauge-invariant momentum $P$ that is
conjugate to the gauge-invariant configuration variable $Q$.  As in
the odd case we write the generating function in the form
\begin{equation}
  \mathcal{S}=\rcp2\sum_{i,j=1}^{4}
  \alpha_{ij}\gamma^i\gamma^j+\sum_{i=1}^{4}\beta_i\gamma^i,
\end{equation}
where $\alpha_{ij}=\alpha_{(ij)}$. The coefficients $\alpha_{ij}$ are
functions of background variables only, while $\beta_i$ depend
linearly upon $P$. For convenience, we redefine the constraints in the
form $P_i=p_i-K_i\approx0$, $(i=1,2,4)$, where $K_i$ is a function of
$\gamma^1$, $\gamma^2$, $\gamma^3$, $\gamma^4$, and $p_3$. (This is
actually possible.) Although this redefinition is not necessary, it
helps us to make computation more transparent and easier.
Let
\begin{equation}
  \{\gamma_*^i\}=(\gamma^{1}_{*},\gamma^{2}_{*},Q,\gamma^{4}_{*})
\end{equation}
be new configuration variables, and let
\begin{equation}
  \{p_{i*}\}=(p_{1*},p_{2*},P,p_{4*})
\end{equation}
be their conjugate momenta.
The equations
coming from the Hamilton-Jacobi-like equations,
$P_1(\gamma^i,p_i=\frac{\del\mathcal{S}}{\del \gamma^i})=p_{1*}$,
$P_2(\gamma^i,p_i=\frac{\del\mathcal{S}}{\del \gamma^i})=p_{2*}$, and
$P_4(\gamma^i,p_i=\frac{\del\mathcal{S}}{\del \gamma^i})=p_{4*}$
are the following:

\textbf{(i)}
From $P_1=c_{10}+c_{11}\gamma^1+\cdots+c_{14}\gamma^4=p_{1*}$:
\begin{eqnarray}
c_{10} \wb \beta_1 +  \Delta_1 \beta_3=p_{1*},\\
   c_{11} \wb
   \alpha_{11} + \Delta_1 \alpha_{13}
   +\frac{\lambda^2}{{\dpi2}\,{\dq1}} + 
   \frac{2\,m^2}{{\dpi2}\,{\dq2}} + 
   \frac{{\dpi2}\,{\dq2}}{2\,{{\dq1}}^2}=0 ,\\
   c_{12} \wb
   \alpha_{12} +  \Delta_1 \alpha_{23}
   +\frac{\lambda\,{\sqrt{2 + \lambda^2}}}
   {{\sqrt{2}}\,{\dpi2}\,{\dq1}}=0 ,\\
   c_{13} \wb
   \alpha_{13} + \Delta_1\alpha_{33}+
   \frac{2\,\left(2(\lambda^2-1) + 
       {\dpi1}\,{\dpi2} \right) \,{\dq1} - 
     3\,{{\dpi2}}^2\,{\dq2}}{4\,{\dpi2}\,
     {\dq1}\,{\dq2}}=0, \\
   c_{14} \wb
   \alpha_{14} + \Delta_1 \alpha_{34}
   -\frac{2\,\lambda\,m}{{\dpi2}\,{\dq2}}=0.
\end{eqnarray}

\textbf{(ii)}
From $P_2=c_{20}+c_{21}\gamma^1+\cdots+c_{24}\gamma^4=p_{2*}$:
\begin{eqnarray}
   c_{20} \wb \beta_2 + \Delta_2\beta_3=p_{2*},\\
   c_{21} \wb
    \alpha_{12} + \Delta_2\alpha_{13}
      + \frac{-2m^2{\dpi1}{\dpi2}{\dq1}^2 + 
        \lambda^2(2\lambda^2q_1q_2+ 4m^2{\dq1}^2 + 
           {\dpi2}^2{{\dq2}}^2 ) }{2{\sqrt{2(2 +\lambda^2)}}
        \lambda{\dpi2}{\dq1}^2{\dq2}}=0, \\
      c_{22} \wb \alpha_{22} + \Delta_2\alpha_{23}+
    \frac{\lambda^2 + {\dpi1}{\dpi2}}
      {2\,{\dpi2}{\dq1}}=0
      ,\\
      c_{23} \wb \alpha_{23} + \Delta_2\alpha_{33}+
      \frac{
        4\,m^2\,{{\dpi2}}^2\,{\dq1} + 
        \lambda^2( 2( 2(\lambda^2-1) + \dpi1\dpi2 )\dq1 + {\dpi2}^2\dq2 ) }{
        4\,{\sqrt{2(2 + \lambda^2)}}\,
        \lambda\,{\dpi2}\,{\dq1}\,
        {\dq2}} =0
      , \\
      c_{24} \wb  \alpha_{24} + \Delta_2\alpha_{34}+
    \frac{m\,\left( -2\,\lambda^2\,{\dq1} + 
          {\dpi2}\,\left( {\dpi1}\,{\dq1} - 
             2\,{\dpi2}\,{\dq2} \right)  \right) }{
        {\sqrt{2(2 + \lambda^2)}}\,{\dpi2}\,
        {\dq1}\,{\dq2}}=0
      .
\end{eqnarray}

\textbf{(iii)}
From $P_4=c_{40}+c_{41}\gamma^1+\cdots+c_{44}\gamma^4=p_{4*}$:
\begin{eqnarray}
  c_{40} \wb \beta_4+\frac{2\,m\,\beta_3}{\lambda} =p_{4*}, \\
   c_{41} \wb
   \alpha_{14}
   + \frac{2\,m\,\alpha_{13}}{\lambda} 
   - \frac{m\,{\dpi1}}{\lambda\,{\dq2}}=0
   ,\\
   c_{32} \wb \alpha_{24}+\frac{2\,m\,\alpha_{23}}{\lambda}=0,\\
   c_{43} \wb 
   \alpha_{34}
   + \frac{2\,m\,\alpha_{33}}{\lambda} 
   +\frac{m\,{\dpi2}}{\lambda\,{\dq2}}=0
   ,\\
   c_{44}\wb 
   \alpha_{44}
   + \frac{2\,m\,\alpha_{34}}{\lambda}
   +\frac{{\dpi1}}{{\dq2}}=0.
\end{eqnarray}
For convenience we have defined
\begin{equation}
  \begin{split}
   \label{eq:defDelta}
   \Delta_1 &\equiv \frac{{\dpi1}}{{\dpi2}} - 
     \frac{{\dq2}}{{\dq1}} 
     =\Sigma-2\frac{m^2}{\lambda^2}, \\
   \Delta_2 &\equiv 
   \frac{\lambda}{\sqrt{2(\lambda^2+2)}}\Sigma,
  \end{split}
\end{equation}
where
\begin{eqnarray}
   \label{eq:defSigma}
  \Sigma\wb \frac{{\dpi1}}{{\dpi2}} - 
     \frac{{\dq2}}{{\dq1}}+2\frac{m^2}{\lambda^2}.
\end{eqnarray}

Let us first look at the set of equations $c_{i0}=p_{i*}$ ($i=1,2$,
and $4$), which consists of three equations for four unknowns
$\beta_i$ ($i=1\sim4$). To determine them we put $\beta_3=p_{3*}\equiv
P$, where $P$ is the gauge-invariant momentum, then we immediately
obtain the following solution
\begin{equation}
  \label{eq:defbetas}
  \begin{split}
    \beta_1 &= p_{1*}-\Delta_1 P, \\
    \beta_2 &= p_{2*}-\Delta_2 P, \\
    \beta_3 &= P, \\
    \beta_4 &= p_{4*}-{\frac{2\,m\,P}{\lambda}}.
  \end{split}
\end{equation}

The remaining 12 equations, $c_{1i}=c_{2i}=c_{4i}=0$ ($i=1\sim4$), for
10 unknowns $\alpha_{ij}=\alpha_{(ij)}$ ($i,j=1\sim4$) look
overdetermined at first, but this system is
\textit{underdetermined}. In fact, given, say, $\alpha_{33}$, we find
we can solve in terms of $\alpha_{33}$ all equations except three
equations $c_{21}=c_{24}=c_{41}=0$, and these three equations are
automatically satisfied as long as the background Einstein equations
are satisfied. So, the equations are an underdetermined system.  In
later calculations we will express $\alpha_{ij}$ in terms of
$\alpha_{33}$ when it is useful. In such a case, $\alpha_{33}$
\textit{is regarded as an arbitrary (prescribed) function of $t$}.
(``Prescribed'' means that $\alpha_{33}$ should be prescribed
\textit{before} solving the perturbation equations.)  As pointed out
in the section for the odd perturbations, this is a result of the fact
that there remains freedom to canonically transform among a given
gauge invariant pair of variables.

We can determine the gauge-invariant variable $Q$ from the generating
function $\mathcal S$ with the solution \reff{eq:defbetas}. We get
\begin{eqnarray}
  \label{eq:evenQ}
  Q\wa \frac{\del\mathcal{S}}{\del P} \nonumber \\
  \wa 
  - \Delta_1\gamma^1
  - \Delta_2\gamma^2
  +\gamma^3
  -{\frac{2\,m\,\gamma^4}{{\lambda}}}.
\end{eqnarray}

Let us denote the remaining gauge configuration variables as
$\gamma_*^i=\del\mathcal S/\del p_{i*}$ ($i=1,2$, and $4$). They are
conjugate to $p_{i*}$. For
coherent notation we also define $\gamma_*^3\equiv Q$. We can easily
find
\begin{equation}
  \label{eq:eg*i=gi}
  \gamma_*^i=\gamma^i, \quad (i=1,2,\mbox{and } 4),
\end{equation}
and the inverse is
\begin{equation}
  \begin{split}
  \gamma^1 &= \gamma_*^1, \\
  \gamma^2 &= \gamma_*^2, \\
  \gamma^3 &= \Delta_1\gamma_*^1+\Delta_2\gamma_*^2
  +Q+\frac{2m}{\lambda}\gamma_*^4, \\
  \gamma^4 &= \gamma_*^4.
  \end{split}
\end{equation}
As for the momenta, from
\begin{equation}
  p_i=\frac{\del\mathcal{S}}{\del\gamma^i}=\alpha_{ij}\gamma^j+\beta_i
\end{equation}
(and the above relations for $\gamma^i$) we find
\begin{equation}
  \begin{split}
  p_1 &=
  (\alpha_{11}+\alpha_{13}\Delta_1)\gamma_*^1
  +(\alpha_{12}+\alpha_{13}\Delta_2)\gamma_*^2
  +\alpha_{13}Q
  +(\alpha_{14}+\alpha_{13}\frac{2m}{\lambda})\gamma_*^4 \\
  & \quad +p_{1*}- \Delta_1 P, \\
  p_2 &=
  (\alpha_{12}+\alpha_{23}\Delta_1)\gamma_*^1
  +(\alpha_{22}+\alpha_{23}\Delta_2)\gamma_*^2
  +\alpha_{23}Q
  +(\alpha_{24}+\alpha_{23}\frac{2m}{\lambda})\gamma_*^4 \\
  & \quad +p_{2*}- \Delta_2 P, \\
  p_3 &=
  (\alpha_{13}+\alpha_{33}\Delta_1)\gamma_*^1
  +(\alpha_{23}+\alpha_{33}\Delta_2)\gamma_*^2
  +\alpha_{33}Q
  +(\alpha_{34}+\alpha_{33}\frac{2m}{\lambda})\gamma_*^4 \\
  & \quad + P, \\
  p_4 &=
  (\alpha_{14}+\alpha_{34}\Delta_1)\gamma_*^1
  +(\alpha_{24}+\alpha_{34}\Delta_2)\gamma_*^2
  +\alpha_{34}Q
  +(\alpha_{44}+\alpha_{34}\frac{2m}{\lambda})\gamma_*^4 \\
  & \quad +p_{4*}- \frac{2m}{\lambda} P.
  \end{split}
\end{equation}
Note that the coefficients of $\gamma_*^i$ ($i=1,2,$ and $4$) in the
above relations can be immediately explicitly expressed in terms of
the background variables using the equations in (i) to (iii).

\subsubsection*{\ssnameGDH}

The Hamiltonian $\Heventot$ for the new variables is obtained from
\begin{equation}
  \Heventot=\Heven+\frac{\del \mathcal{S}}{\del t},
\end{equation}
where $(\del/\del t)$ acts on the background variables only.
The resulting Hamiltonian becomes of the form
\begin{equation}
  \label{eq:Hevendec}
  \Heventot=\HevenGI(Q,P)+J_1p_{1*}+J_2p_{2*}+J_4p_{4*},
\end{equation}
where $\HevenGI(Q,P)$ is the \textit{gauge-invariant Hamiltonian}, and
$J_i=J_i(Q,P,\gamma^1_*,\gamma^2_*,\gamma^4_*,p_{i*})$ ($i=1,2$, and
$4$) are the \textit{\Jname s}. 

From a straightforward computation we find the following
gauge-invariant Hamiltonian: 
\begin{equation}
  \label{eq:HevenGI}
  \begin{split}
    \HevenGI(Q,P) &=
    \bigg\{ 
    \bigg(
      \frac{(11{\lambda}^2+4)\pi_2^2 q_2}{16\,{q_1}} 
      + \frac{(\lambda^2+2)}{4}\pi_1\pi_2 \\
      & \quad\quad
      + \frac{{\lambda}^2({\lambda}^2 -1)^2{q_1}}{\pi_2^2 q_2}
      + \frac{5{\lambda}^4}{2} - {\lambda}^2 -3 
      \bigg) \frac{1}{(\lambda^2+2)q_2} \\
  & \quad \quad
  + \left(
    {\left( \frac{5\lambda^2}{4}+1 \right)\pi_2^2} 
    + \frac{\lambda^2(\lambda^2-1)q_1}{q_2}
  \right)
    \frac{\Sigma}{(\lambda^2+2)q_2} \\
    & \quad \quad
    + \frac{\lambda^2\pi_2^2{q_1}}{4q_2^2}\frac{\Sigma^2}{\lambda^2+2}
    +\cxeven \frac{\alpha_{33}}{2}
    \bigg \} \, \frac{NQ^2}{2\sqrt{q_2}}
  \\
  & \quad
  +\frac{\dot\alpha_{33}}{2}Q^2
  +\frac{N\Omega P^2}{2q_1\sqrt{q_2}}
  +\cxeven\frac{NPQ}{2\sqrt{q_2}},
  \end{split}
\end{equation}
where
\begin{equation}
  \label{eq:evendefOmega}
  \Omega\equiv {q_2^2+2q_1 q_2\Sigma
        +\frac{\lambda^2}{\lambda^2+2}q_1^2\,\Sigma^2} ,
\end{equation}
and
\begin{equation}
  \label{eq:cxeven}
\begin{split}
  \cxeven &\equiv
  2({\lambda}^2-1)\pi_2\inv - 
        {\pi_1}
      + \frac{5}{2}\,{{\pi_2}}q_1\inv{q_2}
    \\
    & \quad
    + \left( \left( \frac{7\lambda^2}{2}+4  \right){\pi_2} 
      + \frac{2{\lambda}^2(\lambda^2-1) \,{q_1}}{{\pi_2}\,
        {q_2}} \right) \frac{\Sigma}{(\lambda^2+2)} 
      \\
      & \quad
      + \frac{{\lambda}^2{\pi_2}{q_1}{\Sigma}^2}{(\lambda^2+2)q_2}
      + \frac{2\Omega \, \alpha_{33}}{q_1}.
\end{split}
\end{equation}
As we noted earlier we have expressed $\alpha_{ij}$ in terms of
$\alpha_{33}$ only, solving the equations $c_{1i}=c_{2i}=c_{4i}=0$
($i=1\sim4$) (except the ``identities'' $c_{21}=c_{24}=c_{41}=0$).
The function $\alpha_{33}$ is therefore now an arbitrary (prescribed)
function of $t$.

As for the \Jname s, we find
\begin{equation}
  \label{eq:evenmulti}
\begin{split}
  J_1 &= -(\pi_2+2q_2 \alpha_{33})\frac{NQ}{2\sqrt{q_2}}
  -N\sqrt{q_2}P \\
  &  +\paren{(\lambda^2-1)\pi_2\inv q_1-\frac14\pi_2 q_2}
  \frac{N \gamma^1_*}{q_1\sqrt{q_2}} \\
  &  +\frac{\lambda}{\sqrt{2(\lambda^2+2)}}
  \paren{(\lambda^2-1)\pi_2\inv q_1+\frac34\pi_2 q_2}
  \frac{ N \gamma^2_*}{
    q_1\sqrt{q_2}} \\
  & -\pi_2\sqrt{q_2}\os_0-\lambda \os_1, \\
   J_2 &= -\frac{\lambda}{\sqrt{2(\lmpp)}}
   \bigg( \frac32\pi_2+2(\lmm)\pi_2\inv q_1q_2\inv \\
   & \hspace{2em}
    +(\pi_2q_1q_2\inv+2q_1\alpha_{33})\,\Sigma \bigg)
   \frac{NQ}{\sqrt{q_2}} \\
   & -\lambda\sqrt{\frac{2}{\lmpp}}\frac{N\Sigma\, q_1P}{\sqrt{q_2}}
    -\frac{\lambda\sqrt{2(\lmpp)}N\gamma^1_*}{\pi_2\sqrt{q_2}} \\
   & -\paren{\lambda^2\pi_2\inv q_1+\pi_2 q_2}
   \frac{N\gamma^2_*}{q_1\sqrt{q_2}} +\frac{Nq_1p_{2*}}{\sqrt{q_2}}\\
   &  +\sqrt{2(\lambda^2+2)}\os_1,  \\
  J_4 &=
  -\frac{m}{\lambda}(\pi_2+2q_2\alpha_{33})\frac{NQ}{\sqrt{q_2}}
   -\frac{2 m N\sqrt{q_2}P}{\lambda} \\
  & +\frac{2\lambda m N\gamma^1_*}{\pi_2\sqrt{q_2}} \\
  &
  +m\sqrt{\frac{2}{\lmpp}}\paren{(\lambda^2-1)\pi_2\inv q_1+\frac34\pi_2 q_2}\frac{N\gamma^2_*}{q_1\sqrt{q_2}}
  \\
  & -\paren{\pi_1q_1-\pi_2q_2}\frac{N\gamma^4_*}{q_1\sqrt{q_2}}
  +\frac{N\sqrt{q_2}p_{4*}}{2} \\
  &  -m\os_1+\lambda\os_3.
\end{split}
\end{equation}
These functions provide the time derivatives of the gauge variables
$\gamma^i_*$ ($i=1,2$, and $4$), i.e., 
\begin{equation}
  \label{eq:egeqs}
 \dot \gamma^i_*=J_i|_0 
\end{equation}
for each $i$, where $|_0$ stands for imposing the constraints
$p_{i*}=0$.

\subsubsection*{The Wave Equation for the Even Perturbation}

Let us choose $\alpha_{33}$ so that it makes the
Hamiltonian a diagonal form. We therefore choose
\begin{equation}
\begin{split}
   \alpha_{33}
   & = -\frac{q_1}{2\Omega} \bigg\{
  2({\lambda}^2-1)\pi_2\inv - 
        {\pi_1}
      + \frac{5}{2}\,{{\pi_2}}q_1\inv{q_2}
    \\
    & \quad
    + \left( \left( \frac{7\lambda^2}{2}+4  \right){\pi_2} 
      + \frac{2{\lambda}^2(\lambda^2-1) \,{q_1}}{{\pi_2}\,
        {q_2}} \right) \frac{\Sigma}{(\lambda^2+2)} 
      \\
      & \quad
      + \frac{{\lambda}^2{\pi_2}{q_1}{\Sigma}^2}{(\lambda^2+2)q_2}
      \bigg\}
\end{split}
\end{equation}
to make $\cxeven=0$.
Then, the Hamilton equations are given by
\begin{equation}
  \label{eq:evenhameqsgi}
  \begin{split}
  \dot Q &= \frac{N\Omega P}{q_1\sqrt{q_2}}, \\
  \dot P &= 
  -
\bigg\{ 
    \bigg(
      \frac{(11{\lambda}^2+4)\pi_2^2 q_2}{16\,{q_1}} 
      + \frac{(\lambda^2+2)}{4}\pi_1\pi_2 \\
      & \quad\quad
      + \frac{{\lambda}^2({\lambda}^2 -1)^2{q_1}}{\pi_2^2 q_2}
      + \frac{5{\lambda}^4}{2} - {\lambda}^2 -3 
      \bigg) \frac{1}{(\lambda^2+2)q_2} \\
  & \quad \quad
  + \left(
    {\left( \frac{5\lambda^2}{4}+1 \right)\pi_2^2} 
    + \frac{\lambda^2(\lambda^2-1)q_1}{q_2}
  \right)
    \frac{\Sigma}{(\lambda^2+2)q_2} \\
    & \quad \quad
    + \frac{\lambda^2\pi_2^2{q_1}}{4q_2^2}\frac{\Sigma^2}{\lambda^2+2}
    \bigg\} \frac{NQ}{\sqrt{q_2}}
    -{\dot\alpha_{33}}Q.
  \end{split}
\end{equation}

It is easy (in principle) to find explicitly the wave
equation for $Q$ from the above Hamilton equations, but instead of
writing down one that has very complicated form, let us present the
wave equation for $Q$ for the exact background solution
\reff{eq:BIIIsol}. It is given by
\begin{equation}
  \label{eq:evenwavek}
  \ddot Q -
  2\bigg(
  {\frac{(t-2k)}{t_+t_-}}
  -\frac{X}{Z}
  \bigg) \dot Q
  + \bigg( 
  m^2\,\frac{t_+^2}{t_-^2}
  +\lambda^2\rcp{t_+t_-}
  +\frac{Y}{Z} \bigg) Q=0,
\end{equation}
where
\begin{equation}
\begin{split}
X & \equiv 
   8\,m^2(t-2k){\tp}^2+
   2\,{{\lambda}}^2\,m^2( 2 t-k ) \,{\tp}^2 +
   2{{\lambda}}^2(2\,t - 3\,k )  +
   \frac{{{\lambda}}^4(2\,t-k )t}{\tp},
   \\
Y & \equiv 
    16\,m^4\,{{t_+}}^4
    -4\,k\,m^2\,{{t_+}} \left({{\lambda}}^2\,\frac{t-4\,k}{\tm}
      - 8 \right)
    +2\,k^2\,{{\lambda}}^4\,\frac{2t +k}{\tm\tp^{2}},
    \\
Z & \equiv 
       4\,m^4\,{{t_+}}^6
      + 8\,m^2\,{{t_-}}\,{{t_+}}^3
      + 4\,{{\lambda}}^2\,m^2\,t\,{{t_+}}^3
      + 2\,{{\lambda}}^2\,{{t_-}}\,{{t_+}}
      + {{\lambda}}^4\,t^2.
\end{split}
\end{equation}

\subsection{Nongeneric ($\lambda=0$) perturbations}
\label{sec:evennongen}

In the following we work with a nongeneric even mode phase space
$\evenGamma_{0,m}$
and suppose that the perturbation canonical tensors are given by (See
Eq.\reff{eq:expantion})
\begin{equation}
  \label{eq:evengamma:0}
  \begin{split}
  \gamma_{ab} &= \gamma^1\E1ab+\gamma^3\E3ab, \\
  p^{ab} &= \mu_{q_0}(f_1p_1\uE1ab+f_3p_3\uE3ab).
  \end{split}
\end{equation}
Note that $\dim\evenGamma_{0,m}=4$.

As we have seen in Sec.\ref{sec:constraints}, when $\lambda=0$ we have
two constraints $\mathcal{D}H^{(\lambda=0)}\approx0$ and
$\mathcal{D}H_{l}^{(\lambda=0)}\approx0$. (See Eqs.\reff{eq:DH:0} and
\reff{eq:DHl:0}) Note, however, that $\mathcal{D}H_{l}^{(\lambda=0)}$
vanishes identically when $m=0$, and thus we are left with a single
nontrivial constraint $\mathcal{D}H^{(\lambda=m=0)}\approx0$. We must
therefore consider the two cases $m=0$ and $m\neq0$ separately.

\subsubsection{The $\lambda=0$ and $m\neq0$ case: Pure gauge}

We have two (nontrivial) constraints
$\mathcal{D}H_{(\lambda=0)}\approx 0$ and
$\mathcal{D}H_{l(\lambda=0)}\approx 0$ when $\lambda=0$ and
$m\neq0$. Since the phase space $\evenGamma_{0,m}$ is four
dimensional, these two constraints define $4-2=2$ dimensional
constraint subspace and the gauge transformations these constraint
functions generate on this subspace contract this subspace to a point:
$\bar\evenGamma_{0,m\neq0}=\{e\}$. In other words, the phase space
$\evenGamma_{0,m\neq0}$ is a space of pure gauge, and therefore we may
not be interested in this space.

\subsubsection{The case $\lambda=0$ and $m=0$: 
Locally homogeneous perturbations}

In this case we have only one constraint
$\mathcal{D}H^{(\lambda=m=0)}\approx0$, so the dimension of the
gauge-invariant phase space $\dim\bar\evenGamma_{0,0}$ is
two($=4-1-1$). An explicit form of the constraint is obtained from
Eq.\reff{eq:DH:0} by putting $m=0$:
\begin{equation}
  \label{eq:DH:00}
  \begin{split}
    \mathcal{D}H^{(\lambda=m=0)}\equiv & -
    {p_1}\,{\dpi2}\,{\dq1}\,{\dq2} - {p_3}\,{\dq2}\, \left(
      {\dpi1}\,{\dq1} -{\dpi2}\,{\dq2} \right) - {\gamma^1}
        \frac{{{\dpi2}}^2\,{\dq2}^2 }{2\,{\dq1}}
    \\
    &+ {\gamma^3}\,\left( {\dq1} - \frac{{\dpi1}\,{\dpi2}\,{\dq1}}{2}
      + \frac{3\,{{\dpi2}}^2\,{\dq2}}{4} \right)  \approx 0.
  \end{split}
\end{equation}

We do not have to perform computations from scratch to decouple
gauge. As easily seen, our rule to reinterpret the computations
presented in Sec.\ref{sec:evennongen} for the $\lambda=m=0$
case is
(1) first to put $\gamma^i=p_i=0$ for $i=2,4$,
(2) then take limit $m\goes0$,
(3) finally take limit $\lambda\goes0$.
Keeping the order of steps (2) and (3) is necessary to drop correctly
the terms like $m/\lambda$. This order is particularly important when
getting the correct wave equation, e.g., from Eq.\reff{eq:evenwavek}.

We find that the gauge-invariant configuration variable becomes
\begin{equation}
  Q=-\paren{\frac{\pi_1}{\pi_2}-\frac{q_2}{q_1}}\gamma^1+\gamma^3.
\end{equation}
The wave equation can be obtained from the generic equation
\reff{eq:evenwavek} by {first} putting $m=0$, {then}
taking the limit $\lambda\goes 0$, as noted. We have
\begin{equation}
  \ddot Q+\frac{2}{\tp}\dot Q=0.
\end{equation}
The solution is
\begin{equation}
  Q=C_1\tp\inv+C_2,
\end{equation}
where $C_1$ and $C_2$ are integration constants.

\section{Gauge Properties of the Odd Perturbations}
\label{sec:prefgauge}
\def\pth{\mathcal{\vartheta}}

In this section we discuss gauge issues for the odd perturbation
system to obtain the perturbed spacetime metrics in a completed form.
Remember that the system contains a gauge(-dependent) variable
$\gamma_*$ and the odd shift function $\os$. Choosing a particular
profile of the shift function $\os(t)$ is called a \textit{gauge
  choice} for the odd perturbations. The evolution of the gauge
variable $\gamma_*$ is determined by the evolution equation (the gauge
equation) \reff{eq:dgam*=J}, depending on a given gauge choice. The
gauge equation also depends (parametrically) on the gauge-invariant
variable $Q$, which is however determined independently of $\gamma_*$
and $\os$. By solving the gauge equation for a given shift function
$\os$ we can express the gauge variable $\gamma_*$ in terms of $Q$,
and thereby we can express the metric functions $\gamma^5$ and
$\gamma^6$ in terms of $Q$. We see below how these functions are
expressed in some gauge choices.

We first need to obtain the map acting on the odd
variables induced by \diffeos\ for later use.
\begin{lemma}[Induced map]
  Let $v(t)$ be an arbitrary function of time. Then, the one-parameter
  \diffeo\ generated by the odd vector
\begin{equation}
  \label{eq:oddspgenvec}
  Y_a=v(t)V_a
\end{equation}
induces the
following map $I_\mathrm{odd}$ on the odd perturbation variables
$(\os,\gamma^5,\gamma^6)$:
\begin{equation}
  \label{eq:oddindN56}
  \begin{split}
    \os & \goes \os+\paren{\dot v-\frac{\dot
    q_1}{q_1}v}, \\
I_\mathrm{odd}[v]:\quad\quad
    \gamma^5 &\goes \gamma^5+\sqrt{2(\lambda^2+2)}v, \\
    \gamma^6 &\goes \gamma^6-mv,
  \end{split}
\end{equation}
where $\os=\os(t)$ is the odd shift function.
\end{lemma}

\proofmark This is a straightforward calculation of the induced map
\begin{equation}
  \delta g_{ab}\goes \delta g_{ab}+2\nabla_{(a}Y_{b)}.
\end{equation}
Here $\delta \dg ab$ is an odd perturbation spacetime metric, and
$\nabla_a$ is the covariant derivative associated with the
background spacetime. A straightforward computation shows
\begin{equation}
  2\nabla_{(a}Y_{b)}=
  2\paren{\dot v-\frac{\dot q_1}{q_1}v}(dt)_{(a}V_{b)}
  +v(\sqrt{2(\lambda^2+2)}\E5ab-m\E6ab),
\end{equation}
from which we can immediately read off the induced map on $\delta
N_a$, $\gamma^5$, and $\gamma^6$. \endofproofmark

This induced map tells us that fixing the odd shift function $\os$
does \textit{not} imply a complete fixing of the freedom of \diffeos.
We obtain the following.
\begin{prop}[Pure gauge solution]
  \label{prop:2}
  Consider the odd perturbation equations \reff{eq:oddeqns}
  with a given odd shift function $\os(t)$.
  Let $C_1$ be a constant parameter. Then, the set of functions
  \begin{equation}
    \gamma^5_\mathrm{(d)}= C_1 q_1,\quad 
    \gamma^6_\mathrm{(d)}=-C_1 \frac{\nu}{\sqrt2}q_1,
  \end{equation}
  provides the one-parameter solution of the equations which is
  generated by spatial \diffeos.
\end{prop}

\textit{Proof}: The zero perturbation $\os=\gamma^5=\gamma^6=0$ is a
trivial solution of the perturbation equations. The \diffeo\ generated
by the spatial vector \reff{eq:oddspgenvec} induces the map
\reff{eq:oddindN56}, which maps the zero solution to a pure gauge
solution. We impose the condition that $\os=0$ be retained, then this
requires that we choose $v(t)=C_1 q_1$, where $C_1$ is a constant. The
pure gauge solution of the claim immediately follows after a
redefinition of the constant $C_1$. This solution can be superposed
with any other solution $(\os,\gamma^5,\gamma^6)$ and does not change
the odd shift function $\os$, from which the claim
holds. \endofproofmark

\medskip We can express the solution of the odd perturbation equations
in terms of the gauge-invariant variable $Q$ for a given gauge
choice. While as we remarked, a gauge choice means a choice of the
odd shift function $\os$, it is often useful to express $\os$ with
another gauge function.
\begin{prop}[Gauge-variable-free general solution]
  \label{prop:oddmostgen}
  Let $Q$ be the gauge-invariant variable for the odd perturbation
  which evolves according to Eq.\reff{eq:ddQ}, and let $\pth=\pth(t)$
  be an arbitrary function of time. Then, the
  solution of the odd perturbation equations \reff{eq:oddeqns}, up to
  \diffeos, is given in terms of $Q$ and $\pth$ by
  \begin{equation}
    \label{eq:odddeltaNaandg56}
    \begin{split}
      \os &= O(Q,\dot Q;\pth), \\
      \gamma^5 &= \sqrt{2(\lambda^2+2)}\, \pth \, Q, \\
      \gamma^6 &= (1-m \pth)Q,
    \end{split}
  \end{equation}
  where the function $O(Q,\dot Q;\pth)$ is defined by
\begin{equation}
  \label{eq:defO(Q,P;gf)2l}
  O(Q,\dot Q;\pth)\equiv \paren{{\dot \pth}-\frac{\dot q_1}{q_1}\pth
    +\frac{m}{\lambda^2+2}\frac{\dot q_1}{ u}}Q
  +\paren{\pth-\frac{m}{\lambda^2+2}\frac{q_1}{ u}}\dot Q.
\end{equation}
\end{prop}

\textit{Proof}: Remember that the gauge variable $\gamma_*$ is
governed by the evolution equation $\dot\gamma_*=J|_0$; see
Eqs.\reff{eq:oddF} and \reff{eq:dgam*=J}. For convenience we
re-express $P$ in terms of $\dot Q$ (and $Q$), using
the Hamilton equations. Similarly, we express $\pi_1$ and $\pi_1$ (the
background momentum variables) in terms of $q_1$, $q_2$ and their
time-derivatives, using the background Hamilton equations.  The
gauge evolution equation $\dot\gamma_*=J|_0$ then becomes
\begin{equation}
  \label{eq:oddgammadifwithOgen}
  \dot\gamma_*
  -\paren{\frac{\dot q_1}{q_1}-\frac{\dot \gf}{\gf}}\gamma_*
  =\rcp\gf \paren{O(Q,\dot Q;\gf)-  s},
\end{equation}
where $O(Q,\dot Q;\gf)$ is the same as that defined in
Eq.\reff{eq:defO(Q,P;gf)2l} (with $\pth$ replaced by $\gf$). Here,
$\gf$ is the mock shift function (the parameter function in
defining the gauge variable $\gamma_*$). From
Eq.\reff{eq:gamma56Q*}, we can think of $\gamma_*$ as a function of
$\gamma^5$ (or similarly of $\gamma^6$),
\begin{equation}
  \gamma_*=Q-\frac{\gamma^5}{\sqrt{2(\lambda^2+2)}\gf}.
\end{equation}
Substituting this into the above equation we obtain the following
evolution equation for $\gamma^5$,
\begin{equation}
  \label{eq:ODEgm5}
  \dot \gamma^5-\frac{\dot q_1}{q_1}\gamma^5=
  \sqrt2\nu\frac{q_1^2}{u}\paren{\frac{Q}{q_1}}\!\dot{\big.}
  +\sqrt{2(\lambda^2+2)}\os.
\end{equation}
The homogeneous solution to this equation is given by
$\gamma^5_\mathrm{(hom)}=C_1 q_1$, which is however found to be the
pure gauge solution (see the previous proposition), so we drop it. To
find the special solution we are interested in it is useful to
transform like $\gamma^5=f/q_1$. Using a new function $\pth$, we
rewrite the shift function $\os$ as
\begin{equation}
  \label{eq:oddsparam}
  \os= O(Q,\dot Q;\pth)
\end{equation}
so that the equation can explicitly be integrated. The
solution for $\gamma^5$ is then given by
$\gamma^5=\sqrt{2(\lambda^2+2)}\pth Q$. $\gamma^6$ follows from the
definition of $Q$, $\gamma^6=Q-(\nu/\sqrt2)\gamma^5$.  \endofproofmark

Note that there is no $\gf$-dependence in Eq.\reff{eq:ODEgm5},
therefore $\gamma^5$ and similarly $\gamma^6$ do not depend on the
mock shift $\gf$.

\remarkmark The above solution can be viewed as the ``zero-gauge''
solution corresponding to the ``gauge-fixing'' $\gamma_*=0$. Remember,
however, that we have freedom of choosing a function of time, the mock
shift $\gf$,  to define the gauge variable $\gamma_*$. Because of this
arbitrariness the ``gauge-fixing'' by $\gamma_*=0$ does not have any
invariant meaning. We can obtain the formula led by this
observation. In fact, from Eq.\reff{eq:oddgammadifwithOgen} the
condition $\gamma_*=0$ corresponds to taking the same relation
\reff{eq:oddsparam} (with $\pth$ replaced by $\gf$), and the same
formula for $\gamma^5$ and $\gamma^6$ is easily found from
Eq.\reff{eq:gamma56Q*} and $\gamma_*=0$. Therefore the function $\pth$
is the same as the mock shift $\gf$ \textit{under the condition}
$\gamma_*=0$. Nevertheless the above formula is valid regardless of
this condition, and in such a general situation the two functions
$\pth$ and $\gf$ have nothing to do with each other; $\pth$ just
parameterizes the odd shift $\os$, while $\gf$ parameterizes the
freedom of defining $\gamma_*$. In particular, $\pth$ can be zero, as
opposed to $\gf$.

\medskip

It is easy to express $\gamma^5$ and $\gamma^6$ directly
in terms of the odd shift function $\os$. To this we solve the
relation $\os= O(Q,\dot Q;\pth)$, which is a simple ODE for
$\pth$. The solution is
\begin{equation}
  \label{eq:gensolgf}
  \pth=\frac{q_1}{Q} {\int\brace{\frac{m}{\lambda^2+2}
      \frac{q_1}{u}
    \paren{\frac{Q}{q_1}}\!\dot{\big.}
    +\frac{\os}{q_1}}dt},
\end{equation}
where we dropped the homogeneous solution
$\pth_\mathrm{(hom)}=C_1q_1/Q$, which corresponds to the pure gauge
solution. If we substitute this solution into
Eqs.\reff{eq:odddeltaNaandg56}, we may obtain the explicit solution
for a given odd shift function $\os$. 

\medskip

One of the particular gauge choices of our interest may be
so-called the synchronous gauge, which is characterized by $\os=0$ for
the odd perturbations.
\begin{prop}[Synchronous gauge]
  \label{coro:odd1}
  Let $Q$ be the gauge-invariant variable for the odd perturbation
  which evolves according to Eq.\reff{eq:ddQ}. Then, the odd perturbed
  metric with vanishing shift is up to \diffeos\ given by
\begin{equation}
  \label{eq:oddsyncsol}
  \begin{split}
    \os &= 0, \\
    \gamma^5 &= \sqrt2\nu q_1\paren{
      \frac{Q}{u}- \int\frac{Q}{q_1}
    \paren{\frac{q_1}{u}}\!\dot{\big.}\,dt }, \\
    \gamma^6 &= \frac{q_2}{u}Q+\nu^2 q_1\int\frac{Q}{q_1}
    \paren{\frac{q_1}{u}}\!\dot{\big.}\,dt.
  \end{split}
\end{equation}
\end{prop}

\textit{Proof}: This can be immediately obtained by setting $\os=0$ in
Eq.\reff{eq:gensolgf};
\begin{equation}
  \label{eq:oddsolofgf}
  \pth=\pth_0\equiv{ \frac{m}{\lambda^2+2}\frac{q_1}{Q}
    \int \frac{q_1}{u}\paren{\frac{Q}{q_1}}\!\dot{\big.}dt }.
\end{equation}
We perform an integration by parts then substitute the resulting
$\pth=\pth_0$ into the general expression \reff{eq:odddeltaNaandg56}.
\endofproofmark

\medskip

Formula \reff{eq:odddeltaNaandg56} provides the most general form of
the odd metric functions expressed in terms of the gauge invariant
variable $Q$. Let us confirm that they are certainly isometric to each
other for any different choices of $\pth$.
\begin{prop}[Equivalence of the solutions]
  \label{prop:oddequiv}
  The odd perturbed spacetime metrics specified in Proposition
  \ref{prop:oddmostgen} with different $\pth(t)$ are all
  isometric to each other (up to the linear order).
\end{prop}

\proofmark We show that the perturbed metric with an arbitrary $\pth$
in Proposition \ref{prop:oddmostgen} is isometric to the one
in Proposition \ref{coro:odd1}, which in turn implies that all the
metrics are isometric to each other.  Let $g^{(0)}+\varepsilon \delta
g+O(\varepsilon^2)$ be the perturbed spacetime metric that the
perturbation metric in Proposition \ref{prop:oddmostgen} defines. Let
$ Y_a=v(t)V_a $ be an odd vector on this perturbed spacetime, and
consider the \diffeo\ generated by this.  From the map
\reff{eq:oddindN56} we find that
\begin{equation}
  \begin{split}
    \gamma^5 = \sqrt{2(\lambda^2+2)}\, \pth\, Q & \goes
    \sqrt{2(\lambda^2+2)}\, \pth\, Q+\sqrt{2(\lambda^2+2)}\, v \\
    & = \sqrt{2(\lambda^2+2)} (\pth\, Q+v), \\
    \gamma^6 = Q-m \pth Q & \goes
    Q-m \pth Q-mv \\ 
    &= Q-m(\pth Q+v),
  \end{split}
\end{equation}
from which we see that the transformation by the \diffeo\ is
equivalent to transforming
\begin{equation}
  \label{eq:gfandv}
  \pth\goes \pth'=\pth Q+v.
\end{equation}
On the other hand to make $\delta N_a=0$, it is found from the map
\reff{eq:oddindN56} that we need to choose $v(t)$ to satisfy
\begin{equation}
  \dot v-\frac{\dot q_1}{q_1}v+ O(Q,\dot Q;\pth)=0.
\end{equation}
We can easily solve this first order ODE, and obtain
\begin{equation}
  \label{eq:solvodd}
  v(t)=-q_1\int\frac{O(Q,\dot Q;\pth)}{q_1}dt.
\end{equation}
With this $v$, we have
\begin{equation}
  \pth '=\pth Q+v=\pth Q-q_1\int\frac{O(Q,\dot Q;\pth )}{q_1}dt.
\end{equation}
What we need to confirm is that this $\pth'$ does not depend upon the
original choice of $\pth$ and coincides with $\pth_0$ defined in
Eq.\reff{eq:oddsolofgf}. To this we first differentiate the above
equation (after an appropriate rearrangement) to delete the
integral. Then, we have the following ODE for $\pth '$:
\begin{equation}
  \dot \pth '+\paren{\frac{\dot Q}{Q}-\frac{\dot q_1}{q_1}}\pth '
  =\dot \pth +
  \paren{\frac{\dot Q}{Q}-\frac{\dot q_1}{q_1}-\frac{O(Q,\dot Q;\pth
  )}{\pth Q}}\pth .
\end{equation}
Substituting the definition \reff{eq:defO(Q,P;gf)2l} of $O(Q,\dot
Q;\pth )$, we obtain
\begin{equation}
  \dot \pth '+\paren{\frac{\dot Q}{Q}-\frac{\dot q_1}{q_1}}\pth '
  =\frac{m}{\lambda^2+2}\frac{q_1}{u}
  \paren{\frac{\dot Q}{Q}-\frac{\dot q_1}{q_1}}.
\end{equation}
Now, we can see that the apparent dependence upon the original
choice of $\pth $ has disappeared. Moreover this equation is the same
as $O(Q,\dot Q; \pth')=0$, which implies $\pth'=\pth_0$. Hence, all
the perturbed metrics with different $\pth$ are isometric to the
synchronous perturbed metric given in Corollary \ref{coro:odd1}.
\endofproofmark

Finally, we introduce a particularly useful gauge choice in the
present perturbation system.  With this choice the perturbed metric
does not contain, in contrast to the synchronous form given in
Corollary \ref{coro:odd1}, the time-derivative of $Q$ and can be
written in the simple multiplicative form $\delta g=Q\times G$, where
$G$ is a symmetric tensor that depends only on the background
variables.
\begin{prop}[Standard gauge]
  \label{prop:3}
  Let $Q$ be the gauge-invariant variable for the odd perturbation
  which evolves according to Eq.\reff{eq:ddQ}. When the perturbation
  of the spacetime metric is written simply in the product form
  $\delta \dg ab=QG^\mathrm{(odd)}_{ab}$ for a symmetric tensor
  $G^\mathrm{(odd)}_{ab}$ which is independent of $Q$, the tensor
  $G^\mathrm{(odd)}_{ab}$ is given by
  \begin{equation}
    \label{eq:oddrelsimpmetric}
    G^\mathrm{(odd)}_{ab}=
     {
      \sqrt2 \nu \frac{q_1}{u} \E5ab+
      {\frac{q_2}{u}} \E6ab+
      2\frac{m}{\lambda^2+2}\paren{\frac{q_1}{u}}\!
      \dot{\big.}\, V_{(a}dt_{b)}  }.
  \end{equation}
\end{prop}

\textit{Proof}:  Let us choose the function $\pth$ in formula
\reff{eq:odddeltaNaandg56} so as to eliminate the $\dot Q$-dependence
of the perturbed metric. When $m\neq 0$, we can eliminate the $\dot
Q$-dependence in $O(Q,\dot Q;\pth)$ by choosing (see
Eq.\reff{eq:defO(Q,P;gf)2l})
\begin{equation}
  \pth=\bar\pth\equiv\frac{m}{\lambda^2+2}\frac{q_1}{u},
\end{equation}
and in this case we have
\begin{equation}
  \label{eq:oddnicesol}
  \begin{split}
    \delta N_a &=
    \frac{m}{\lambda^2+2}\paren{\frac{q_1}{u}}\!\dot{\big.}\, QV_a, \\
    \gamma^5 &= \sqrt2 \nu \frac{q_1}{u} Q, \\
    \gamma^6 &= {\frac{q_2}{u}} Q.
  \end{split}
\end{equation}
This completes the proof for the case $m\neq 0$. Since it is possible
to take limit $m\goes0$ in the above expression, the same formula is
correct in the case $m=0$, as well. [In fact, when $m=0$ the solution
of \reff{eq:ODEgm5} is given by
\begin{equation}
  \gamma^5=\sqrt{2(\lambda^2+2)}q_1\int\frac{\os}{q_1}dt
\end{equation}
for a given $\os$. From the assumption, we must have $\os=fQ$ for a
function $f$ which is independent of $Q$, but in this case the above
$\gamma^5$ is not of the assumed form unless $f=0$, which
reproduces Eqs.\reff{eq:oddnicesol} for $m=0$.]  \endofproofmark

This gauge choice is similar to so-called the \textit{longitudinal
  gauge} known in the perturbation theory of isotropic and homogeneous
spacetimes. We call this gauge choice the \textit{standard gauge} for
the odd perturbation, due to its simplicity.

Other possible choices of gauge which are of some interest would be
those by $\gamma^5=0$ (``solid base'' gauge), and $\gamma^6=0$
(``orthogonality-preserving'' or simply ``orthogonal'' gauge),
which, respectively, correspond to $\pth=0$ and $\pth=1/m$.
(``Orthogonality'' here refers to the one between the base $\Sigma_g$
and fiber $S^1$.)

\section{Gauge Properties of the Even Perturbations}
\label{sec:evendiffeo}

Let us first see how \diffeos\ act on the even perturbations.

\begin{lemma}[Induced map]
\label{le:eveninduce}
Let $a_i$ ($i=0,1,3$) be arbitrary functions of time. Then,
the one-parameter \diffeo\ generated by an even vector
\begin{equation}
  \label{eq:even4vector}
  Y_a=a_0S(dt)_a+a_1 S_a+a_3\bar S\ss 3a
\end{equation}
induces the
following map $I_\mathrm{even}$ on the even perturbation variables:
\begin{equation}
  \label{eq:evendifftrans}
  \begin{split}
    \os_0 &\goes 
    \os_0 -\frac{1}{N}\paren{\dot a_0+\frac{\dot N}{N}a_0}, \\
    \os_1 &\goes \os_1+\dot a_1-q_1\inv\dot q_1a_1+\lambda a_0, \\
    \os_3 &\goes \os_3+\dot a_3-q_2\inv\dot q_2a_1-m a_0, \\
I_\mathrm{even}[a_0,a_1,a_3]:\quad\quad
\gamma^1 &\goes \gamma^1-\frac{\dot q_1}{N^2}a_0-\lambda a_1, \\
    \gamma^2 &\goes \gamma^2+\sqrt{2(\lambda^2+2)}a_1, \\
    \gamma^3 &\goes \gamma^3-\frac{\dot q_2}{N^2}a_0+2m a_3, \\
    \gamma^4 &\goes \gamma^4-m a_1+\lambda a_3,
  \end{split}
\end{equation}
where $\os_0$, $\os_1$, and $\os_3$ are, respectively, the even lapse
function, even base shift function, and even fiber shift function.
\end{lemma}

\proofmark This is a straightforward computation of
$2\nabla_{(a}Y_{b)}$.  For our background metric $\dg
ab=-N^2(dt)_a(dt)_b+q_{ab}$, it holds that
\begin{equation}
  \begin{split}
  \nabla_a(dt)_b &= -\frac1{2N^2}\dot \ggg_{ab} \\
  &= \frac{\dot N}{N}(dt)_a(dt)_b-\rcp{2N^2}\dot q_{ab}.
  \end{split}
\end{equation}
Also, for a spatial vector $A_a$ such that $A_a(\del_t)^a=0$, we
find
\begin{equation}
  \nabla_aA_b=D_aA_b-q^{cd}A_c\dot q_{d(a}(dt)_{b)}.
\end{equation}
Using these relations we can find
\begin{equation}
  \begin{split}
    2\nabla_{(a}Y_{b)} &= 
    2S\paren{\dot a_0+\frac{\dot N}{N}a_0}(dt)_a(dt)_b \\
    & + 2\paren{\dot a_1-q_1\inv\dot q_1a_1+\lambda a_0}
    (dt)_{(a}S_{b)} \\
    & + 2\paren{\dot a_3-q_2\inv\dot q_2a_1-m a_0}
    \bar S(dt)_{(a}\ss3{b)} \\
    & -\frac{a_0}{N^2}\paren{\dot q_1\E1ab+\dot q_2\E3ab} \\
    & -a_1\paren{\lambda\E1ab-\sqrt{2(\lambda^2+2)}\E2ab+m\E4ab} \\
    & +a_3\paren{2m\E3ab+\lambda\E4ab},
  \end{split}
\end{equation}
from which we can read off the map claimed. \endofproofmark

It is straightforward to confirm that the even gauge-invariant
perturbation variable $Q= -\Delta_1\gamma^1-\Delta_3\gamma^2
+\gamma^3-\frac{2m}{\lambda}\gamma^4$ is certainly invariant under
this map, $Q\goes Q$.  (To confirm this we may need the relation
$\Delta_1=\dot q_2\dot q_1\inv$, which is obtained from the definition
\reff{eq:defDelta} using the background Einstein equations
\reff{eq:backeom}.)

\medskip

As in the odd case let us find the pure gauge solution that can be
superposed on any solution with given even lapse and shift functions.
\begin{prop}[Pure gauge solution]
  Consider the even perturbation equations with a given even lapse
  function $\os_0$ and even shift functions $\os_1$ and $\os_3$.
  Let $C_0$, $C_1$, and $C_2$ be constants and let
  \begin{equation}
    \begin{split}
      a_0 &= \frac{C_0}{N}, \\
      a_1 &= -\lambda q_1 C_0\int\frac{dt}{Nq_1}+C_1q_1, \\
      a_3 &= m q_2 C_0\int\frac{dt}{Nq_2}+C_2q_2.
    \end{split}
  \end{equation}
  Then, the set of functions
\begin{equation}
  \begin{split}
    \gamma^1_\mathrm{(d)} &= -\frac{\dot q_1}{N^2}a_0-\lambda a_1, \\
    \gamma^2_\mathrm{(d)} &= \sqrt{2(\lambda^2+2)}a_1, \\
    \gamma^3_\mathrm{(d)} &= -\frac{\dot q_2}{N^2}a_0+2m a_3, \\
    \gamma^4_\mathrm{(d)} &= -m a_1+\lambda a_3,
  \end{split}
\end{equation}
represents the three parameter solution of the equations
which is generated by \diffeos.
\end{prop}

\textit{Proof}: Use the same logic of Proposition
\ref{prop:2}. Consider the \diffeos\ generated by the even vector of
the form \reff{eq:even4vector}. From Eqs.\reff{eq:evendifftrans} we
find that to keep $\delta N_a=0$
\begin{equation}
  \begin{split}
    \dot a_1-q_1\inv\dot q_1a_1 &= -\lambda a_0, \\
    \dot a_3-q_2\inv\dot q_2a_1 &= m a_0
  \end{split}
\end{equation}
should hold. The solution is easily found, provided that $a_0$ is given:
\begin{equation}
  \begin{split}
      a_1 &= -\lambda q_1 \int\frac{a_0dt}{q_1}+C_1q_1, \\
      a_3 &= m q_2\int\frac{a_0dt}{q_2}+C_2q_2.
  \end{split}
\end{equation}
On the other hand, we also find that to keep $\delta N=0$
\begin{equation}
  {\dot a_0+\frac{\dot N}{N}a_0}=0.
\end{equation}
The solution is
\begin{equation}
  a_0 =C_0 N\inv.
\end{equation}
The claim follows by substituting this into the above $a_1$ and $a_3$,
and applying the corresponding $I_\mathrm{even}$ map to the trivial
(i.e., zero) solution.
\hfill\rule{.5em}{.5em}

\medskip

\def\bigdotone{\paren{\frac{q_1^2\Sigma}{\Omega}}\!\dot{\big.}}
\def\bigdottwo{\paren{\frac{q_1q_2}{\Omega}}\!\dot{\big.}}
\def\bigpatthree{\paren{\bigdotone+\frac{q_1^2\dot\Sigma}{\Omega}}}
As in the odd case, we can find the longitudinal gauge-like solution,
where the gauge-invariant variable $Q$ appears in a simple
multiplicative form.
\begin{prop}[Standard gauge]
  \label{prop:evenlon}
  Let $Q$ be the gauge-invariant (configuration) variable for the even
  perturbation which evolves according to
  Eqs.\reff{eq:evenhameqsgi}. When the perturbation of the spacetime
  metric is written in the product form $\delta \dg
  ab=QG^\mathrm{(even)}_{ab}$ for a symmetric tensor
  $G^\mathrm{(even)}_{ab}$ which is independent of $Q$ and depend only
  on background quantities, the tensor $G^\mathrm{(even)}_{ab}$ is
  given by
  \begin{equation}
    \label{eq:evenrelsimpmetric}
    \begin{split}
    G^\mathrm{(even)}_{ab} &= \frac{-1}{\Omega}\bigg[
     q_1q_2\E1ab+2q_1^2\Delta_2\E2ab \\
    & \hspace{3em}
    -q_2(q_1\Delta_1+q_2)\E3ab
    +2\frac{m}{\lambda}q_1q_2\E4ab \\
    & \hspace{3em} 
    + \frac{N^2}{\lambda^2+2}
    \paren{4(\lambda^2-1)q_1\pi_2^{-2}+3q_2} S(dt)_a(dt)_b \bigg] \\
    & \hspace{1em} 
    +\frac{2N^2}{\dot q_1}
    \paren{\Upsilon_1+\frac{\lambda^2}{\lambda^2+2}\Upsilon_2}
    S(dt)_a(dt)_b \\
     & \hspace{1em} 
     -\frac{2}{\lambda^2+2}(\Upsilon_2+\Upsilon_3)\paren{
     \lambda S_{(a}(dt)_{b)}+m \bar S\ss3{(a}(dt)_{b)} } \\
    &  \hspace{1em} 
    - \frac{4m}{\lambda^2}\Upsilon_1 \, \bar S\ss3{(a}(dt)_{b)},
    \end{split}
  \end{equation}
  where 
  \begin{equation}
      \Upsilon_1 \equiv \bigdottwo, \quad
      \Upsilon_2 \equiv \bigdotone, \quad
      \Upsilon_3 \equiv \frac{q_1^2\dot\Sigma}{\Omega},
  \end{equation}
  and symbols $\Delta_1$, $\Delta_2$, $\Sigma$, and $\Omega$ are those
  defined in Eqs.\reff{eq:defDelta}, \reff{eq:defSigma}, and
  \reff{eq:evendefOmega}.
\end{prop}

\textit{Sketch of the derivation of \reff{eq:evenrelsimpmetric}}: The
assumption of the gauge leads us to put $\gamma^i=\pth_i Q$
($i=1,2,4$), where $\pth_i$ are functions of the background
quantities. ($\gamma^3$ can be determined from the definition of $Q$.)
From Eqs.\reff{eq:eg*i=gi}, this also means $\gamma_*^i=\pth_i Q$
($i=1,2,4$). We substitute these relations into the gauge equations
\reff{eq:egeqs}. We then eliminate $P$ and rewrite the gauge equations
in terms of $Q$ and $\dot Q$ by using (one of) the Hamilton equations
\reff{eq:evenhameqsgi}, and determine the functions $\pth_i$ so that the
coefficients of $\dot Q$ in those equations vanish. Those equations
are simple algebraic equations which can be immediately solved, and
the solution is written in terms of the background quantities only. We
have
\begin{equation}
  \begin{split}
    \pth_1 &= -\frac{q_1q_2}{\Omega}, \\
    \pth_2 &= -\frac{2q_1^2\Delta_2}{\Omega}, \\
    \pth_4 &= -\frac{2m}{\lambda}\frac{q_1q_2}{\Omega}.
  \end{split}
\end{equation}
From these, $\gamma^i$ ($i=1,2,4$), and therefore $\gamma^3$ are
determined. The even lapse and shift functions $\os_i$ ($i=0,1,3$) are
determined from the gauge equations with the above choice of
$\pth_i$. They are also algebraic (linear) equations for $\os_i$, so
it is easy to solve, and the solution automatically satisfies the
gauge assumption. We have
\begin{equation}
  \begin{split}
    \os_0 &= {N} \bigg\{
      \frac{\Omega\inv}{\lambda^2+2}
    \paren{2(\lambda^2-1)q_1\pi_2^{-2}+\frac32q_2} \\
    & \hspace{3em} -\frac{1}{\dot q_1}
    \bra{\frac{\lambda^2}{\lambda^2+2}\bigdotone+\bigdottwo}
  \bigg\} \, Q, \\
    \os_1 &= -\frac{\lambda}{\lambda^2+2} \bigpatthree Q, \\
    \os_3 &= -m\bra{ \frac2{\lambda^2}\bigdottwo
      +\frac{1}{\lambda^2+2}\bigpatthree }Q. \\
  \end{split}
\end{equation}
(We have used the background constraint $\mathcal{H}=0$ to simplify
the result, e.g., eliminating one of the background variables $\pi_1$
in $\os_0$.)  From these we obtain \reff{eq:evenrelsimpmetric}.
\endofproofmark

Due to the complexity (or lengthiness) of the even gauge equations it
may be impractical to write down a general solution for an arbitrary
gauge. Instead, we can derive any solution with a desired gauge by
acting the map $I_\mathrm{even}$ on the above \textit{standard gauge
  solution}.  In this case, the functions which are considered free
are the three functions $a_i$ ($i=0,1,3$).

\section{Summary and Discussions}
\label{sec:summary}

A brief summary first. We have seen that mode functions on a closed
orthogonal Bianchi III space are specified by two kinds of
eigenvalues; the fiber eigenvalue $m$ and the base eigenvalue
$\lambda$. Using the metrics and covariant derivatives on the base
$\Sigma_g$ and the fiber $S^1$, we have constructed the ``even'' mode
vectors and tensors from the mode functions. Using the area two-form
$\varepsilon_{ab}$ on the base in addition to the same metrics and
covariant derivatives, we have constructed the ``odd'' mode vectors
and tensors. The ``harmonic'' mode vectors and tensors are associated
with the (Hodge) harmonic vectors on the base and the mode functions
on the fiber. The ``\TT'' mode tensors are associated with the \TT\
tensors on the base and the mode functions on the fiber.  This way the
mode quantities can naturally be categorized into the \textit{even,
  odd, harmonic}, and \TT\ ones. Perturbations of vacuum solutions
associated with different kinds (i.e., even, odd, $\cdots$) and/or
different eigenmodes are all independent of each other, in other
words, they all decouple, as stated in Theorem \ref{th:1}. This makes
it possible to treat them separately.

We have defined mode tensors $\E iab$ ($i=1\sim 9$) for a given set of
$m$ and $\lambda$ ($i=1\sim6$) or for a given $m$ ($i=7\sim 9$), which
are eigentensors of the Laplacian and are $L^2$-orthogonal to each
other with respect to the standard $L^2$-norm. To separate variables
we expand the first variation of the spatial metric in terms of the
mode tensors, resulting in getting the coefficients
$\gamma^i=\gamma^i(t)$ as the fundamental perturbation configuration
variables. The canonical conjugates can also be expanded by the same
mode tensors and we obtain $p_i=p_i(t)$ as the conjugates of
$\gamma^i$. The Hamiltonians for these perturbation variables are
provided in Eq.\reff{eq:Hoddtilde} (odd part) and \reff{eq:Heventilde}
(even). That the harmonic part can be obtained from the odd one by
formally taking limit $\lambda\goes0$ was pointed out in
Sec.\ref{sec:ham}. The Hamilton equations for the \TT\ part are given
in Eqs.\reff{eq:TTeom}.

A set of Hamilton (i.e., evolution) equations for perturbation
variables is generally a system of complicatedly coupled linear ODEs
which depend on choice of gauge, i.e., depend on choice of variations
of lapse function $\delta N$ and shift vector $\delta N_a$. One of our
main tasks in our course of analysis is to decouple the system into
gauge-dependent and independent parts for each one of the even and odd
perturbation systems. (The harmonic one can be thought of as a limit
of the odd system, as already remarked. Since the \TT\ system is not
constrained, the fundamental variables for the \TT\ system are
themselves gauge-invariant.) One of the advantages of employing
Hamiltonian formalism is that this task amounts to finding a certain
kind of canonical transformation.

We have used the \textit{method of generating function} to find such
desired canonical transformations. This systematic method was crucial
for our achievement of the task. The decoupled-form Hamiltonians are
given in Eq.\reff{eq:oddtotHam} with \reff{eq:HoddGI} and
\reff{eq:oddF} (odd) and Eq.\reff{eq:Hevendec} with \reff{eq:HevenGI}
and \reff{eq:evenmulti} (even). The wave equations for gauge-invariant
variables $Q$ (the definition of $Q$ is different depending on the
type of perturbations), or the Hamilton equations for $Q$ and its
conjugate $P$ are given in Eq.\reff{eq:ddQ} (odd) and
\reff{eq:evenhameqsgi} (even). These evolution equations for the
gauge-invariant variables are the ``heart'' of the perturbation
dynamics.

To recover an entire perturbed spacetime metric, however, it is
unavoidable to choose a particular gauge. As a useful choice, we have
defined the \textit{standard gauge}, in which the gauge-invariant
variable $Q$ appears in the perturbed metric in the simplest manner
where the metric is expressed in terms of $Q$ only and free from $\dot
Q$ or $P$. The forms of metrics in this gauge are given in
Propositions \ref{prop:3} and \ref{prop:evenlon}.

We have thus resolved all the fundamental issues concerning the
technical aspects of the perturbations of closed orthogonal Bianchi
III vacuum spacetimes, including separation of variables, decoupling
into gauge-dependent and independent parts, derivation of wave
equations, and choice of gauge.

Our primary interest concerning properties of the perturbations of the
Bianchi III spacetime may be their future asymptotic behaviors.  For
this not only do we need to perform a further careful analysis of the
wave equations but we need to clarify how to subtract the effect of
the anisotropic expansion of the background and thereby get the
``net'' effect of the perturbations. This nontrivial task is
indispensable to see under what condition the system should be called
``stable,'' or to say at what rate the perturbations decay or
grow. These subjects will be discussed in a separate paper.  In the
remaining part, let us discuss the scalar and electromagnetic field
equations and the \TT\ perturbation equations, which are simpler than
the other perturbation equations.

In Appendix \ref{apx:scalar} we have dealt with the wave equation for
massless scalar fields. When the background is Minkowski (with a
slicing that is Bianchi III), we can obtain the exact solutions for
the (mode-decomposed) wave equations for all possible eigenvalues.
The solutions tell us that the fields with $m\neq0$ all decay as
$t\goes \infty$ at the same rate (See Eq.\reff{eq:scdecratemneq0}),
while the $U(1)$-symmetric (i.e., $m=0$) fields decay at several
different rates depending on the value of $\lambda$ (See
Eq.\reff{eq:esswk0n0asy}). It is however true that all modes except
the zero mode are decaying. It seems therefore plausible that all the
modes are decaying (except the zero mode) for any general orthogonal
Bianchi III backgrounds, too.

The (source-free) electromagnetic field equation has been dealt with
in the subsequent appendix. Because of the vector-like nature of the
electromagnetic fields, the mode-decomposed wave equations belong to
either the even type or odd type (or the harmonic type, which can be
obtained as the limit $\lambda\goes 0$ of the odd type). Interestingly
enough, the wave equations for both even and odd fields coincide with
those for the scalar field (after a normalization if necessary) when
the background is Minkowski. They are therefore decaying again on this
background (if we think of the normalizations involved natural).

Among the gravitational perturbations the simplest are the \TT\
perturbations. This kind of perturbation is, as noted in
Sec.\ref{sec:TT}, the only exceptional case among the four kinds where
it is clear how the variable should be normalized to see the
stability.  We have seen in the same section that the wave equations
for the (normalized) \TT\ perturbation variable, and therefore their
asymptotic behaviors, are exactly the same as those for the scalar
fields (for $\lambda=0$). The background is therefore expected to be
stable against the \TT\ perturbations.

As noted, detailed studies of dynamical properties of the
perturbations will be reported in a separate paper\cite{YT}.

\appendix

\section{Spatially Closed Bianchi III Spacetimes}
\label{apx:compact}

In this section we discuss how we can construct the spatially closed
Bianchi III spacetimes. The so-called \Teich\ parameters are the key
ingredient to understand the compactifications involved. \textit{The
  \Teich\ space of a manifold $M$ modeled on a homogeneous space $X$}
is the quotient space $\mathcal{M}/\mathcal{D}_0$, where
$\mathcal{M}=\mathcal{M}(M,X)$ is the space of the locally homogeneous
metrics on $M$ which are locally isometric to $X$, and
$\mathcal{D}_0=\mathcal{D}_0(M)$ stands for the \diffeos\ of $M$ that
are connected to the identity. \textit{\Teich\ parameters} are defined
as coordinates in a \Teich\ space \footnote{Beware that several
  different definitions of ``\Teich\ space'' are used in the
  literature. In general, they are equivalent for two dimensional
  surfaces, but not for higher dimensional manifolds, including
  three-dimensional ones. Ours follow \cite{KLR,Oh,KTH,TKH1,TKH2}.}
and they are interpreted as the configuration variables that emerge as
a result of compactification.  We note that \Teich\ parameters can
also be understood as the parameters in the covering group acting on
$X$. In the following we first show how we can compactify Bianchi III
homogeneous manifolds by explicitly finding covering groups acting on
them, and then generalize the compactification to ones for spacetimes.

Let $H^2=(\R^2,\tilde h_{ab})$ be the standard hyperbolic plane, and
$E^1=(\R,\tilde l_{ab})$ be the standard Euclidean line. We introduce
the standard coordinate systems with which these metrics are
expressed\footnote{The hyperbolic plane model with this metric defined
  in the region $y>0$ is known as the \textit{upper-half plane model},
  and the plane is often denoted like $\R^2_{+}$ to emphasize it is
  defined on the upper-half plane. In this section however we
  understand that $\R^2_{+}$ is homeomorphic to $\R^2$ and do not
  distinguish those two for notational simplicity.}
\begin{equation}
  \tilde h_{ij}dx^idx^j= \frac{dx^2+dy^2}{y^2},
\end{equation}
and
\begin{equation}
  \tilde l_{ij}dx^idx^j= dz^2.
\end{equation}
The coordinates are defined in the region $x\in (-\infty,\infty)$,
$y\in (0,\infty)$, and $z\in (-\infty,\infty)$, and they span the
whole $H^2$ and $E^1$.

As well known $H^2$ can be compactified to a higher genus surface
$\Sigma_g$ ($g\geq 2$) by the left-action of a discrete subgroup
$\Gamma^H$
of the orientation-preserving isometry group $\Isom_+ H^2$:
\begin{equation}
  \pi_H: \quad H^2\goes (\Sigma_g,\dh ab)=H^2/\Gamma^H.
\end{equation}
The topology of the closed surface $\Sigma_g$ can be specified by its
fundamental group $\pi_1(\Sigma_g)$, which is an infinite group
generated by $2g$ generators $\alpha_i$ and $\beta_i$ ($i=1\sim g$)
which satisfy a single relation $[\alpha_1,\beta_1]\cdots
[\alpha_g,\beta_g]=1$. That is, in the standard notation,
\begin{equation}
  \pi_1(\Sigma_g)=\angl{\alpha_1,\cdots,\alpha_g,
    \beta_1,\cdots,\beta_g|
    [\alpha_1,\beta_1]\cdots [\alpha_g,\beta_g]=1},
\end{equation}
where $[\alpha_i,\beta_i]\equiv
\alpha_i\beta_i\alpha_i\inv\beta_i\inv$ is the commutator of group
elements. The covering group $\Gamma^H$ is a representation of this
infinite group into $\Isom_+ H^2$.

The group $\Isom_+ H^2$ is isomorphic to the projective real special
linear group of second rank, $\Isom_+ H^2\simeq \PSL{2,\R}$. Suppose
we embed the generators $\alpha_i$ and $\beta_i$ into $\PSL{2,\R}$ so
that they satisfy the relation for $\pi_1(\Sigma_g)$.  (That is, we
identify $\alpha_i$ and $\beta_i$ with elements in $\PSL{2,\R}$ so
that they satisfy the relation for $\pi_1(\Sigma_g)$ with respect to
the multiplications in $\PSL{2,\R}$.)  However, we do not need an
explicit representation of such $\alpha_i$ and
$\beta_i\in\PSL{2,\R}$. We just note they generate $\Gamma^H$, which
we denote
\begin{equation}
  \Gamma^{H}=\brace{\alpha_1,\cdots,\alpha_{g},
    \beta_1,\cdots,\beta_{g}},
\end{equation}
where $\alpha_i,\, \beta_i \in
\PSL{2,\R}$ ($i=1\sim g$).

Note that, since $\PSL{2,\R}$ is three dimensional, we need (at most)
$2g\times 3=6g$ parameters to represent $\Gamma^{H}$. However, since
the generators must satisfy the relation $[\alpha_1,\beta_1]\cdots
[\alpha_g,\beta_g]=1$, the number of independent parameters reduces by
three, because this relation has three components. Furthermore,
remember the fact that different $\Gamma^H$ and $\Gamma^H{}'$ can
result in the same closed surface, i.e., $H^2/\Gamma^H$ and
$H^2/\Gamma^H{}'$ are isometric to each other, if $\Gamma^H=a\inv
\circ \Gamma^H{}'\circ a$ for an $a\in \Isom H^2$. (This map is called
\textit{conjugation}. See, e.g., \cite{TKH2}.)  Since $\Isom H^2$ is
three dimensional, these conjugation maps reduce the number \textit{at
  most} by another three, and we know they do reduce it by the maximal
number, hence resulting in $6g-6$ independent parameters in
$\Gamma^H$. This count corresponds to the well-known fact
(e.g.\cite{Tro}) that a higher genus surface $\Sigma_g$ with genus
$g\geq2$ has $6g-6$ \Teich\ parameters.

Let us generalize the covering group $\Gamma^H$ of the closed surface
$\Sigma_g$ to that of the trivial $S^1$-fiber bundle $M\simeq
\Sigma_g\times S^1$. The topology of this closed manifold can be
uniquely specified by its fundamental group, which can be represented
by
\begin{equation}
  \pi_1(M)=\angl{\alpha_1,\cdots,\alpha_g,
    \beta_1,\cdots,\beta_g,\varphi \,|
    [\alpha_1,\beta_1]\cdots [\alpha_g,\beta_g]=
    [\alpha_i,\varphi]=[\beta_i,\varphi]=1},
\end{equation}
where the new generator $\varphi$ corresponds to the loop along the
$S^1$-fiber. (This expression of $\pi_1$ follows from
the general theory of $S^1$-bundles over a surface
or more general \textit{Seifert fiber spaces}. See, e.g., \cite{Hem}. )

We consider, as the universal cover, the direct product $H^2\times
E^1$ with the rescaled metric
\begin{equation}
  \label{eq:rescdm}
  \tilde q_{ab}=q_1\tilde h_{ab}+q_2\tilde l_{ab},
\end{equation}
where $q_1$ and $q_2$ are constants.  The isometry group of this
manifold is the direct product of those of $H^2$ and $E^1$, i.e.,
$\Isom H^2\times E^1\simeq \Isom H^2\times \Isom E^1$, where the
orientation-preserving component of $\Isom E^1$, $\Isom_+E^1$, is the
one-dimensional translation group. So, an element $\vec{\alpha}\in
\Isom_0 H^2\times E^1\simeq\Isom_+ H^2\times \Isom_+ E^1$ can
be\footnote{The group $\Isom_0 H^2\times E^1\simeq\Isom_+ H^2\times
  \Isom_+ E^1$ is the identity component of $\Isom_+H^2\times E^1$,
  the orientation-preserving isometry group of $H^2\times E^1$. We
  focus on this component to realize a compactification. To
  generalize the argument to the full $\Isom_+H^2\times E^1$, one may
  need to consider the orientation-reversing isometries of $H^2$ and
  $E^1$, as well.}  expressed as
\begin{equation}
  \vec{\alpha}=\svector{\alpha}{s},
\end{equation}
using $\alpha\in\Isom_+ H^2$ and $s\in\Isom_+ E^1$. This acts on
points $\x=(x,y,z)\in M$, viewing $\x=(x,y)\times \{z\}\in H^2\times
E^1$. The multiplication for two isometries $\alpha$, $\alpha'\in
\Isom_0 H^2\times E^1$ is defined by
\begin{equation}
  \vec{\alpha}\cdot\vec{\alpha}'
  =\svector{\alpha}{s}\cdot\svector{\alpha'}{s'}
  =\svector{\alpha\alpha'}{s+s'},
\end{equation}
where $\alpha\alpha'$ is understood as the multiplication in
$\PSL{2,\R}$.

\begin{prop}
\label{prop:Gamma}
Let
\begin{equation}
  \Gamma^{H}=\brace{\alpha_1,\cdots,\alpha_{g},
    \beta_1,\cdots,\beta_{g}}
\end{equation}
be the covering group for $(\Sigma_g,\dh ab)$ acting on $H^2$.  Let
$M$ be the product $\Sigma_g\times S^1$ and let $(M,q_{ab})$ be a
locally homogeneous manifold modeled on $H^2\times E^1=(\R^3,\tilde
q_{ab})$, where $\tilde q_{ab}$ is the rescaled metric
\reff{eq:rescdm}. Let us represent $(M,q_{ab})$ with the following
covering map
\begin{equation}
  \label{eq:cptpi}
  \pi:\quad (\R^3,\tilde q_{ab})\goes
  (M,q_{ab})=(\R^3,\tilde q_{ab})/\Gamma.
\end{equation}
Then the covering group $\Gamma\subset\Isom_0 H^2\times E^1$ can be
represented as
\begin{equation}
  \label{eq:generalGamma}
  \Gamma=\brace{
    \svector{\alpha_1}{v_1},\cdots,\svector{\alpha_g}{v_g},
    \svector{\beta_1}{w_1},\cdots,\svector{\beta_g}{w_g},
    \svector{1}{\varphi} },
\end{equation}
where $v_i\in\R$, $w_i\in\R$ and $\varphi>0$ are arbitrary
constants. Here, the $8g-5$ independent parameters in $\Gamma$, i.e,
$6g-6$ parameters in $\Gamma^H$ plus the $2g+1$ parameters $v_i$,
$w_i$ and $\varphi$, span the \Teich\ space of $M$ modeled on $(\R^3,
\tilde q_{ab})$.
\end{prop}

\textit{Proof}: Let us denote the generators of $\Gamma$ as in
\begin{equation}
  \Gamma=\brace{\vec\alpha_1,\cdots,\vec\alpha_g,
    \vec\beta_1,\cdots,\vec\beta_g,\vec \varphi},
\end{equation}
where $\vec\alpha_i,\vec\beta_i,\vec\varphi\in \Isom_0 H^2\times E^1$.
From the relation $[\vec\alpha_1,\vec\beta_1]$ $\cdots$
$[\vec\alpha_g,\vec\beta_g]=1$, it is obvious that the $\Isom_+
H^2$-part of $\vec \alpha_i$ and $\vec \beta_i$ should satisfy the
unarrowed same relation, from which we have
\begin{equation}
  \label{eq:cptab}
  \vec\alpha_i=\svector{\alpha_i}{v_i},\quad
  \vec\beta_i=\svector{\beta_i}{w_i},\quad (i=1\sim g)
\end{equation}
where $v_i$ and $w_i$ are real constants, and $\alpha_i$ and $\beta_i$
are the generators of $\Gamma^H$. Conversely, if
$\vec\alpha_i$ and $\vec\beta_i$ are given in this form, 
from the fact that $\Isom_+ E^1$ is abelian, we see
\begin{equation}
 [\vec\alpha_1,\vec\beta_1]\cdots [\vec\alpha_g,\vec\beta_g]=
 \svector{1}{0}=1,
\end{equation}
that is, the $2g$ generators \reff{eq:cptab} satisfy
the required relation for $\pi_1(M)$ for arbitrary $v_i$ and $w_i$.

For an isometry $\vec a\in\Isom_0 H^2\times E^1$ of the form
\begin{equation}
  \vec a=\svector{a}{s},
\end{equation}
the conjugation map acts on $\vec\alpha_i$ (and similarly on
$\vec\beta_i$) as
\begin{equation}
  \vec a\inv\circ\vec\alpha_i\circ\vec a
  =\svector{a\inv\alpha_ia}{v_i}.
\end{equation}
That is, the conjugation is effective only on the $\Isom_+ H^2$ part.
From the discussion on $\Gamma^H$, this fact merely implies that the
$2g$ generators $\vec\alpha_i$ and $\vec\beta_i$ contain $6g-6$
independent parameters.

It is also easy to check that to meet the remaining commutativity
relations $[\vec\alpha_i,\vec\varphi]=[\vec\beta_i,\vec\varphi]=1$
the $S^1$-fiber generator $\vec\varphi$ should be of the form
\begin{equation}
  \vec\varphi=\svector{1}{\varphi},
\end{equation}
where $\varphi>0$ is an arbitrary positive parameter. ($\varphi$
cannot be zero for $M$ not to degenerate. Moreover,
$\vec\varphi=(1,\varphi)$ is conjugate to
$\vec\varphi_{-}=(1,-\varphi)$ by the reflection isometry
$r: (x,y,z)\goes (x,y,-z)$. So, it is enough to consider positive
$\varphi$.) The conjugation maps with respect to the isometries
connected to the identity do not affect this generator. This completes
the proof.  \hfill\rule{.5em}{.5em}

The \Teich\ space of $M$ was first discussed in \cite{KLR,Oh}. 

\begin{figure}[hbtp]
  \centering
  \scalebox{.5}{\includegraphics{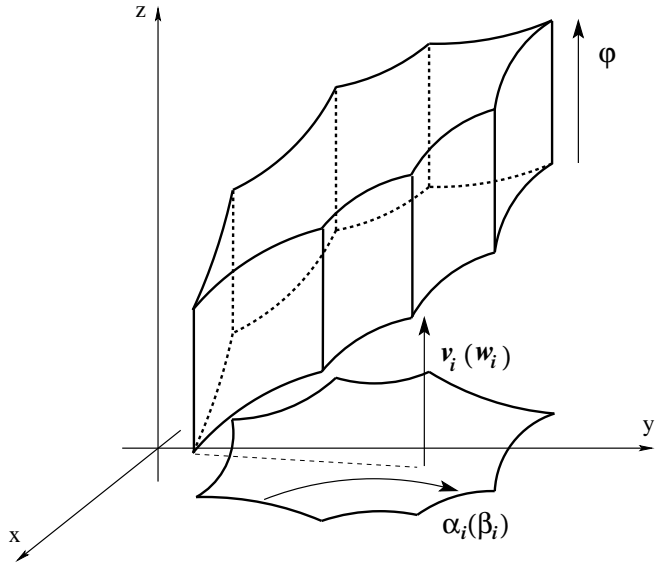}}
  \caption{A fundamental domain of a compactified Bianchi III manifold
    (the decahedron in the figure) and its projection onto a
    $H^2$-plane (the octagon in the $x$-$y$ plane). (For convenience
    of presentation, the $x$-$y$ coordinates in this figure are
    assumed to be those of the Poincar\'e model rather than the upper
    half plane model.)}
  \label{fig:fdomain}
\end{figure}

Figure \ref{fig:fdomain} pictures a fundamental domain of a
compactified Bianchi III manifold and its projection onto the
$H^2$-plane of $z=0$.

Note that the universal cover $H^2\times E^1$ has two natural
structures; the foliation by $E^1$, and the foliation by $H^2$. It is
easy to see that the foliation by $E^1$, which is equivalent to the
foliation by our $z$-axes, descends to a foliation by $S^1$ after the
compactification we discussed (cf. Theorem 4.13 in \cite{Sco}).  In
contrast, the above proposition tells us that the $H^2$-foliation does
\textit{not} descend to a $\Sigma_g$-foliation unless $v_i=w_i=0$ for
all $i$.  (Note that the $H^2$-foliation can be expressed in terms of
the coordinates as the foliation by $z=\mathrm{constant}$
surfaces. So, the condition that the $H^2$-foliation descends to
a $\Sigma_g$-foliation is that \textit{each} $z=\mathrm{constant}$
surface is compactified to $\Sigma_g$. This does not happen unless
$v_i=w_i=0$, as seen from the previous proposition.)  We call these
$2g$ parameters $v_i$ and $w_i$ the \textit{distortion parameters}.
\footnote{We do not use the word ``twist'' as in \cite{KTH} to avoid
  confusions with those in the so-called pants-decompositions of
  hyperbolic higher genus surfaces.}  It may be natural to consider
the subclass where the inheritance $H^2$-foliation $\goes$
$\Sigma_g$-foliation holds:
\begin{definition}
  When the distortion parameters all vanish, i.e., when the covering
  group $\Gamma$ in Proposition \ref{prop:Gamma} is of
  the form
\begin{equation}
  \label{eq:gammaorth}
  \Gamma_\mathrm{orth}=\brace{
    \svector{\alpha_1}{0},\cdots,\svector{\alpha_g}{0},
    \svector{\beta_1}{0},\cdots,\svector{\beta_g}{0},
    \svector{1}{\varphi} },
\end{equation}
we say that the resulting closed manifold $(M,q_{ab})$ is
{orthogonal}. 
\end{definition}
Note that the closed hyperbolic surface $(\Sigma_g,q_1\dh ab)$ does
\textit{not} exist (or, is \textit{not} embedded) in $(M,q_{ab})$,
unless $(M,q_{ab})$ is orthogonal.  \footnote{It is apparent that even
  if $(M,q_{ab})$ is not orthogonal there exist a foliation of $M$ by
  $\Sigma_g$, since $M$ is (topologically) the product $\Sigma_g\times
  S^1$. The induced metric $h'_{ab}$ on each leaf of such a foliation
  is however not hyperbolic, $h'_{ab}\neq q_1\dh ab$ ($q_1$ is a
  constant).}

When $(M,q_{ab})$ is orthogonal, the manifold admits both the
foliation by $(\Sigma_g,q_1\dh ab)$ and foliation (or fibration) by
$(S^1,q_2l_{ab})$. The metric $q_{ab}$ which is induced from the
universal cover one $\tilde q_{ab}$ can be written in the form
\begin{equation}
  \label{eq:themetricq}
  q_{ab}=q_1 \dh ab+q_2 l_{ab}.
\end{equation}
A tangent vector
$v^a=v^1\del_x+v^2\del_y+v^3\del_z$ on $p\in M$ can be uniquely
decomposed as
\begin{equation}
  \label{eq:vecdecomp}
  v^a=v_H^a+v_S^a,
\end{equation}
where $v_H^a=v^1\del_x+v^2\del_y$ is the part tangent to the
$(\Sigma_g,q_1 \dh ab)$-leaf to which the point $p$ belongs, and
$v_S^a=v^3\del_z$ is the part tangent to the $(S^1,q_2 l_{ab})$-fiber
to which $p$ belongs.  $v_H^a$ and $v_S^a$ are orthogonal to each
other with respect to the metric $q_{ab}$.  We can also check from the
above form of the metric that the covariant derivative operator $D_a$
associated with $q_{ab}$ can be written in the form
\begin{equation}
  \label{eq:D=hatD+chi}
  D_a=\hat D_a+\ss3a\c3,
\end{equation}
where $\hat D_a$ is the covariant derivative operator associated with
$\dh ab$ and $\c3=\del_z$ is the derivative operator along the circle
fiber. It is easy from this formula to see that the Laplacian
$\Lap_q=q^{ab} D_a D_b$
with respect to $q_{ab}$ is given by the sum
\begin{equation}
  \Lap_q=q_1\inv \Lap_h+q_2\inv (\c3)^2,
\end{equation}
where $\Lap_h=\uh ab\hat D_a\hat D_b$ is the Laplacian with respect to
$\dh ab$. Therefore the product of eigenfunctions for $(\Sigma_g,\dh
ab)$ and for $(S^1,l_{ab})$ becomes an eigenfunction for $(M,q_{ab})$.

Our next step is to generalize our compactification to one for
\textit{spacetimes}. Let $\tilde Y\equiv (\R^4,\udg ab)$ be the
universal covering spacetime of a spatially closed Bianchi III
spacetime $Y\equiv(M\times\R,q_{qb})$. Let $t$ be a time function of
$(\R^4,\udg ab)$ defined in such a way that each $t=\mathrm{constant}$
surface coincides with $(\R^3,\tilde q_{ab})$, where the scale factors
$q_1$ and $q_2$ in $\tilde q_{ab}$ are regarded as functions of
$t$. Then, we assume that the universal covering spacetime metric is
of the synchronous form
\begin{equation}
  \label{eq:cudg}
  \begin{split}
  \udg ab &= -N^2 (dt)_a(dt)_b+\tilde q_{ab} \\
  &= -N^2 (dt)_a(dt)_b+q_1 \tilde h_{ab}+q_2 \tilde l_{ab},
  \end{split}
\end{equation}
where $N=N(t)$ and $(dt)_a\equiv \nabla_a t$. Apparently, $\Isom\tilde
Y=\Isom H^2\times \Isom E^1$. $\Isom\tilde Y$ is also equivalent to
the isometry group of the spatial manifold, $\Isom\tilde Y=\Isom
\tilde M\equiv\Isom (\tilde M,\tilde q_{qb})$. In terminology adopted
in \cite{TKH1,TKH2}, this fact can be rephrased as
$\mathrm{Esom}\tilde M=\Isom \tilde M$, where $\mathrm{Esom}\tilde M$
is the extendible isometry group of $(\R^3,\tilde
q_{ab})$. \footnote{The isometry group of a spacetime with time
  function $t$ has the subgroup whose action retains the foliation by
  $t=\mathrm{constant}$. This subgroup can also be identified with a
  subgroup of the spatial manifold specified by $t=\mathrm{constant}$,
  and called the \textit{extendible isometry group} of this spatial
  manifold \cite{TKH1,TKH2}. Note that in general to compactify a
  spatially homogeneous spacetime $\tilde Y$ to a spatially locally
  homogeneous spacetime $Y$ with spatial manifold $M$, we need to
  embed $\pi_1(M)$ \textit{not} into $\Isom \tilde M$ but into
  $\mathrm{Esom} \tilde M$. (This is apparent if we think of
  $\mathrm{Esom}\tilde M\subset \Isom\tilde Y$.)}  In general,
$n=\dim\Isom\tilde M-\dim \mathrm{Esom}\tilde M$ gives the degrees of
freedom of choosing the velocities of the \Teich\ parameters of $M$ at
an initial surface $t=\mathrm{constant}$ \cite{TKH1,TKH2}. In the
present case where $n=0$ however there is no such freedom, and the
compactification of the spatial manifold straightforwardly generalizes
to the compactification of the spacetime manifold, i.e., our spatially
closed spacetime solution can be represented as
\begin{equation}
  (M\times\R,\dg ab)=(\R^4,\udg ab)/\Gamma,
\end{equation}
where $\Gamma$ is of the form \reff{eq:generalGamma} and acts on $\R^4$
as
\begin{equation}
  \Gamma \cdot (t,\x)=(t,\Gamma\x).
\end{equation}
$\udg ab$ may be assumed to satisfy Einstein's equation on the
universal cover $\R^4$.  We can write the spacetime metric $\dg ab$ as
\begin{equation}
  \dg ab=-N^2(dt)_a(dt)_b+q_{ab}
\end{equation}
in both cases where $(M,q_{ab})$ is orthogonal and non-orthogonal.
It is however only for the orthogonal case that the spatial metric
$q_{ab}$ can be decomposed as Eq.\reff{eq:themetricq}.

The vacuum universal cover metrics $\udg ab$ form a one parameter
family up to isometry.\cite{Pe} On the other hand, $\Gamma$ has $8g-5$
parameters, which are related to the \Teich\ parameters of the spatial
section. The general solution is therefore a $8g-4(=8g-5+1)$ parameter
solution. The orthogonal solution, where
$\Gamma=\Gamma_\mathrm{orth}$, has only $6g-4$ parameters. However,
even if the background solution is of orthogonal type, i.e., even if
the distortion parameters of the background solution are all zero, it
is worth remarking that the distortion parameters can be perturbed off
the zeros (through the zero mode harmonic perturbations).

\section{Wave Equation for the Scalar Field}
\label{apx:scalar}

Consider the scalar wave equation
\begin{equation}
  \ug ab\nabla_a\nabla_b\Psi=0.
\end{equation}

This equation can be derived from the action $I=(-1/2)\int \ug
ab(\nabla_a\Psi)(\nabla_b\Psi) d\mu_g$, from which we can read the
following Lagrangian for our background
\begin{equation}
  L=-\rcp2 Nq_1\sqrt{q_2}\int_M(-\rcp{N^2}(\dot\Psi)^2+q^{ab}
  (D_a\Psi)(D_b\Psi) )d\mu_{q_0}.
\end{equation}
The momentum $p_\Psi$ conjugate to $\Psi$ is
\begin{equation}
  p_\Psi=\frac{\del L}{\del\dot\Psi}=\frac{q_1\sqrt{q_2}}{N}\dot\Psi,
\end{equation}
so the Hamiltonian is given by
\begin{eqnarray}
  H\wa \int p_\Psi\dot\Psi-L \nonumber \\
  \wa \frac{N}{2}\int_M
  \paren{\frac{p_\Psi^2}{q_1\sqrt{q_2}}
    +{q_1\sqrt{q_2}}\,q^{ab}(D_a\Psi)(D_b\Psi)}d\mu_{q_0}.
\end{eqnarray}

Let us expand the wave function using the mode function
$S=S_{\lambda,m}$ as
\begin{equation}
  \Psi=\psi S, \quad p_\Psi=p_\psi S,
\end{equation}
where $\psi=\psi(t)$ and $p_\psi=p_\psi(t)$ are functions of time,
since each mode of this form decouples from the others. Substituting
this form into the above Hamiltonian and performing integration by
parts (and dropping the constant $\int S^2\d\mu_{q_0}=1$), we have a
reduced Hamiltonian
\begin{equation}
  H=\frac{N}{2}
  \paren{\frac{p_\psi^2}{q_1\sqrt{q_2}}
    +{q_1\sqrt{q_2}}(\lambda^2q_1\inv+m^2q_2\inv)\psi^2}.
\end{equation}
The Hamilton equations derived from this are
\begin{equation}
  \begin{split}
    \dot \psi = & \frac{Np_\psi}{q_1\sqrt{q_2}}, \\
    \dot p_\psi = & 
    -{N}{q_1\sqrt{q_2}}(\lambda^2q_1\inv+m^2q_2\inv)\psi,
  \end{split}
\end{equation}
or eliminating $p_\psi$ we have a wave equation for $\psi$,
\begin{equation}
  \label{eq:swaveeq}
  \ddot\psi+\paren{\frac{\dot q_1}{\dq1}+\rcp2
  \frac{\dot q_2}{\dq2}-\frac{\dot N}{N}}\dot\psi
  +N^2(\lambda^2\dq1\inv+m^2\dq2\inv)\psi=0.
\end{equation}

Upon substituting the exact solution \reff{eq:BIIIsol} we obtain an
explicit equation
\begin{equation}
  \label{eq:swaveeqn}
  \ddot\psi+\frac{2t}{\tp\tm}\dot\psi
  +\paren{\lambda^2\rcp{t_+t_-}+m^2\,\frac{t_+^2}{t_-^2}}\psi=0.
\end{equation}
Another practical form is obtained by transforming $\psi\goes
\zeta=(\tp\tm/t)^{1/2}\psi$:
\begin{equation}
  \label{eq:swaveeqn:zeta}
  \ddot\zeta+\frac{1}{t}\dot \zeta +Z_{sw}\zeta=0,
\end{equation}
where the coefficient $Z_{sw}$ of $\zeta$ is
\begin{equation}
  Z_{sw}\equiv \rcp{\tp^2\tm^2}
  \paren{m^2t^2((t+2k)^2+2k^2)-({\rcp4-\lambda^2})t^2
  +{k^2}\paren{km^2(4t+k)-\frac{k^2}{4t^2}-\lambda^2+\frac32}}.
\end{equation}

\textbf{Solutions for the Minkowski background ($k=0$)}

Let us consider the case when $k=0$ (the Minkowski background). In
this case the wave equation \reff{eq:swaveeqn:zeta} becomes
Bessel's equation
\begin{equation}
  \label{eq:swk0zetaeqn}
  \ddot\zeta+\rcp t\dot \zeta
  +\paren{m^2-\frac{1-4\lambda^2}{4\,t^2}}\zeta=0.
\end{equation}
The original field $\psi$ and $\zeta$ are related through
$\psi=\zeta/\sqrt t$. We can express the general solution to the above
equation for $\zeta$ with Bessel functions (when $m\neq0$):
\begin{equation}
  \label{eq:esswk0}
  \zeta= 
  C_1J_{\rcp2{\sqrt{1-4\lambda^2}}}(mt)+
  C_2N_{\rcp2{\sqrt{1-4\lambda^2}}}(mt),
  \quad (m\neq 0)
\end{equation}
where $C_1$ and $C_2$ are integration constants. $J$ and $N$ are,
respectively, the Bessel function of the first kind and that of the
second kind. It is worth noting that when $\lambda=0$ this solution
can be written only with elementary functions:
\begin{equation}
  \zeta= 
  C_1\sqrt{\frac{2}{\pi mt}}{\sin mt}-
  C_2\sqrt{\frac{2}{\pi mt}}{\cos mt}.
  \quad (m\neq 0,\lambda=0)
\end{equation}
The asymptotic form of the solution \reff{eq:esswk0} as $t\goes\infty$
is of our particular interest, which is, as found in any appropriate
formula table, given by
\begin{equation}
  \label{eq:esswk0asy}
  \zeta\sim
  C_1\sqrt{\frac{2}{\pi mt}}
  {\cos\paren{mt-\frac{(1+\sqrt{1-4\lambda^2})\pi}{4}}}+
  C_2\sqrt{\frac{2}{\pi mt}}
  {\sin\paren{mt-\frac{(1+\sqrt{1-4\lambda^2})\pi}{4}}}.
\end{equation}
Hence, in the range $\lambda^2<1/4$ the changes of $\lambda$ result in
changes of phase in the asymptotic form, while for $\lambda^2>1/4$
they change the amplitude for a fixed $t$ (since, e.g.,
$\Re(\cos(a+bi))= \cosh b \cos a$). The decaying rate (of the
amplitude) of the field $\psi=\zeta/\sqrt t$ is, however, universal,
which is given by
\begin{equation}
  \label{eq:scdecratemneq0}
  |\psi|=O(\rcp t). \quad (t\goes\infty, m\neq 0)
\end{equation}

The equation for $m=0$ should be considered separately, but finding
the exact solution is not difficult, which is given by
\begin{equation}
  \label{eq:esswk0n0}
  \zeta=
  \begin{cases}
  C_1t^{\rcp2\sqrt{1-4\lambda^2}}+
  C_2t^{-\rcp2\sqrt{1-4\lambda^2}},
  & (m=0,\lambda^2<1/4) \\
  C_1+
  C_2\log t,
  & (m=0,\lambda^2=1/4) \\
  C_1\cos\paren{\rcp2\sqrt{4\lambda^2-1}\log t}+
  C_2\sin\paren{\rcp2\sqrt{4\lambda^2-1}\log t},
  & (m=0,\lambda^2>1/4)
  \end{cases}
\end{equation}
where $C_1$ and $C_2$ are integration constants. The decaying rate is
therefore
\begin{equation}
  \label{eq:esswk0n0asy}
  |\psi|=
  \begin{cases}
  O(t^{\rcp2(\sqrt{1-4\lambda^2}-1)}),
  & (m=0,\lambda^2<1/4) \\
  O(\frac{\log t}{\sqrt t}),
  & (m=0,\lambda^2=1/4) \\
  O(\rcp{\sqrt t}).
  & (m=0,\lambda^2>1/4)
  \end{cases}
\end{equation}
A critical point appears at $\lambda^2=1/4$ in the present $m=0$ case,
as opposed to the $m\neq0$ generic cases. (We remark that the critical
eigenstate for $\lambda^2=1/4$ does not necessarily exit. Also
$\lambda$ such that $\lambda^2<1/4$ (so-called ``small eigenvalue'')
may not exist. The spectrum varies depending upon the compactification
of the background.)

\section{Electromagnetic equations}
\label{apx:elemag}

Consider the source-free Maxwell equation
\begin{equation}
  \label{eq:Mx1}
  \ug cb\nabla_cF_{ab}=0.
\end{equation}
Here, the electromagnetic tensor $F_{ab}$ is obtained from the vector
potential $A_a$ through
\begin{equation}
  \label{eq:Mx2}
  F_{ab}=\del_aA_b-\del_bA_a.
\end{equation}

In local coordinates Eq.\reff{eq:Mx1} reduces to
\footnote{Greek indices run 0 to 4, and small Latin ones 1 to
  3. $g$ is the determinant of
  $\dg\mu\nu$.}
\begin{eqnarray}
  0\wa \rcp{\sqrt{-g}}
  \del_\nu(\sqrt{-g}\ug\mu\alpha\ug\nu\beta F_{\alpha\beta}) \nnpar
  \wa  \rcp{N\dq1\sqrt{\dq2}\mu_h}
  \del_\nu(N\dq1\sqrt{\dq2}\mu_h\ug\mu\alpha\ug\nu\beta
  F_{\alpha\beta}) \nnpar
  \wa  \frac{\del_0(N\dq1\sqrt{\dq2})}{N\dq1\sqrt{\dq2}}
  \ug\mu i\ug00 F_{i0}
  +\del_0 (\ug\mu i\ug00 F_{i0})
  +\rcp{\mu_h}\del_i(\mu_h\ug\mu\alpha\ug ij F_{\alpha j}). 
  \label{eq:redMx}
\end{eqnarray}

\subsection*{The odd part}
\def\ov{{a}}

For the odd part we can put
\begin{equation}
  A_a=\ov(t)V_a,
\end{equation}
where $V_a\equiv c_m\hat V_a=c_m \epsilon_a{}^b\hat D_b \hat S$.
Equation \reff{eq:redMx} is found to be trivial for $\mu=0$, and
$\mu=3$, so we are left with $\mu=1,2$. In this case,
Eq.\reff{eq:redMx} reduces to \footnote{Capital Latin indices run 1
  and 2. Any hatted object is lowered and raised by the standard
  hyperbolic metric $\dh AB$ and its inverse $\uh AB$.}
\begin{equation}
\begin{split}
  0 =& \frac{\del_0(N\dq1\sqrt{\dq2})}{N\dq1\sqrt{\dq2}}
  \ug AB\ug00 F_{B0}
  +\del_0 (\ug AB\ug00 F_{B0}) \\
  & +\rcp{\mu_h}\del_D(\mu_h\ug AB\ug DC F_{BC})
  +\del_3(\ug AB\ug 33 F_{B3}).
\end{split}
\end{equation}
Using
\begin{eqnarray}
  F_{B0}\wa -\dot a c_m\hat V_B, \\
  F_{B3}\wa m a \bar c_m \hat V_B, \\
  F_{BC}\wa ac_m (\hat D_B\hat V_C-\hat D_C\hat V_B),
\end{eqnarray}
and
\begin{eqnarray}
  \Lap_h \hat V_A\wa -(\lambda^2+1)\hat V_A, \\
  \hat D^B\hat D_A\hat V_B \wa -\hat V_A,
\end{eqnarray}
we obtain the following wave equation:
\begin{equation}
  \ddot a+\paren{\rcp2\frac{\dot q_2}{q_2}-\frac{\dot N}{N}}\dot a
  +N^2(\lambda^2\dq1\inv+m^2\dq2\inv)a=0.
\end{equation}

Upon substituting the exact solution we have
\begin{equation}
  \label{eq:wvodd}
  \ddot a+\frac{2k}{\tp\tm}\,\dot a
  +\paren{\lambda^2\frac{1}{\tp\tm}+m^2\frac{\tp^2}{\tm^2}}a=0.
\end{equation}
To compare with the scalar wave equation we transform the variable as
$a\goes \zeta=a \sqrt{\tm/(\tp t)}$. We have the following equation
for $\zeta$,
\begin{equation}
  \ddot\zeta+\rcp t\dot \zeta+Z_{oe}\zeta=0,
\end{equation}
where
\begin{equation}
  \begin{split}
  Z_{oe} &\equiv \rcp{\tp^2\tm^2}
  \bigg\{ m^2t^2((t+2k)^2+2k^2)-({\rcp4-\lambda^2})t^2 \\
  &\hspace{4em}
  +{k}\paren{ 2t(1 + 2k^2m^2)- \frac{k^3}{4t^2}
    -k(\frac{1}{2}+{\lambda}^2)+ k^3m^2 } \bigg\}.
  \end{split}
\end{equation}
It is straightforward to see that \textit{when $k=0$ the above
  equation for $\zeta$ becomes exactly the same as the scalar wave
  equation  \reff{eq:swk0zetaeqn} for $k=0$}. Therefore in particular
the decaying properties for the scalar field and odd electromagnetic
field are exactly the same in this case.

\subsection*{The even part}

The even part of vector potential can be put in the form
\begin{equation}
  A_a=a_0 S (dt)_a+a_1 S_a+a_3\bar S\ss3a,
\end{equation}
where $a_I=a_I(t)$, ($I=0,1$, and $3$). It is straightforward to
calculate the components of the electromagnetic tensor through
Eq.\reff{eq:Mx2}. Noting the basic relations for our eigenfunctions
$\del_A S=\lambda S_A$ and $\chi_3 S=-m\bar S$, we have \footnote{As
  in the odd case, capital Latin indices $A$, $B$  stand for 1 or 2.}
\begin{eqnarray}
  \nonumber
  F_{0A}\wa (\dot a_1-\lambda a_0)S_A \\
  \wb \psi_1 S_A, \\
  \nonumber
  F_{03}\wa (\dot a_3+ma_0)\bar S \\
  \wb \psi_2 \bar S, \\
  \nonumber
  F_{A3}\wa (\lambda a_3+m a_1)\bar S_A \\
  \wb \psi_3 \bar S_A, \\
  F_{AB}\wa 0.
\end{eqnarray}
We have defined field strengths of the electric field ($\psi_1$ and
$\psi_2$) and magnetic field ($\psi_3$) above, which are all functions
of time.  Note that they are constrained by definition to satisfy
\begin{equation}
  \label{eq:relpsi}
  \dot\psi_3=m\psi_1+\lambda\psi_2.
\end{equation}

The strengths $\psi_1$, $\psi_2$, and $\psi_3$ are invariant under the
(mode-decomposed) gauge transformation
\begin{equation}
  A_a\goes A_a+\del_a(fS),
\end{equation}
where $f=f(t)$ is an arbitrary function of time. It is easy to see
this invariance using the following induced transformation:
\begin{equation}
  \begin{split}
    a_0 &\goes a_0+\dot f, \\
    a_1 &\goes a_1+\lambda f, \\
    a_3 &\goes a_3-mf.
  \end{split}
\end{equation}

Next, look at the components of the Maxwell equation \reff{eq:redMx},
which are given by

(i) $\mu=0$:
\begin{equation}
  \label{eq:constrpsi12}
  -\lambda q_1\inv\psi_1+mq_2\inv \psi_2=0.
\end{equation}

(ii) $\mu=A$:
\begin{equation}
  \label{eq:psi1psi3d}
  \dot\psi_1=
  \paren{
    \frac{\dot N}{N}-\frac{\dot q_1}{q_1}
    -\rcp2\frac{\dot q_2}{q_2}}\psi_1
  +\frac{\dot q_1}{q_1}\psi_1-N^2q_2\inv m\psi_3.
\end{equation}

(iii) $\mu=3$:
\begin{equation}
  \dot\psi_2=
  \paren{
    \frac{\dot N}{N}-\frac{\dot q_1}{q_1}
    -\rcp2\frac{\dot q_2}{q_2}}\psi_2
  +\frac{\dot q_2}{q_2}\psi_2-N^2q_1\inv \lambda\psi_3.
\end{equation}

Note that from the two constraints \reff{eq:relpsi} and
\reff{eq:constrpsi12} functions $\psi_1$ and $\psi_2$ are solved in
terms of
$\dot\psi_3$ as
\begin{eqnarray}
  \label{eq:psi12d3}
    \psi_1 \wa (\lambda^2q_1\inv+m^2q_2\inv)\inv
    mq_2\inv\dot\psi_3, \\
  \label{eq:psi12d3'}
    \psi_2 \wa (\lambda^2q_1\inv+m^2q_2\inv)\inv
    \lambda q_1\inv\dot\psi_3.
\end{eqnarray}
Equations \reff{eq:psi12d3} and \reff{eq:psi1psi3d} can virtually be
thought of as an unconstrained gauge-invariant system of Hamiltonian
equations for the variable $\psi_3$ and its ``momentum'' $\psi_1$. If
we eliminate $\psi_1$ we obtain a wave equation
\begin{equation}
  \ddot\psi_3
  -\paren{
    \frac{\dot N}{N}-\frac{\dot q_1}{q_1}
    -\rcp2\frac{\dot q_2}{q_2}
    +\frac{m^2\dot q_1+\lambda^2\dot q_2}{m^2q_1+\lambda^2 q_2}
    }\dot\psi_3
  +N^2(\lambda^2q_1\inv+ m^2q_2\inv)\psi_3 =0.
\end{equation}

If using the exact solution we have
\begin{equation}
  \label{eq:wvpsi3}
  \ddot\psi_3
  +2 \paren{
    \frac{t}{\tp\tm}
    -\frac{m^2\tp +\lambda^2 k \tp^{-2}}{m^2\tp^2+\lambda^2 \tm\tp\inv}
    }\dot\psi_3
  +\paren{\lambda^2\rcp{\tp\tm}+ m^2\frac{\tp^2}{\tm^2}}\psi_3 =0.
\end{equation}
Instead of getting an equation for $\psi_3$, we can also get one
for $\psi_1$, which is
\begin{equation}
  \ddot\psi_1
  +\frac{6k}{\tp\tm} \dot\psi_1
  +\paren{\lambda^2\rcp{\tp\tm}+ m^2\frac{\tp^2}{\tm^2}
    -\frac{4k(t-2k)}{\tp^2\tm^2}}\psi_1 =0.
\end{equation}
Since $\psi_1$ and the ``derivative'' of $\psi_3$ are related with
Eq.\reff{eq:psi12d3}, we can reproduce the equation for $\psi_1$ from
that of $\psi_3$ by differentiating Eq.\reff{eq:wvpsi3} (with a
further conformal transformation). On the other hand,  there is no
function $f(t)$ such that only the conformal transformation
$\psi_3\goes \tilde\psi=f\psi_3$ reproduces the equation for
$\psi_1$. We remark that it is the equation for $\psi_1$ that has a
direct resemblance to the odd equation \reff{eq:wvodd}, since they
coincide when $k=0$.

\section*{Acknowledgments}

The authors gratefully acknowledge support for this research from the
Erwin Schr\"odinger International Institute for Mathematical Physics
(ESI), the American Institute of Mathematics (AIM) and the Albert
Einstein Institute (AEI) which provided their facilities and
hospitality while much of the work was carried out. They also
acknowledge support from the Japan Society for the Promotion of
Science (JSPS) and from the National Science Foundation through its
grant PHY-0098084 to Yale University.

\end{document}